\documentclass[11pt,a4paper]{article}

\usepackage[T1]{fontenc}
\usepackage[latin1]{inputenc}
\usepackage[nosort]{cite}
\usepackage{color}
\usepackage[bulletsep]{collref}
\usepackage{graphicx}
\usepackage{bbm}
\usepackage{amsmath}
\usepackage{amssymb}
\usepackage{subfig}
\usepackage{verbatim}
\usepackage{multirow}
\usepackage{tikz}
\usepackage{amsthm}
\usepackage{fancyhdr}
\usepackage{wrapfig}
\usepackage{hyperref}

\fancypagestyle{firststyle}
{
\lfoot{}
\cfoot{}
\rfoot{}
\rfoot{
\raisebox{0mm}[0mm][0mm]{
\begin{picture}(0,0)
\put(0.0,-1.0){\includegraphics[scale=0.1]{fig\thepage.png}}
\end{picture}}}
}

\makeatletter \@addtoreset{equation}{section} \makeatother

\makeatletter
\let\old@startsection=\@startsection
\let\oldl@section=\l@section
\renewcommand{\@startsection}[6]{\old@startsection{#1}{#2}{#3}{#4}{#5}{#6\mathversion{bold}}}
\renewcommand{\l@section}[2]{\oldl@section{\mathversion{bold}#1}{#2}}
\makeatother

\makeatletter
\let\old@makecaption=\@makecaption
\def\@makecaption{\small\old@makecaption}
\makeatother

\let\oldPhi=\Phi
\let\oldPsi=\Psi
\let\oldGamma=\Gamma
\let\oldDelta=\Delta
\let\oldSigma=\Sigma
\let\oldTheta=\Theta
\let\oldPi=\Pi
\let\oldUpsilon=\Upsilon
\renewcommand{\Phi}{\mathnormal{\oldPhi}}
\renewcommand{\Psi}{\mathnormal{\oldPsi}}
\renewcommand{\Gamma}{\mathnormal{\oldGamma}}
\renewcommand{\Sigma}{\mathnormal{\oldSigma}}
\renewcommand{\Delta}{\mathnormal{\oldDelta}}
\renewcommand{\Theta}{\mathnormal{\oldTheta}}
\renewcommand{\Pi}{\mathnormal{\oldPi}}
\renewcommand{\Upsilon}{\mathnormal{\oldUpsilon}}
\newcommand{\ft}[2]{{\textstyle\frac{#1}{#2}}}

\newcommand{\superN}{\mathcal{N}}

\newcommand{\Lagr}{\mathcal{L}}

\newcommand{\tr}{\mathop{\mathrm{tr}}}

\newcommand{\trans}{{\scriptscriptstyle\mathrm{T}}}

\newcommand{\Integers}{\mathbb{Z}}
\newcommand{\Reals}{\mathbb{R}}
\newcommand{\Complex}{\mathbb{C}}

\newcommand{\Sphere}{\mathrm{S}}  % {\mathbbm{S}}
\newcommand{\AdS}{\mathrm{AdS}}

\makeatletter
\newlength{\apb@width}
\newcommand{\autoparbox}[2][c]{\settowidth{\apb@width}{#2}\parbox[#1]{\apb@width}{#2}}
\newcommand{\includegraphicsbox}[2][]{\autoparbox{\includegraphics[#1]{#2}}}
\makeatother
\ifx\genfrac\sdflkaj

\else

\fi
\newcommand{\sfrac}[2]{{\textstyle\frac{#1}{#2}}}
\newcommand{\half}{\sfrac{1}{2}}

\newcommand{\matr}[2]{\left(\begin{array}{#1}#2\end{array}\right)}
\newcommand{\alg}[1]{\mathfrak{#1}}
\newcommand{\grp}[1]{\mathrm{#1}}

\newcommand{\grSU}{\grp{SU}}

\newcommand{\grSL}{\grp{SL}}

\newcommand{\lrbrk}[1]{\left(#1\right)}
\newcommand{\bigbrk}[1]{\bigl(#1\bigr)}

\newcommand{\biggsbrk}[1]{\biggl[#1\biggr]}

\newcommand{\comm}[2]{[#1,#2]}

\newcommand{\lrabs}[1]{\left|#1\right|}
\newcommand{\abs}[1]{{|#1|}}

\newcommand{\lreval}[1]{\left.#1\right|}

\newcommand{\nn}{\nonumber}

\newcommand{\earel}[1]{\mathrel{}&\hspace{-2\arraycolsep}#1\hspace{-2\arraycolsep}&\mathrel{}}
\newcommand{\eq}{\earel{=}}

\def\[{\begin{equation}}
\def\]{\end{equation}}

\makeatletter
\def\mr@ignsp#1 {\ifx\:#1\@empty\else #1\expandafter\mr@ignsp\fi}%
\newcommand{\multiref}[1]{\begingroup%\let\protect\string%
\xdef\mr@no@sparg{\expandafter\mr@ignsp#1 \: }%
\def\mr@comma{}%
\@for\mr@refs:=\mr@no@sparg\do{\mr@comma\def\mr@comma{,}\ref{\mr@refs}}%
\endgroup}
\makeatother

\newcommand{\hypref}[2]{\ifx\href\asklfhas #2\else\href{#1}{#2}\fi}
\newcommand{\Secref}[1]{Section~\multiref{#1}}
\newcommand{\secref}[1]{section~\multiref{#1}}

\newcommand{\appref}[1]{appendix.~\multiref{#1}}

\newcommand{\tabref}[1]{table~\multiref{#1}}

\newcommand{\figref}[1]{figure~\multiref{#1}}
\renewcommand{\eqref}[1]{(\multiref{#1})}

\ifx\href\asklfhas\newcommand{\href}[2]{#2}\fi

\newcommand{\comma}{\quad,\quad}
\newcommand{\unit}{\mathbf{1}}

\newcommand{\eps}{\varepsilon}
\newcommand*{\diff}{{\mathrm d}}

\newcommand{\be}{\begin{eqnarray}}
\newcommand{\ee}{\end{eqnarray}}

\newcommand{\Dbizz}{\mathcal{D}}

\newcommand{\delC}{\widehat\delta}
\newcommand{\master}{\widehat\delta}

\newcommand{\chargeC}{\mathbb{J}}
\newcommand{\levz}{\mathrm{J}}

\newcommand{\chargeY}{\mathrm{J}}

\newcommand{\chargeV}{\mathbb{J}}

\begin{document}

\thispagestyle{firststyle}

\begin{flushright}\footnotesize
%\texttt{arXiv:xxxx.xxxx}\\
\texttt{HU-EP-16/30}%
\end{flushright}
\vspace{1cm}

\begin{center}%
{\huge\textbf{\mathversion{bold}%
Nonlocal Symmetries, Spectral Parameter and Minimal Surfaces in AdS/CFT
}\par}

\vspace{1.2cm}

\large \textsc{Thomas Klose, Florian Loebbert, Hagen M\"{u}nkler} \vspace{8mm} \\
\large\textit{%
Institut f\"{u}r Physik, Humboldt-Universit\"{a}t zu Berlin, \\
Zum Gro{\ss}en Windkanal 6, 12489 Berlin, Germany
} \\

\texttt{\\ \{klose,loebbert,muenkler\}@physik.hu-berlin.de}

%%%%%%%%
\par\vspace{10mm}

\includegraphicsbox{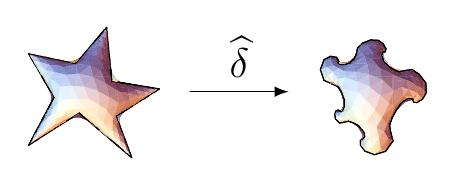}
\vspace{10mm}

\textbf{Abstract} \vspace{5mm}

\begin{minipage}{12.4cm}

We give a general account of nonlocal symmetries in symmetric space models and their relation to the AdS/CFT correspondence. In particular, we study a master symmetry which generates the spectral parameter and acts as a level-raising operator on the classical Yangian generators. The master symmetry extends to an infinite tower of symmetries with nonlocal Casimir elements as associated conserved charges. We discuss the algebraic properties of these symmetries and establish their role in explaining the recently observed one-parameter deformation of holographic Wilson loops. Finally, we provide a numerical framework, in which discretized minimal surfaces and their master symmetry deformation can be calculated.

\end{minipage}
\end{center}

%%%%%%%%%%%%%%%%%%%%%%%%%%%%%%%%%%%%%%%%%%%%%%%%%%%%%%%%%%%%%%%%%%%%%%%%%%%
%%%%%%%%%%%%%%%%%%%%%%%%%%%%%%%%%%%%%%%%%%%%%%%%%%%%%%%%%%%%%%%%%%%%%%%%%%%
\newpage

\tableofcontents

\bigskip
\noindent\hrulefill
\bigskip

%%%%%%%%%%%%%%%%%%%%%%%%%%%%%%%%%%%%%%%%%%%%%%%%%%%%%%%%%%%%%%%%%%%%%%%%%%%
%\newpage
\section{Introduction}

The conjectural duality between type IIB string theory on $\mathrm{AdS}_5\times \mathrm{S}^5$ and $\superN=4$ super Yang--Mills (SYM) theory in four dimensions is arguably one of the most important discoveries of modern theoretical physics \cite{Maldacena:1997re}. It relates a four-dimensional quantum gauge theory to gravity, and thus gives hope for a deep understanding of the connection among all fundamental interactions. While gauge theories feature prominently in the description of phenomenological results, the framework underlying the string side of this duality rests on the theoretical concept of symmetric coset models. The latter are particularly popular with regard to the AdS/CFT duality, though the study of generic coset models by itself provides a beautiful arena for exploring the interconnection between symmetry and physics. It is the aim of this paper to better understand the symmetries of symmetric space models and their relation to the AdS/CFT correspondence.
\medskip

The Lie algebra symmetry of the Lagrangian of $\superN=4$ SYM theory and the isometry algebra of the super-string background $\mathrm{AdS}_5\times \mathrm{S}^5$ are both given by the superconformal algebra $\alg{psu}(2,2|4)$ \cite{Beisert:2010kp}.
Despite this large amount of symmetry, the two theories involved in the above gauge/gravity duality are highly nontrivial. This makes explicit computations difficult and generically only possible at leading orders in perturbation theory. The duality becomes computationally most accessible in the planar limit, where the string and gauge theory are believed to be completely integrable. 
The mathematical structure underlying this integrability is provided by the Yangian algebra which enhances the manifest $\alg{psu}(2,2|4)$ Lie algebra symmetry to a quantum group \cite{Beisert:2010jq, Torrielli:2011gg, Spill:2012qe,Loebbert:2016cdm}. The Yangian is well known to be the basic concept for the description of the class of rational integrable models. As such it provides a typical feature of integrable quantum field theories in two dimensions, where it is realized via nonlocal conserved charges. The two-dimensional nature of the Yangian becomes apparent in the classical (worldsheet) string theory describing the AdS/CFT duality in the limit of large 't~Hooft coupling~$\lambda$.  Generically, and in particular at weak coupling, this symmetry reveals itself via nonlocal Ward identities realized on various classes of observables.
In the case of the AdS/CFT duality, the Yangian also has an alternative formulation via the combination of the ordinary and the so-called dual superconformal symmetry \cite{Drummond:2008vq,Drummond:2010km}. The latter was found in the context of the duality between scattering amplitudes and Wilson loops \cite{Alday:2007hr,Brandhuber:2007yx,Drummond:2008aq} which serves as the motivation to introduce a dual space of coordinates. The existence of this dual formulation is another curious feature of the AdS/CFT correspondence which appears to be tightly related to its planar integrability.
\medskip

Wilson loops represent a particularly interesting type of observable in the above context --- not only because of their duality to scattering amplitudes. In particular, they form part of an intriguing observation lying at the heart of the AdS/CFT correspondence \cite{Maldacena:1998im,Rey:1998ik}: The renormalized area $A_\text{ren}$ of a minimal surface in $\mathrm{AdS}_5$ (a string worldsheet) encodes the strong-coupling expectation value of a gauge theory Wilson loop on the four-dimensional boundary of this space. That is, for large 't~Hooft coupling $\lambda\gg 1$ one finds the relation
\begin{equation}\label{eq:MWL}
\left \langle W(\gamma) \right \rangle \simeq \exp { \left[-\sfrac{\sqrt{\lambda}}{2 \pi}  A_{\mathrm{ren}}(\gamma) \right] }.
\end{equation}
The so-called Maldacena--Wilson loop entering this equation is defined on the boundary contour $\gamma$ of the associated minimal surface and takes the form
\begin{equation}\label{eq:MaldaWL}
W(\gamma)=\frac{1}{N}\tr \mathcal{P} e^{i\oint_\gamma d\sigma \big[A_\mu(x)\dot x^\mu+\Phi_i(x)\abs{\dot x}n^i\big]}.
\end{equation}
Here $A_\mu$ and $\Phi_{i=1,\dots,6}$ denote the gauge field and scalars of $\superN=4$ SYM theory with gauge group $\grp{SU}(N)$, respectively, and the $n^i$ correspond to coordinates on $\mathrm{S}^5$ that obey $\vec{n}^2=1$. Since its strong-coupling description is simply given by the string action in $\mathrm{AdS}_5$, the Maldacena--Wilson loop inherits all symmetries of the underlying sigma model. In particular, integrability manifests itself in the fact that the Wilson-loop expectation value \eqref{eq:MWL} at strong coupling is invariant under a representation of the Yangian generators \cite{Muller:2013rta,Munkler:2015gja}. The form of these generators will be rederived below from the underlying model.
In the past, Wilson loops in $\superN=4$ SYM theory have been an origin for great insights and computational progress, in particular via methods from the integrability toolbox, see e.g.\ \cite{Alday:2009dv,Alday:2010vh,Alday:2010ku,Drukker:2012de,Basso:2013vsa,Toledo:2014koa}.
\medskip

Let us now come to a curious observation which, extending the discussion of \cite{Klose:2016uur}, we aim to understand in the present paper. Restricting the string coset model to the subspace $\mathrm{AdS}_3$, it was observed in \cite{Ishizeki:2011bf} that the above Wilson loop expectation value is invariant under a one-parameter family of deformations, see also \cite{Kruczenski:2013bsa,Kruczenski:2014bla,Cooke:2014uga,Dekel:2015bla,Huang:2016atz}. This one-parameter transformation deforms the contour $\gamma$ into a contour $\gamma_u$ and thus modifies the shape of the associated minimal surface. At the same time, it leaves the surface area invariant. The degree of freedom associated with this one-parameter family is in fact the spectral parameter $u$, which enters the standard integrability framework describing the model. Notably, the above observation shows that in this case the spectral parameter is not merely an auxiliary quantity but parametrizes physical string worldsheets. What is the symmetry behind this family of solutions? Is it related to the Yangian?
\medskip

In principle, the spectral parameter is not needed to see the integrability of a field theory in two dimensions. Given a flat and conserved current $j$, the procedure introduced by Br{\'e}zin, Itzykson, Zinn--Justin and Zuber (for short BIZZ) makes it possible to obtain an infinite tower of conserved charges as follows \cite{Brezin:1979am}: 
Introducing a covariant derivative $\Dbizz=\diff- j$, one may define a conserved current $j^{(n+1)}=\Dbizz \chi^{(n)}$ of level $n+1$, provided the auxiliary potential $\chi^{(n)}$ is associated with a conserved current $j^{(n)}=\ast \diff \chi^{(n)}$ of level $n$. The above flat and conserved Noether current $j$ represents the first element of this tower of conserved currents and thus the start of the induction; see \secref{sec:ssm} for more details. Integration then yields an infinite tower of conserved (nonlocal) charges
\begin{equation*}
\levz^{(n)}\simeq\int \ast j^{(n)}.
\end{equation*}
We demonstrate in \secref{sec:Algebra} that in the case of symmetric space models the Poisson-algebra of these charges forms the classical analogue of the Yangian.
An alternative way to obtain the above Yangian charges employs the spectral parameter $u$ which enters the definition of the Lax connection
$\ell_u = \frac{u}{1+u^2} \big (u\, j +  \ast j \big)$. Expanding the monodromy for this Lax connection then yields the charges $\levz^{(n)}$, which are conserved for appropriate boundary conditions. Here, the spectral parameter represents an auxiliary quantity that allows us to efficiently package the conserved Yangian charges into the monodromy. 
\medskip

In the present paper we demonstrate how the above mentioned deformation of holographic Wilson loops, the potentials $\chi^{(n)}$ entering the BIZZ procedure, and the spectral parameter are related to each other.
We will see that in the case of symmetric coset models the  \emph{a priori auxiliary} potentials of the BIZZ procedure in fact induce a tower of \emph{physical} symmetries. In particular, the potential $\chi^{(0)}$ represents the generator of the spectral parameter, which may be lifted from the auxiliary role in the definition of the Lax connection to a degree of freedom parametrizing a one parameter-family of physical solutions of the group-valued field $g(\tau,\sigma)$ defined on the worldsheet:
\begin{equation}
\master: g\stackrel{\chi^{(0)}}{\longrightarrow}g_u.
\end{equation}
As such, this symmetry induces a one-parameter family of holographic Wilson loops, wheras the area of the associated minimal surfaces or the Wilson loop expectation value, respectively, is independent of the spectral parameter. Let us emphasize that for a generic integrable field theory in two dimensions, the auxiliary BIZZ potentials $\chi^{(n)}$ define a recursion for the conserved charges encoding the integrability. In the particular case of symmetric space models, this recursion turns into a physical master symmetry of the equations of motion, which is parametrized by the spectral parameter. In particular, this master symmetry acts as a raising operator for the Yangian charges, i.e.\ we have $\master\, \levz^{(n)}=\levz^{(n+1)}$. Curiously, the conserved charge assoicated to the symmetry $\master$ is given by the (nonlocal) quadratic Casimir of the underlying Lie algebra.
\medskip

As discussed in more detail in \secref{sec:prevlit}, the above one-parameter family of transformations has been observed in various different settings. In particular, Eichenherr and Forger found this transformation in the context of symmetric space models as early as 1979 and referred to it as a \emph{dual transformation} \cite{Eichenherr:1979ci}. Since then this or related symmetry transformations have been discussed by several authors, see e.g.\ \cite{Schwarz:1995td,Schwarz:1995af,Beisert:2012ue}. It seems worthwhile to further elaborate on this curious nonlocal symmetry and to better understand its role within the AdS/CFT correspondence.

\medskip
This paper is organized as follows:
In \secref{sec:ssm} we review the setup of symmetric space models which underlies large parts of this work. In \secref{sec:master} we introduce the nonlocal master transformation that generates the spectral parameter of the model and furnishes the central symmetry we wish to understand here. We continue by demonstrating in \secref{sec:nonlocalsymm} how this master symmetry can be used to construct the tower of Yangian symmetries from the basic Lie algebra symmetry as well as a tower of Virasoro-like symmetries building on the master symmetry itself. We discuss the algebra of the various symmetry variations in \secref{sec:Algebra}, and demonstrate explicitly that the Poisson-algebra of the nonlocal charges building on the Lie algebra symmetry of symmetric space models corresponds to a classical Yangian. We then come to the explicit application of the above symmetries to holographic Wilson loops in \secref{sec:WilsonLoops}; in particular we show how the master symmetry generates the spectral-parameter transformation observed in the past. In  \secref{sec:prevlit} we briefly comment on the connection to some of the previous literature. \Secref{sec:numerics} is concerned with a numerical approach to the computation of minimal surfaces which allows us to evaluate the master symmetry on nontrivial examples. We end with a brief conclusion and outlook in \secref{sec:conclusions}.

%%%%%%%%%%%%%%%%%%%%%%%%%%%%%%%%%%%%%%%%%%%%%%%%%%%%%%%%%%%%%%%%%%%%%%%%%%%
\section{Symmetric Space Sigma Models}
\label{sec:ssm}

This section is to introduce the models under consideration and to define our notations. We are interested in open strings or minimal surfaces on symmetric coset spaces $\mathrm{M} = \grp{G}/\grp{H}$ such as, e.g.\
\begin{align}
&\AdS_3 \simeq\frac{\grp{SO}(2,2)}{\grp{SO}(1,2)}
\, ,
&
 & \AdS_5\times \Sphere ^5 \simeq \frac{\grp{SU}(2,2)\times\grp{SU}(4)}{\grp{SO}(1,4)\times\grp{SO}(5)} \, .
\end{align}
In particular, we will discuss the explicit example of Euclidean anti-de-Sitter space $\mathrm{EAdS}_5\simeq  \frac{\grp{SO}(1,5)}{\grp{SO}(5)}$ in \secref{sec:cosetAdS5}. For the moment, we consider the properties that are common to all of these models. 

\paragraph{Symmetric spaces.}

In the study of symmetric spaces it is convenient to describe them as general homogeneous spaces $\mathrm{M} = \grp{G}/\grp{H}$ which allow for a $\Integers_2$ grading. Via this grading, the algebra $\alg{g} = \mathfrak{h} \oplus \mathfrak{m}$ of the group $\grp{G}$ splits into an even and an odd part respectively, such that
\begin{align}
  \left[\mathfrak{h} \, , \, \mathfrak{h} \right] &\subset \mathfrak{h} \, , 
&
   \left[\mathfrak{h} \, , \,   \mathfrak{m} \right] &\subset \mathfrak{m} \, , 
  &
   \left[\mathfrak{m} \, , \, \mathfrak{m} \right] &\subset \mathfrak{h} \, .
\end{align}
The grading is defined with respect to an involutive automorphism $\Omega: \alg{g} \to \alg{g}$ according to
\begin{align}
  \alg{h} &= \left \lbrace \mathrm{g} \in \alg{g} \, \vert \, \Omega(\mathrm{g}) =  \mathrm{g} \right \rbrace \, , &
  \alg{m} &= \left \lbrace \mathrm{g} \in \alg{g} \, \vert \, \Omega(\mathrm{g}) = -\mathrm{g} \right \rbrace \, .
\end{align}

We denote the map from the worldsheet $\Sigma$ to the target space $\mathrm{M}$ by $g = g(\tau,\sigma) \in\grp{G}$ and the associated Maurer--Cartan current by $U = g^{-1}\diff g \in \alg{g}$. The $\alg{h}$ and $\alg{m}$ parts of this current are defined as the projections
\begin{align}
A &:= P_{\alg{h}}(U) = \half \bigbrk{ U + \Omega(U) } \, , 
&
a &:= P_{\alg{m}}(U) = \half \bigbrk{ U - \Omega(U) } \, , 
&
U &= A + a \, . \label{eqn:Ucomp}
\end{align}
The Maurer--Cartan current is flat by construction, $\diff U + U \wedge U = 0$, which implies that
\begin{align}
\diff A + A \wedge A +  a \wedge a &= 0 \, , 
&
 \diff a + a \wedge A +  A \wedge a &= 0 \, . \label{Uflatcomp}
\end{align}
The action is given by\footnote{The worldsheet has Euclidean signature and we use complex worldsheet coordinates $z = \sigma + i \tau$, which implies $\partial = \frac{1}{2} (\partial_\sigma - i \partial_\tau)$, $h_{z\bar{z}} = h_{\bar{z}z} = \frac{1}{2}$ and $h^{z\bar{z}} = h^{\bar{z}z} = 2$. Hodge duality acts as $*\diff\sigma = \diff\tau$ and $*\diff\tau = - \diff\sigma$, or $*\diff z = -i \diff z$ and $*\diff\bar{z} = i\diff\bar{z}$. This implies $(*a)_z = -i a_z$, $(*a)_{\bar{z}} = i a_{\bar{z}}$ and $*a\wedge b = - a\wedge *b$.
}
\begin{align}
S = \int \tr \left( a \wedge \ast a \right) = \int \diff^2 \sigma \, \sqrt{h} \, h^{\alpha\beta} \tr \left(a_\alpha a_\beta \right) = 2i \int \diff z \wedge \diff \bar{z} \tr \left( a_z a_{\bar{z}} \right) \; , \label{Action}
\end{align}
where  we have introduced the worldsheet metric~$h$ and the last version is written in conformal gauge. Varying the action with respect to $g$ gives the equations of motion. As the trace is invariant under the automorphism $\Omega$, i.e.\ 
$\tr ( \mathrm{g}_1\mathrm{g}_2 ) = \tr ( \Omega(\mathrm{g}_1) \Omega(\mathrm{g}_2))$, we have $\tr( A_\alpha a_\beta ) = 0$ for $A_\alpha\in\alg{h}$ and $a_\beta\in\alg{m}$. This simplifies the derivation of the equations of motion as it implies that $\delta \tr \left( a \wedge \ast a \right) = 2 \tr \left( \delta U \wedge \ast a \right)$. Now using that $\delta U = \diff ( g^{-1} \delta g ) + [ U, g^{-1} \delta g ]$, we obtain
\begin{align}
\delta S = 2 \int \tr \left[ \lrbrk{ \diff ( g^{-1} \delta g ) + [ U, g^{-1} \delta g ] } \wedge \ast a  \right]
         = - 2 \int \tr \left[ \lrbrk{ \diff \ast a + \ast a \wedge U + U \wedge \ast a } g^{-1} \delta g \right] \, .    
\end{align}
Inserting $U = A+a$, we see that the $a$-part drops out, leaving us with the equations of motion
\begin{align}
\diff \ast a + \ast a \wedge A + A \wedge \ast a &= 0 \, . \label{EOM}
\end{align}

Varying with respect to the worldsheet metric, we get the Virasoro constraints, which in conformal gauge read
\begin{align}
\tr \left(a_z a_z \right) = \tr \left(a_{\bar{z}} a_{\bar{z}} \right) &= 0 \label{Virasoro} \, .
\end{align}
The action is invariant under local gauge transformations given by right multiplication of $g$ by $R(\tau,\sigma)\in \grp{H}$. The components of the Maurer--Cartan current transform as
\begin{align*}
&A \;\;\mapsto\;\; R^{-1} A R + R^{-1} \diff R \, , 
&
 &a \;\;\mapsto\;\; R^{-1} a R \, ,
\end{align*}
such that the action is invariant, because $A$ does not enter. Global $\grp{G}$-symmetry acts by left multiplication of $g$ by a constant element $L\in \grp{G}$. The current $U$ and hence its components $A$ and $a$ are invariant since $g^{-1}\diff g \mapsto g^{-1}L^{-1}\diff(Lg) = g^{-1}\diff g$.
\paragraph{Flat Noether current.}
 In order to derive the Noether current associated to this symmetry, we consider the infinitesimal transformation 
 \begin{equation}\label{eq:Liesymmetry}
 \delta g = \epsilon g
 \end{equation} 
 with $\epsilon \in \alg{g}$. The common trick is to pretend that $\epsilon$ depends on the worldsheet coordinates and to read off the Noether current from the variation of the action $\delta S = \int \ast j \wedge \diff\epsilon$. Using that the variation of $g$ implies $\delta U = g^{-1} \diff \epsilon g$,we obtain
\begin{align}
  \delta S = - 2 \int \tr ( \ast a \wedge g^{-1} \diff\epsilon g ) = -2 \int \tr (   g\ast a g^{-1} \wedge  \diff\epsilon).
\end{align}
We can read off the Noether current and the associated charge as
\begin{align} \label{eqn:Noethercurrent}
j &= -2 g a g^{-1} \, , 
&
 \chargeY &= \int \ast j \, .
\end{align}
The Noether current is both conserved and flat. In order to verify these properties, note that from the definition of $U$, we have for some one-form $\omega$:
\begin{align}
\diff ( g \omega g^{-1} ) = g \left( \diff \omega + U \wedge \omega  + \omega \wedge U \right) g^{-1}.
\end{align}
Correspondingly we obtain
\begin{align}
\diff \ast j = -2 g \left( \diff \ast a + A \wedge \ast a + \ast a \wedge A \right) g^{-1} = 0 \, , \label{jcons}
\end{align}
by using the equations of motion \eqref{EOM} and 
\begin{align}
\diff j + j \wedge j = -2 g \left( \diff  a + A \wedge  a +  a \wedge A + 2 a \wedge a - 2 a \wedge a  \right) g^{-1} = 0 \, , 
\end{align}
due to the flatness of $U$. The flatness of the conserved current indicates the model's integrability, since it allows to construct an infinite hierarchy of conserved charges. These charges may be obtained from an iterative procedure introduced by Br{\'e}zin, Itzykson, Zinn--Justin and Zuber (BIZZ) in 1979 \cite{Brezin:1979am}.

In order to briefly review the BIZZ recursion, we begin by defining the covariant derivative $\Dbizz$ which contains the Noether current $j$ as a connection and which acts on (matrix-valued) functions $f$ and one-forms $\omega$ as
\begin{align} \label{eqn:Dbizz}
\Dbizz f &= \diff f - f j \, , &
\Dbizz \, \omega &= \diff \omega + \omega \wedge j \, .
\end{align} 
With these definitions, we have (on-shell)
\begin{align}
\diff \ast \Dbizz f &=  \Dbizz \ast \diff f \, , & \Dbizz \, \Dbizz f &= 0 \, .
\end{align}
Suppose now that we have a conserved current $j^{(n)}$ that can be written as $j^{(n)} = \Dbizz \chi^{(n-1)}$ for some $\chi^{(n-1)}$. As this current is conserved, we may calculate its potential $\chi^{(n)}$ defined by $j^{(n)} = \ast \diff \chi^{(n)}$. From this potential, we define a new current by
\begin{align}
 j^{(n+1)} = \Dbizz \chi^{(n)} \, .
\end{align}
The new current is conserved:
\begin{align}
  \diff \ast j^{(n+1)} = \diff \ast \Dbizz \chi^{(n)} = \Dbizz \ast \diff \chi^{(n)} = \Dbizz j^{(n)} = \Dbizz \, \Dbizz \chi^{(n-1)} = 0 \, . 
\end{align}
In this way, we can construct an infinite set of conserved currents starting from $j^{(0)} = -j$ which can indeed be expressed as $j^{(0)} = \Dbizz \chi^{(-1)}$ with $\chi^{(-1)} = \unit$. The first few currents obtained from this recursion take the form
\begin{align}\label{eq:bizzcurrents}
 j^{(0)} &= -j \, , &
 j^{(1)} &= \ast j - \chi^{(0)} j \, , &
 j^{(2)} &= j + \chi^{(0)} \ast j - \chi^{(1)} j \, , & \ldots.
\end{align}

\paragraph{Lax connection.}

A different method to see the integrability of the model is to introduce a flat Lax connection depending on a spectral parameter $u\in\Complex$, which can also be used to derive an infinite set of conserved charges. One commonly starts out by deforming the Maurer--Cartan form $U$ into a flat connection  using the components $A$, $a$, and $*a$, which satisfy the flatness condition \eqref{Uflatcomp} as well as the equations of motion \eqref{EOM}. The most general deformation that preserves the flatness leads to the Lax current
\begin{align} \label{eqn:definition-Lax}
L_u = A + \frac{1-u^2}{1+u^2} \, a - \frac{2u}{1+u^2} \ast a \, ,
\end{align}
which reduces to the undeformed case by setting $u$ to zero. We note that the $\mathfrak{h}$-part of the Maurer--Cartan current remains unaltered, while the transformation of the $\mathfrak{m}$-part is reminiscent of a worldsheet rotation. This can be made explicit by writing the deformed field as
\begin{align} \label{eqn:deform-a}
a \;\; \mapsto \;\; a_u := \frac{1-u^2}{1+u^2} \, a - \frac{2u}{1+u^2} \ast a = e^{-i\theta} a_z \, \diff z + e^{i\theta} a_{\bar{z}} \, \diff \bar{z} 
\end{align}
using the parametrization
\begin{equation}\label{eq:theta}
e^{i\theta} = \frac{1-iu}{1+iu}.
\end{equation}
Depending on the context, either the $u$- or the $\theta$-parametrization will be more convenient.
It is worth noting that the transformation \eqref{eqn:deform-a} is not an honest rotation, from which it differs in two respects: First, only the $\alg{m}$-valued part of the connection, $a_z$, is transformed and not the $\alg{h}$-valued part $A_z$. Second, if the transformation was induced by a worldsheet rotation $z \mapsto e^{i\vartheta} z$, then the transformation had an extra term 
\begin{equation}
  \delta a_z = -i \vartheta a_z - i \,\vartheta (z\partial - \bar{z}\bar{\partial})a_z \; ,
\end{equation}
as compared to the infinitesimal version of \eqref{eqn:deform-a} given by
\begin{equation}
  \delta a_z = -i\theta a_z \; .
\end{equation}
The additional term is due to the fact that the argument of $a_z(z)$ transforms as well under a rotation.

%%%%%%%%%%%%%%%%%%%%%%%%%%%%%%%%%%%%%%%%%%%%%%%%%%%%%%%%%%%%%%%%%%%%%%%%%%%
\section{Master Symmetry and Spectral Parameter}
\label{sec:master}
In this section, we introduce the \emph{master symmetry}, which maps a solution $g$ of the equations of motion to another solution $g_u$ of the equations of motion. The symmetry is based on the obseration that the action, the equations of motion and the Virasoro constraints remain unaltered upon replacing $U \to L_u$, see also \cite{Eichenherr:1979ci}.
For the action and Virasoro constraints this follows immediately from the relation \eqref{eqn:deform-a}: 
\begin{align}
S_u &= \frac{(1-u^2)^2 + 4 u^2}{(1+u^2)^2} \int  \tr \left( a \wedge \ast a \right) = S \, , &
\tr \left( a_{u,z} a_{u,z} \right) &= e^{-2i \theta} \tr \left( a_{z} a_{z} \right) = 0 \, . 
\end{align}
The invariance of the equations of motion follows by making use of the flatness condition \eqref{Uflatcomp}: 
\begin{align}
\diff \ast a_u + \ast a_u \wedge A_u + A_u \wedge \ast a_u =& \frac{1-u^2}{1+u^2} \left( \diff \ast a + \ast a \wedge A + A \wedge \ast a \right) = 0 \, .
\end{align}
The transformation $U \mapsto L_u$ can be carried over to the fundamental fields $g$ by imposing the differential equation
\begin{align} \label{eqn:def-deformed-sol}
g_u^{-1} \diff g_u = L_u \, , 
\end{align}
along with the boundary condition $g_u (z_0) = g (z_0)$ for some fixed point $z_0$ on the worldsheet. Since $L_u$ is flat, there exists a unique solution for simply connected worldsheets. In the following, we will refer to this transformation as a \emph{master symmetry} due to its property to map conserved charges to conserved charges and to generate infinite towers of nonlocal symmetries of the model; this is explicitly demonstrated below. 

We note that in order to think of the deformation \eqref{eqn:deform-a} as a symmetry transformation of physical solutions, we need to impose a reality condition on $L_u$ which leads to the restriction that $u \in \Reals$.  

\paragraph{The potential $\chi$.}
A convenient description of the transformed solution is obtained by setting 
\begin{equation}
g_u = \chi_u g.
\end{equation}
 As a consequence of \eqref{eqn:def-deformed-sol}, the $\mathrm{G}$-valued function $\chi_u$ that mediates between the original and the transformed solution has to satisfy the equation $\diff \chi_u = \chi_u \ell_u$, where the flat connection $\ell_u$ appearing in this equation is given by
\begin{align}
\ell_u = g L_u g^{-1} - \diff g \, g^{-1} = g (L_u - U) g^{-1} \, .
\end{align}
The flatness of $\ell_u$ follows directly from the flatness of $U$ and $L_u$:
\begin{align*}
\diff \ell_u &= 
\diff g \wedge (L_u-U) g^{-1} + g (\diff L_u- \diff U) g^{-1} -  g (L_u-U) \wedge \diff g^{-1}  \\ 
& = g \left( U \wedge (L_u-U) - L_u \wedge L_u + U \wedge U + (L_u-U) \wedge U \right) g^{-1} \\
& = - g \left( (L_u -U) \wedge (L_u -U) \right) g^{-1} = - \ell_u \wedge \ell_u \, .
\end{align*}
Inserting the explicit form of $L_u$ from \eqref{eqn:definition-Lax}, we see that the Lax connection $\ell_u$ represents a flat deformation of the Noether current $j$:
\begin{equation} \label{eqn:deformed-Noether}
\ell_u = \frac{u}{1+u^2} \big (u\, j +  \ast j \big)\, .
\end{equation}
The boundary condition $g_u (z_0) = g (z_0)$ translates into the condition $\chi_u(z_0) = \unit$ for an arbitrarily chosen base point $z_0$ on the worldsheet. Changing the base point, where the initial condition is imposed, from $z_0$ to $z_1$ corresponds to a global $\mathrm{G}$-transformation of the solution from the left by $\chi_u(z_1)^{-1}$. In summary, we have seen that the transformation described by equation \eqref{eqn:def-deformed-sol} can be rewritten as
\begin{align}
g_u &= \chi_u \, g \, , 
&
\diff \chi_u &= \chi_u \, \ell _u \, ,  
&
\chi_u(z_0) &= \unit \, .
\label{eqn:dchi}
\end{align}
In this form, the transformation was described for general symmetric space models in \cite{Eichenherr:1979ci}, see \secref{sec:prevlit} for a brief description. The quantity $\chi_u$ is in fact the generating function for the BIZZ potentials $\chi^{(n)}$ introduced in \secref{sec:ssm}; this was established in \cite{Wu:1982jt}. Concretely, we have 
\begin{align} \label{eqn:Taylor-chi}
\chi_u = \sum_{n=0}^\infty \chi^{(n-1)} u^n \, .
\end{align}
In order to prove this relation, it is helpful to express the BIZZ recursion as a local recursion relation for the potentials in the form of    
\begin{align}
  \ast \diff \chi^{(n+1)} = \Dbizz  \chi^{(n)} \, . \label{eqn:BIZZ-recursion}
\end{align}
It is then easy to see that the defining equation \eqref{eqn:dchi} implies the above recursion relation. We invert \eqref{eqn:deformed-Noether} to obtain
\begin{align} \label{eqn:jfroml}
j = \ell_u - u^{-1} \ast \ell_u  \quad \Rightarrow \quad \ast \ell_u = u ( \ell_u - j ) \, , 
\end{align}
which shows that $\chi_u$ satisfies
\begin{align}
\ast \diff \chi_u = \chi_u \, \ast \ell_u = u \, \chi_u \left( \ell_u - j \right) = u \, \left( \diff\chi_u - \chi_u j \right) = u \, \Dbizz \chi_u \, .
\end{align}
Expanding this equation in powers of $u$ results in the BIZZ recursion \eqref{eqn:BIZZ-recursion}. Noting that $\ell_{u=0} = 0$ implies $\chi_{u=0} = \unit$ concludes the proof.

The infinitesimal version of the master symmetry transformation is given by
\begin{equation} \label{eqn:C-variation}
\delC g = \lreval{\frac{\diff g_u}{\diff u} }_{u=0} = \lreval{\dot{\chi}_u}_{u=0} g = \chi^{(0)} g \, ,
\end{equation}
where we introduced the dot to denote $u$-derivatives. The symbol $\chi^{(0)}$ represents the coefficient of the linear term in the Taylor expansion \eqref{eqn:Taylor-chi}, which satisfies
\begin{align}
\diff \chi ^{(0)} =  \ast j \, .
\label{eqn:dchi0}
\end{align}
We have thus identified the function $\chi^{(0)}$, which generates the master symmetry as the potential of the G-symmetry Noether current. 

\paragraph{Conserved charge.}
In order to derive the conserved current associated to the master symmetry itself,%
\footnote{Note that the master symmetry represents an on-shell symmetry such that the Noether procedure does not apply in a strict sense. Still we may formally apply it. Further comments on this issue can be found at the end of \secref{sec:gensyms}.}
we utilize the same trick as above and introduce a coordinate-dependent transformation parameter $\rho=\rho(z,\bar{z})$ into the variation, $\delC g = \rho \chi^{(0)} g$, such that we can read off a conserved current $ \mathbbm{j}$ from $\delC S = \int \ast \mathbbm{j} \wedge \diff \rho$. As a first step we need the variation
\begin{align}
\delC \,U = g^{-1} \diff(\rho\chi^{(0)}) g = -2 \rho \ast a + \diff \rho \, g^{-1} \chi^{(0)} g \, ,
\end{align}
where we used \eqref{eqn:dchi0} and \eqref{eqn:Noethercurrent}. The first term leaves the action invariant while the second term produces
\begin{align}
  \delC S =  2 \int \tr ( \diff \rho \, g^{-1} \chi^{(0)} g \wedge \ast a )
          = -2 \int \tr ( g \ast a g^{-1} \chi^{(0)} ) \wedge \diff \rho
          = \int \tr ( \ast j \chi^{(0)} ) \wedge \diff \rho \; ,
\end{align}
such that we find the conserved current 
\begin{equation}\label{eq:bbcurrent}
\mathbbm{j} = \tr \big( j \chi^{(0)} \big) \, . 
\end{equation}
The associated conserved charge takes the form
\begin{equation}\label{eq:chargeC}
\chargeC = \int \ast \mathbbm{j}  = \frac{1}{2} \tr( \chargeY \chargeY )\, ,
\end{equation}
and is recognized as the quadratic Casimir of $\mathrm{G}$. What is interesting about this result is not that $\tr (\chargeY \chargeY)$ is conserved, which is obvious since $\chargeY$ is conserved, but the fact that there is a symmetry transformation that has the Casimir as a conserved charge. 
In the following sections, we will discuss the role of the master symmetry for the algebra of nonlocal symmetries. Before, however, let us discuss the elementary properties of the master transformation, which will prove to be helpful for the later discussion.

\paragraph{Group structure.}
Before turning to the discussion of infinitesimal symmetries, we consider the properties of large master symmetry transformations, in particular the relation to $\grp{G}$-symmetries as well as the concatenation of two master symmetry transformations. For this purpose, it is convenient to denote the map $g \mapsto g_u$ by $g_u = M_u (g)$. 
Let us first note that the master symmetry commutes with the $\grp{G}$-symmetry transformations of the model, i.e.\ we have
\begin{align}
M_u(L \cdot g) = L \cdot M_u(g) \, .
\end{align}
This may be concluded from the fact that the Maurer--Cartan current $U = g^{-1} \diff  g$ is invariant under $g \mapsto L \cdot g$ and from the uniqueness of the solution to \eqref{eqn:def-deformed-sol}, once a boundary condition is specified. 

We can proceed similarly for the concatenation of two master symmetry transformations. The structure is particularly clear in terms of the angle spectral parameter $\theta$ introduced in equation \eqref{eq:theta}. If we take $L_{\theta_1}$ to be the Maurer--Cartan current of a transformed solution $g_{\theta_1}$ and calculate the Lax connection for a different angle $\theta_2$, we obtain the Lax connection $L_{\theta_1 + \theta_2}$. This structure can again be carried over to the solutions $g_\theta$ by using the uniqueness of the solution of the defining relation \eqref{eqn:def-deformed-sol}, for which we correspondingly obtain the relation
\begin{align}
\left( M_{\theta_1} \circ M_{\theta_2} \right) (g) = M_{\theta_1 + \theta_2} (g) \, .
\end{align}
In order to express this relation in terms of the parameter $u$,  
\begin{align}
 M_{u_1} \circ M_{u_2} = M_{u_3} \, ,
\end{align}
we need to solve
\begin{align}
e^{i \theta_1 + i \theta_2} = \frac{1-iu_1}{1+iu_1}\frac{1-iu_2}{1+iu_2} = \frac{1-iu_3}{1+iu_3} = e^{i\theta_3}
\end{align}
to find the following composition rule for the spectral parameter $u$:
\begin{align} \label{eqn:addition-theorem}
u_3 = u_1 \oplus u_2 = \frac{u_1+u_2}{1-u_1 u_2} \; .
\end{align}
From this rule, we obtain a formula for the variation of the deformed solution $g_u$,
\begin{align} \label{eqn:addition-theorem-infinitesimal}
  \master g_u = \lreval{ \frac{\diff}{\diff u'} \, g_{u\oplus u'} }_{u'=0} 
  = \frac{\diff g_u}{\diff u} \lreval{ \frac{\diff}{\diff u'} \frac{u+u'}{1-u u'} }_{u'=0} 
  = (1+u^2) \frac{\diff}{\diff u} \, g_u \; .
\end{align}
This translates into an expression for the variation of $\chi_u$ under $\delC$ according to 
\begin{align}
 \master \left( \chi_u \, g \right) = (1+u^2) \frac{\diff}{\diff u} \, \left( \chi_u \, g \right)
 \qquad \Rightarrow \qquad 
 \delC \chi_u = \left(1 + u^2 \right) \frac{\diff}{\diff u} \chi_u - \chi_u \cdot \chi^{(0)} \, . \label{eqn:delchi}
\end{align}

%%%%%%%%%%%%%%%%%%%%%%%%%%%%%%%%%%%%%%%%%%%%%%%%%%%%%%%%%%%%%%%%%%%%%%%%%%%
\section{Integrable Completion via Master Symmetry}
\label{sec:nonlocalsymm}
As outlined in \secref{sec:ssm}, symmetric space models are integrable and feature an infinite number of conserved charges. Different methods exist which allow to obtain this tower of charges. One way is provided by the above BIZZ procedure \cite{Brezin:1979am} that  consists of a recursion built on a flat and conserved current $j$. Equivalently, infinitely many charges may be obtained as the Taylor coefficients in the spectral-parameter expansion of the monodromy over the Lax connection $\ell_u$ \cite{Wu:1982jt}, see \secref{sec:master}. 
In the case of symmetric space models, however, there exists another option, namely to employ the master symmetry discussed in \secref{sec:master} to construct towers of conserved charges \cite{Eichenherr:1979ci}. We derive the relation between these charges and the ones obtained by the BIZZ procedure below. Moreover, we demonstrate explicitly that the master symmetry can be employed to deform \emph{any symmetry variation} $\delta_0$ into a one-parameter family of symmetries $\delta_{0,u}$. Suppose we are given some symmetry variation $\delta_0$ with associated conserved current~$j_0$. 
\begin{enumerate}
\item We may obtain a tower of conserved currents and corresponding charges by iterative application of the master symmetry $\master$ to the conserved current $j_0$.  In this way, the master symmetry induces a one-parameter family of conserved currents $j_{0,u}$ associated with the spectral parameter $u$. 
\item Analogously, the corresponding symmetry variation $\delta_0$ may be deformed into a one-parameter family of symmetries $\delta_{0,u}$, with associated conserved current $\bar{j}_{0,u}$.
\end{enumerate}
We will demonstrate that the resulting one-parameter families of conserved currents $j_{0,u}$ and symmetry transformations $\delta_{0,u}$ are formally related by the Noether procedure:
\begin{center}
\includegraphicsbox{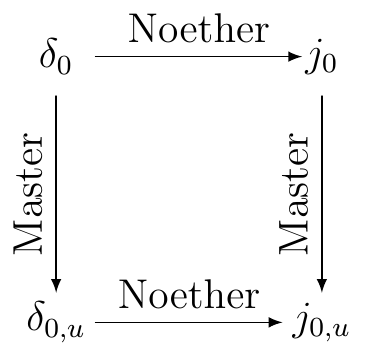}
\end{center}
Since the parameter $u$ generated by the master symmetry is the spectral parameter underlying the integrability of the model, we refer to this procedure as the \emph{integrable completion} of the symmetry $\delta_0$ and its associated current $j_0$, respectively.
In particular, this procedure applies to the master symmetry itself, which results in a one-parameter family of master symmetries $\master_u$ with associated Casimir charges of the G-symmetry introduced above.

%%%%%%%%%
\subsection{Generic Symmetries}
\label{sec:gensyms}

In this subsection we introduce the notion of the \emph{integrable completion} of a symmetry. We show that the integrable completion of symmetry variations and conserved charges via the master transformation furnishes again a symmetry or conserved charge, respectively.

\paragraph{Completion of conserved currents and charges.}
The symmetry variation of a Noether current must also be a conserved current. This is due to the fact that the equations of motion are invariant under the variation and that the conservation of the Noether current is equivalent to the equations of motion. In particular this holds for the master variation $\master$ applied to any conserved current $j_0$, i.e.\ we have the conservation equation
\begin{equation}
\diff \ast \master j_0=0,
\end{equation} 
which implies that the charge $\levz_0=\int  \ast \master j_0$ associated with the current $\master j_0$ is time-independent. We may also apply a large master transformation to the respective currents and charges by making the replacement $g \to g_u=\chi_u g$ (cf.\ \eqref{eqn:dchi}) in the definition of any current $j_0$ or charge $\levz_0$, respectively:
\begin{align}\label{eq:onepjJ}
j_0\to j_{0,u}&=j_0|_{g\to g_u},
&
\levz_0\to \levz_{0,u}&=\levz_0|_{g\to g_u}.
\end{align}
Hence, the master symmetry may be employed to obtain the one-parameter families \eqref{eq:onepjJ} of conserved currents and charges, respectively. We will refer to these as the integrable completions of the current $j_0$ or the charge $\levz_0$, respectively.
\paragraph{Completion of symmetry variations.}
We prove the general statement that for any given symmetry $\delta_0 g$ of the field $g$, the variation $\delta_{0,u}$ defined by 
\begin{equation}\label{eq:defvariation}
\delta_{0,u} g = \chi_u^{-1} \delta_0 ( \chi_u g)
\end{equation}
 is a symmetry as well. This allows us to turn any symmetry $\delta_0$ into a one-parameter family $\delta_{0,u}$ of symmetries. 

Note that the quantity $\chi_u$ is inherently on-shell since its definition requires the Lax connection $\ell_u$ to be flat, which is equivalent to the equations of motion. Consequently, we are not in a position to study the invariance of the action. Instead we show that the variation \eqref{eq:defvariation} provides a symmetry of the equations of motion by demonstrating that 
\begin{align}
  \diff \ast \delta_{0,u} j = 0 \, .
\end{align}
Here, $j$ is the $\grp{G}$-symmetry Noether current.

\paragraph{General symmetry criterion.}
We begin by deriving a necessary and sufficient criterion for a given variation $\delta g$ to be a symmetry. We define $\eta = \delta g\, g^{-1}$ so that we can write the variation as $\delta g = \eta g$ and the induced change of the Maurer--Cartan current as $\delta U = g^{-1} \diff \eta \, g$. In order to calculate the variation of the Noether current, it is convenient to write it in the form (cf.\ \eqref{eqn:Ucomp})
\begin{align}
j = - 2 g P_{\alg{m}}(U) g^{-1} = g \bigbrk{\Omega(U) - U } g^{-1} \, .
\end{align}
Since $\Omega$ is a linear map on $\alg{g}$, its action and the variation commute and we find
\begin{align}
\delta j =  -\diff \eta - [j, \eta ] + g \, \Omega \bigbrk{ g^{-1} \diff \eta \, g } g^{-1} \, .
\label{eqn:delta-j}
\end{align} 
Hence, we have a symmetry of the equations of motion if and only if
\begin{align}
\diff \ast \bigbrk{ \diff \eta + [j, \eta ] } = \diff \bigbrk{ g \, \Omega \bigbrk{ g^{-1} \ast \diff \eta \, g } g^{-1} } \, .
\label{eqn:step}
\end{align}
In order to rewrite this, it is helpful to note that since $ \omega_1 \wedge \omega_2 + \omega_2 \wedge \omega_1$ leads to a commutator and $\Omega$ is an involutive automorphism on $\alg{g}$, we can rewrite e.g. 
\begin{align*}
U \wedge \Omega \bigbrk{ g^{-1} \ast \diff \eta \, g } 
+ \Omega \bigbrk{ g^{-1} \ast \diff \eta \, g } \wedge U
= \Omega \left( \Omega(U) \wedge g^{-1} \ast \diff \eta \, g 
+ g^{-1} \ast \diff \eta \, g  \wedge \Omega(U) \right) \, .
\end{align*}
In this way, the right hand side of equation \eqref{eqn:step} can be expressed as
\begin{align*}
 \diff \bigbrk{ g \, \Omega \bigbrk{ g^{-1} \ast \diff \eta \, g } g^{-1} } &=
 g\, \Omega \bigbrk{ g^{-1} \lrbrk{ \diff \ast \diff \eta + j \wedge \ast \diff \eta + \ast \diff \eta \wedge j } g} g^{-1} \\
 & =  g\, \Omega \bigbrk{ g^{-1} \diff \ast \lrbrk{ \diff \eta + \left[ j , \eta \right]  } g} g^{-1}.
\end{align*}
Thus we find the condition
\begin{align}
 g^{-1} \diff \ast \lrbrk{ \diff \eta + \left[ j , \eta \right]  } g
 =  \Omega \bigbrk{ g^{-1} \diff \ast \lrbrk{ \diff \eta + \left[ j , \eta \right]  } g} \, ,
\end{align}
which states that
\begin{align}
  g^{-1} \diff \ast \lrbrk{ \diff \eta + \left[ j , \eta \right]  } g \in \alg{h} \, .
 \label{eqn:criterion}
\end{align}
This is the sought after necessary and sufficient condition for $\delta g = \eta g$ to be a symmetry of the model. In most of the cases (but not all) that we consider, this condition is actually satisfied in the form
\begin{align}
  \diff \ast \lrbrk{ \diff \eta + \left[ j , \eta \right] } = 0 \, .
 \label{eqn:criterion-strong}
\end{align}
%%%

\paragraph{Integrable completion is a symmetry.}
With this criterion at hand, we return to the variation \eqref{eq:defvariation} given by $\delta_{0,u} g = \chi_u^{-1} \delta_0 ( \chi_u g)$. By Leibniz' rule we obtain
\begin{equation}
  \delta_{0,u} g = \delta_0 g + (\chi_u^{-1} \delta_0 \chi_u ) g \, .
\end{equation}
Hence, the total variation splits into a part $\delta_0 g$, which is a symmetry  by assumption, and the part $\delta'_{0,u} g = \eta g$ with $\eta = \chi_u^{-1} \delta_0 \chi_u$. We now demonstrate that also this second part is a symmetry by showing that \eqref{eqn:criterion-strong} is satisfied. To this end, we calculate
\begin{equation} \label{eqn:conjug-deta}
  \diff \eta = \diff (\chi_u^{-1} \delta_0 \chi_u) = -\chi_u^{-1} \diff \chi_u \, \chi_u^{-1} \delta_0 \chi_u + \chi_u^{-1} \delta_0 \diff \chi_u = \delta_0 \ell_u + [\eta,\ell_u] \, ,
\end{equation}
where we used $\diff \chi_u = \chi_u \ell_u$ as given in \eqref{eqn:dchi}. Then, it follows that
\begin{equation}\label{eq:zwischenschritt}
  \diff \eta + [ j , \eta ] = \delta_0 \ell_u + [\eta,\ell_u-j] = \delta_0 \ell_u + \frac{1}{u} [\eta,\ast\ell_u]  \, ,
\end{equation}
using $\ast \ell_u = u ( \ell_u - j )$ in the last step. Taking the divergence of this equation yields the terms
\begin{equation} \label{eqn:conjug-test1}
  \diff \ast \lrbrk{ \diff \eta + [ j , \eta ] }
   = \diff \ast \delta_0 \ell_u - \frac{1}{u} \diff [\eta,\ell_u]
   = \diff \ast \delta_0 \ell_u - \frac{1}{u} \lrbrk{ \diff\eta\wedge\ell_u + \ell_u\wedge\diff\eta } - \frac{1}{u} [\eta,\diff\ell_u].
\end{equation}
Within the middle terms on the right hand side of \eqref{eqn:conjug-test1}, we replace $\diff\eta$ again with the help of \eqref{eqn:conjug-deta}, which gives
\begin{align}\label{eq:expthree}
  \diff\eta\wedge\ell_u + \ell_u \wedge \diff\eta & = \delta_0 \ell_u \wedge \ell_u + \ell_u \wedge \delta_0 \ell_u + [\eta,\ell_u] \wedge \ell_u + \ell_u \wedge [\eta,\ell_u] \nn \\
  &= \delta_0 \lrbrk{ \ell_u \wedge \ell_u } + [\eta,\ell_u \wedge \ell_u ] 
   = - \delta_0 \, \diff \ell_u - [\eta,\diff\ell_u ]\,.
\end{align}
Using once more that $\ast \ell_u = u ( \ell_u - j )$, we then find
\begin{align}
\diff \ast \lrbrk{ \diff \eta + [ j , \eta ] } = \diff \, \delta_0 \left( \ast \ell_u + \frac{1}{u} \, \ell_u \right) = \frac{1}{u} \, \diff \ast \delta_0 j = 0 \, .
\end{align}
Thus, the condition \eqref{eqn:criterion-strong} is satisfied and the integrable completion \eqref{eq:defvariation} of a symmetry variation $\delta_0$ indeed furnishes a one-parameter family of symmetry variations.

%%%

\paragraph{Noether procedure and on-shell symmetries.}

We would now like to derive conserved charges, which are associated to the nonlocal symmetry transformations discussed above.  Let us point out that due to the definition of $\chi_u$ all of the higher symmetry transformations are inherently on-shell. Carrying out Noether's procedure strictly would require to continue the symmetry variations to off-shell symmetries of the action. This was done in \cite{Dolan:1980kz,Hou:1981hn} for the Yangian-type symmetries of principal chiral models and it seems plausible that it could also be done for symmetric space models. Here, we will be satisfied with deriving on-shell expressions for conserved currents. 
Let us clarify, how these currents are related to the currents one would derive from a (hypothetical) off-shell continuation of the underlying symmetry via Noether's procedure.   

Suppose we had found a way to extend the symmetry transformations discussed above off-shell. This would involve finding off-shell expressions for the quantity $\chi_u$, which can e.g.\ be done as in \cite{Dolan:1980kz,Hou:1981hn} by fixing specific paths from any point on the worldsheet to a common starting point and by defining $\chi_u$ to be the solution of $\diff \chi_u = \chi_u \ell_u$ along this path. This implies that the continued $\chi_u$ satisfies
\begin{align*}
\diff \chi_u = \chi_u \ell _u + f_u \, , 
\end{align*}
where $f_u$ is some one-form which vanishes on-shell. By assumption, the variation of the Lagrangian can be written as a total derivative, $\delta \Lagr = \diff \ast k $. Hence, $k$ represents the contribution to the Noether current which would follow from the off-shell symmetry.

Since we do not have an off-shell continuation of the above symmetries at hand, we simply perform a formal calculation where we use $\diff \chi_u = \chi_u \ell_u$, but we will not use the equations of motion otherwise. For the symmetries we consider, one can show that in this way we obtain $\delta \Lagr = \diff \ast k^\prime$. That is, $k^\prime$ represents the on-shell contribution to the current following from our formal procedure.
It is then clear that $\diff \ast \left( k - k ^\prime \right)$ will be proportional to $f_u$, which vanishes on-shell. Hence, $\left( k - k ^\prime \right)$ vanishes up to the usual freedom in reading off $\ast k$ from $\diff \ast k$. We thus see that the conserved current we derive agrees on-shell with the Noether current one would find if one had continued the symmetry off-shell and carried out Noether's procedure. We note that also the conserved current associated to the variation $\delC$ is derived in this way in \secref{sec:master}. 

%%%%%%%%%
\subsection{Yangian Symmetry}

In this subsection we discuss the first nontrivial example of the above integrable completion via the master symmetry. In particular, we demonstrate that the completion of the Lie algebra symmetry \eqref{eq:Liesymmetry} of symmetric space models yields a tower of nonlocal Yangian symmetries. 

%%%

\paragraph{Completion of conserved currents and charges.}
Applying the master variation to the conserved current $j$ introduced in \eqref{eqn:Noethercurrent}, we find%
\footnote{Note that the difference to the level-1 current given in \eqref{eq:bizzcurrents} merely results from a re-ordering of symmetries, see also the paragraph at the end of this subsection.}
\begin{align}
  \delC j = -2 \ast j + [\chi^{(0)}, j ],
\end{align}
which is indeed a nonlocal conserved current. The corresponding charge 
\begin{align}
\chargeY^{(1)} := \int \ast \delC \,j =2 \int j + \int \limits_{\sigma_1 < \sigma_2} [\ast j_1, \ast j_2 ] 
\end{align}
takes the standard form of a level-1 Yangian charge. In \secref{sec:Algebra} we will demonstrate that these charges indeed obey the Yangian commutation relations. Higher conserved currents and charges can be constructed by repeated application of $\delC$. However, since we know the large transformation generated by $\delC$, we need not carry out this cumbersome procedure. Exponentiating the variation $\delC$ essentially transforms $g$ into $g_u$ and all derived quantities like the Noether current $j$ transform accordingly. The higher charges to be constructed from $\chargeY$ by repeated application of $\delC$ should thus be contained in the one-parameter family of conserved charges obtained from the transformed solutions~$g_u$: 
\begin{align}
\label{def:chargeY}
\chargeY_u &= \int \ast j_u \, ,  
&
j_u &= - 2 g_u \, a_u \, g_u ^{-1} \, . 
\end{align}
A more precise relation can be established by applying equation \eqref{eqn:addition-theorem-infinitesimal} to obtain a recurrence relation for the coefficients of the Taylor expansion of $\chargeY_u$. Since $\Omega$ is a linear map on $\alg{g}$ the variation $\delC$ acts on $a_u = P_\alg{m} ( g_u ^{-1} \diff g_u) $ in the same way as on $g_u$, such that 
\begin{equation}
\delC \,\chargeY_u= (1+u^2) \frac{\diff}{\diff u} \chargeY_u \, .
\end{equation}
The relation takes a simpler form for the angle spectral parameter $\theta$ introduced in \eqref{eq:theta} via the relation $e^{i\theta} = \frac{1-iu}{1+iu}$. In terms of this parameter, the relation reads
\begin{align}
\master \,\levz_\theta=\frac{\diff }{\diff \theta}\levz_\theta  \, ,
\end{align}
which also makes it manifest that the master symmetry generates the spectral parameter. Defining the charges $\chargeY^{(n)}$ to be the coefficients in the Taylor expansion 
\begin{align*}
\chargeY_\theta = \sum \limits _{n=0} ^{\infty}  \frac{\theta^n}{n!} \, \chargeY ^{(n)} \, ,
\end{align*}
we find the recurrence relation
\begin{align}
\master\, \levz^{(n)}=\levz^{(n+1)}.
\end{align}

%%%%

\paragraph{Completion of symmetry variations.}

It is in general not straightforward to obtain nonlocal symmetries associated to nonlocal charges, as they are non-linearly generated and are hence not determined by the Poisson brackets via the ordinary symplectic action. However, the integrable completion of the symmetry variation $\delta_{\epsilon}$ provides a natural candidate for the symmetry variations associated to the charges discussed above. The integrable completion of the symmetry variation $\delta_\epsilon$ is given by
\begin{equation}\label{eq:Yangvars}
\delta_{\epsilon,u}  \, g := \chi_u ^{-1} \delta_\epsilon \left( \chi_u \, g \right) 
= \chi_u ^{-1} \epsilon \chi_u \, g 
= \eta _{\epsilon,u} \, g .
\end{equation}
Similar variations were considered in \cite{Schwarz:1995td}, see \secref{sec:prevlit} for more details on the relation to the present work. We will now show that the Noether charges associated to the above symmetry transformations \eqref{eq:Yangvars} are related to the conserved charges $\chargeY_u$ up to a $u$-dependent factor.  

%%%%%

\paragraph{Noether procedure for Yangian symmetry.}

We turn to the derivation of the conserved current for the nonlocal symmetries of Yangian type, which are given by \eqref{eq:Yangvars}.
As for the derivation of the $\grp{G}$-symmetry Noether current, we allow $\epsilon$ to vary over the worldsheet. We thus find the variation of the Maurer--Cartan current to be
\begin{align}
\delta_{\epsilon,u} U 
= g^{-1} \left( \left[ \eta_{\epsilon,u} , \ell_u \right] + \chi_u ^{-1} \diff \epsilon \chi_u \right) g\,.
\end{align}
For the variation of the action we obtain
\begin{align}
\delta_{\epsilon,u} S 
&= \int \tr \left( \ast j \wedge \left( \left[\eta_{\epsilon,u} , \ell_u \right] + \chi_u ^{-1} \diff \epsilon \, \chi_u \right) \right) 
= - \int \tr \left( \left( \ast j \wedge \ell_u + \ell_u \wedge \ast j \right) \eta_{\epsilon,u} - \chi_u  \ast j \,  \chi_u ^{-1} \wedge \diff \epsilon \right) \nonumber \\
& =  - \int \tr \left( \frac{2 u}{1+u^2} ( j \wedge j ) \, \eta_{\epsilon,u} - \chi_u  \ast j \,  \chi_u ^{-1} \wedge \diff \epsilon \right) \nonumber\\
&= \int \tr \left( \frac{2 u}{1+u^2} \left(\chi_u \,  \diff j \,  \chi_u ^{-1}  \right) \epsilon + \chi_u  \ast j \,  \chi_u ^{-1} \wedge \diff \epsilon \right) , 
\end{align}
where we have inserted the explicit expression \eqref{eqn:deformed-Noether} for $\ell_u$. Note now that 
\begin{align}
\diff \left( \chi_u \,  j \,  \chi_u ^{-1} \right) 
= \chi_u \left( \ell_u \wedge j + j \wedge \ell_u + \diff j \right) \chi_u ^{-1} 
= \frac{1-u^2}{1+u^2} \, \chi_u \,  \diff j \,  \chi_u ^{-1} \, .   
\end{align}
Consequently we have
\begin{align}
\delta_{\epsilon,u}S = \int \tr \left( \frac{2u}{1-u^2} \, \diff \left( \chi_u \,  j \,  \chi_u ^{-1} \right) \, \epsilon + \chi_u  \ast j \,  \chi_u ^{-1} \wedge \diff \epsilon \right) \, ,
\end{align}
and dropping boundary terms, we find
\begin{align}
\delta_{\epsilon,u}  S = \int \tr \left( \chi_u \left(\ast j + \frac{2u}{1-u^2} \, j \right) \chi_u ^{-1} \wedge \diff \epsilon \right).
\end{align}
The Noether current associated to the nonlocal symmetry $\delta_{\epsilon,u} $ is thus given by
\begin{align}
\bar j_u = \chi_u \left( j - \frac{2u}{1-u^2} \, \ast j \right) \chi_u ^{-1} \, .
\label{eqn:jY}
\end{align}
Comparing this expression with the $\grp{G}$-symmetry Noether current of the transformed solution, 
\begin{align}
j_u = - 2 g_u a_u g_u ^{-1} = \chi_u \left( \frac{1-u^2}{1+u^2} \, j - \frac{2u}{1+u^2} \, \ast j \right) \chi_u ^{-1} \, , 
\end{align}
we find the relation
\begin{align}
\bar j_u = \frac{1+u^2}{1-u^2} \, j_u \, .
\end{align}
The symmetry variations associated to the conserved charges $\chargeY_u$ are thus given by
\begin{align}
\bar \delta_{\epsilon,u} g = \frac{1-u^2}{1+u^2} \, \chi_u ^{-1} \epsilon \chi_u  \, g =\frac{1-u^2}{1+u^2} \delta_{\epsilon,u}g\, .
\end{align}

\paragraph{Relation to BIZZ charges.} As explained in the beginning of this section, the charges $\chargeY_u$ discussed above differ from the charges derived from the BIZZ procedure. However, since both charges are based on the quantity $\chi_u$, one should expect that the charges are related to each other. In fact, such a relation can be established using the variation $\delC$. The BIZZ charges are given by the monodromy over the Lax connection $\ell_u$, which amounts to considering the quantity $\chi_u$ over a closed contour. The relation of these charges to those derived from $j_u$ is given by
\begin{align}
\left(1 + u^2 \right) \dot{\chi}_u (z, \bar{z}) \chi_u ^{-1} (z, \bar{z}) = \int \limits _{z_0} ^z \ast j_u \, . \label{eqn:chiandjt}
\end{align}
Note first that the two sides of this equation are the same for $u=0$ since $\diff \chi^{(0)}= \ast j$. In order to prove equality for any value of $u$, we can employ the variation $\delC$ to construct a recurrence relation for the Taylor coefficients on either side of equation \eqref{eqn:chiandjt}. We have already seen that 
\begin{align}
\delC \, j_u = \left( 1 + u^2 \right) \frac{\diff}{\diff u} \, j_u 
\end{align}
and a simple application of equation \eqref{eqn:delchi} shows that 
\begin{align*}
\delC \left[ \left(1 + u^2 \right) \dot{\chi}_u  \chi_u ^{-1} \right] = \left(1 + u^2 \right) \frac{\diff}{\diff u} \left[ \left(1 + u^2 \right) \dot{\chi}_u  \chi_u ^{-1} \right] \, , 
\end{align*}
which proves the relation \eqref{eqn:chiandjt} and hence shows that the charges $\chargeY^{(n)}$ carry the same information as those obtained from the BIZZ procedure.

%%%%%%%%%
\subsection{Master Symmetry}
\label{sec:completionmaster}

Here we consider the second nontrivial example of an integrable completion via the master symmetry. In fact, we apply the completion to the master variation $\master$ itself, yielding a one-parameter family of master transformations $\master_u$ and associated charges $\chargeC_u$.

%%%
\paragraph{Completion of conserved currents and charges.}
As in the case of the Lie algebra and Yangian charges discussed above, we can proceed for the Noether current associated to the symmetry $\delC$ itself. Acting with $\master$ on the current $\mathbbm{j}$ of \eqref{eq:bbcurrent} gives the conserved master current of level one:
\begin{align}
\master \, \mathbbm{j} = 
	\tr \big( j \big( 2 \chi^{(1)} - \chi^{(0)\, 2} \big) - 2 \ast j \chi^{(0)}  \big) \, .
\end{align}
The structure of the conserved quantities, however, turns out to be more transparent if one considers the charges directly. Acting with $\master$  on $\chargeV$ defined in \eqref{eq:chargeC} gives the conserved master charge of level one:
\begin{align}
\label{eq:lev1master}
\chargeV ^{(1)} = \tr \big( \chargeY \, \chargeY ^{(1)} \big) \, . 
\end{align}
Employing a large master transformation provides the generating function 
\begin{equation}\label{eq:masterchargeu}
\chargeV_u = \sfrac{1}{2} \tr \big( \chargeY_u \chargeY_u \big) \, ,
\end{equation}
and switching to the angle spectral parameter $\theta$, we again find the relation
\begin{align*}
\delC \chargeV_\theta = \frac{\diff}{\diff \theta} \chargeV_\theta \, 
\end{align*}
such that the Taylor coefficients $\chargeV ^{(n)}$ of $\chargeV_\theta$ satisfy the recurrence relation
\begin{align}
\delC \, \chargeV ^{(n)} = \chargeV ^{(n+1)} \, .
\end{align} 

\begin{table}
\renewcommand{\arraystretch}{1.4}
\setlength{\tabcolsep}{1.45mm}
\begin{tabular}{| l || c | c || c | c |}\cline{2-5}
\multicolumn{1}{c||}{}&\multicolumn{2}{c||}{Yangian}&\multicolumn{2}{c|}{Master}\\\cline{2-5}
\multicolumn{1}{c||}{}&Variation &Charge&Variation &Charge\\\hline\hline
Level-0 & $\delta_\epsilon g =\epsilon g$& $\levz^{(0)}\equiv \levz= \int \ast j$ & $\master g=\chi^{(0)} g$&$\chargeV ^{(0)} =\half \tr \big( \chargeY \, \chargeY  \big)$
\\\hline
Level-1 &$\delta_\epsilon^{(1)}g=\comm{\epsilon}{\chi^{(0)}}g$ &$\levz^{(1)} =2 \int j + \int [\ast j_1, \ast j_2 ] $&\footnotesize$\master^{(1)} g=[2\chi^{(1)}-(\chi^{(0)})^2]g$ &$\chargeV ^{(1)} = \tr \big( \chargeY \, \chargeY ^{(1)} \big)$
\\\hline
Completion &$\delta_{\epsilon,u} g =  \chi_u ^{-1} \epsilon \chi_u \, g  $&$\levz_u= \int \ast j_u$ &$\master_ug=\chi_u ^{-1} \dot{\chi}_u \, g$&$\chargeV_u = \sfrac{1}{2} \tr \big( \chargeY_u\chargeY_u\big)$
\\\hline
\end{tabular}
\caption{Overview of Yangian and master symmetries with $j_u=-2g_u \, a_u \, g_u ^{-1}$ and $j=j_0$.} 
\label{tab:towers}
\end{table}

%%%%

\paragraph{Completion of symmetry variations.}

 In analogy to \eqref{eq:Yangvars} for the Lie algebra symmetry, we conjugate the master symmetry $\master$ with $\chi_u$ and define the variation
\begin{equation}
\master_u g := \frac{1}{1+u^2} \, \chi_u ^{-1} \master \left( \chi_u \, g \right) =  \chi_u ^{-1} \dot{\chi}_u \, g \, . \label{eq:mastervars}
\end{equation}
Again, similar variations were considered in \cite{Schwarz:1995td}, cf.\ \secref{sec:prevlit}. We will now establish the relation of these symmetry variations to the conserved charges discussed above by deriving the conserved charges associated to the variations \eqref{eq:mastervars}. 

%%%

\paragraph{Noether procedure.}
We derive the Noether current associated to the one-parameter family of master symmetries:
\begin{align}
\delC_{u} g = \rho \, \eta _u \, g = \rho\,\chi_u^{-1} \dot{\chi}_u \, g \, .
\end{align}
Here, we have again introduced a coordinate-dependent transformation parameter $\rho$ in order to derive the Noether current. Making use of $\delC U = g^{-1} (  \diff \eta_u  \rho + \eta_u \diff \rho ) g$ we find the variation of the action to be
\begin{align}
\delC S = \int \left \lbrace \tr \left( \ast j \wedge \diff \eta _u \right) \rho + \tr \left( \ast j \eta _u \right) \wedge \diff \rho\right \rbrace \, .
\end{align}
By using that $\diff \eta _u= [ \eta _u , \ell ] + \dot{\ell}$, we can recast
\begin{align*}
\tr \left( \ast j \wedge \diff \eta_u\right) 
= - \tr \left( \left( \ast j \wedge \ell + \ell \wedge \ast j \right) \eta_u \right) + \tr \left( \ast j \wedge \dot{\ell} \right) 
= \frac{2u}{1+u^2} \tr \left( \diff j \, \eta_u + \frac{1}{1+u^2} \ast j \wedge j \right) \, ,  
\end{align*}
and comparing with
\begin{align*}
\diff \tr \left( j \eta_u \right) = \frac{1-u^2}{1+u^2} \, \tr \left( \diff j \, \eta_u+ \frac{1}{1+u^2} \, \ast j \wedge j \right) 
\end{align*}
yields
\begin{align}
\delC S = \int \left \lbrace \frac{2u}{1-u^2} \, \diff \tr \left( j \, \eta_u \right) \rho+ \tr \left( \ast j \, \eta_u \right) \wedge \diff\rho \right \rbrace 
= \int \tr \left[ \left( \ast j + \frac{2u}{1-u^2} \, j \right) \eta_u \right] \wedge \diff \rho \, .
\end{align}
From this we read off the Noether current 
\begin{align}
\bar{\mathbbm{j}}_u = \tr \left( \bar{j}_u \, \dot{\chi}_u \chi_u^{-1} \right)
= \frac{1+u^2}{1-u^2} \tr \left( j_u \, \dot{\chi}_u \chi_u^{-1} \right) \, .
\end{align}
By virtue of equation \eqref{eqn:chiandjt}, we conclude that
\begin{align}
\bar{\mathbbm{j}}_u \left( z , \bar{z} \right) = \frac{1}{1-u^2} \tr \left( j_u \left( z , \bar{z} \right) \int \limits _{z_0} ^z \ast j_u \right) \, , 
\end{align}
such that the Noether charge is identified as
\begin{align}
\bar{\mathbb{J}}_u = \frac{1}{1-u^2} \, \chargeV_u \, .
\end{align}
Note that $\master_u$ yields the charge $\chargeV_u$ up to a $u$-dependent factor.

%%%%%%%%%
\subsection{Spacetime Symmetry}

The principle to conjugate a known symmetry $\delta_0$ of the model with $\chi_u$ can also be applied to spacetime or worldsheet symmetries. A general spacetime symmetry $\delta_0=\delta_\text{ST}$ is described by
\begin{align}
\delta_\text{ST} \,g = b^\alpha (\tau , \sigma) \partial_\alpha g \, ,
\end{align}
where the specific form of $b^\alpha$ depends on the chosen spacetime symmetry.
The conjugation with $\chi_u$ then leads to the variations
\begin{align}
\label{eqn:conjvar}
\delta_{\text{ST},u} \,g = b^\alpha l_{u , \alpha} g = b^\alpha  \left( \frac{u^2}{1+u^2} \, j_\alpha + \frac{u}{1+u^2} \, \eps_{\alpha \beta} \, h^{\beta \delta} j_\delta \right) g \, ,
\end{align}
where for convenience we write the currents $j$ and $*j$ in terms of their components.
Note now that
\begin{align}
j_\alpha \cdot g = - 2 g P_\alg{m} \left( g^{-1} \partial_\alpha g \right) g^{-1} g = - 2 \partial_\alpha g + 2 g P_\alg{h} \left( g^{-1} \partial_\alpha g \right) \, .
\end{align}
We thus observe that the variations \eqref{eqn:conjvar} are merely $u$-dependent linear combinations of the original spacetime symmetries and gauge transformations.

%%%%%%%%%%%%%%%%%%%%%%%%%%%%%%%%%%%%%%%%%%%%%%%%%%%%%%%%%%%%%%%%%%%%%%%%%%%
\section{Symmetry Algebra}
\label{sec:Algebra}
In this section, we discuss the algebra of symmetries introduced above. We explicitly evaluate the commutators of symmetry variations as well as the Poisson algebra of the conserved charges. 

%%%

\subsection{Algebra of Yangian and Master Variations}
We provide the commutation relations for the variations $\delta_{\epsilon, u}$ and $\master_u$ given in \eqref{eq:Yangvars} and \eqref{eq:mastervars}, respectively. The commutation relations of similar nonlocal symmetries were derived by Schwarz in \cite{Schwarz:1995td} and we follow the methods explained there. Note, however, that Schwarz' discussion does not include the symmetry $\master\equiv\master^{(0)}$.

Let us begin by considering two generic variations $\delta_1$, $\delta_2$, which are the infinitesimal variations associated to some transformations $g \mapsto F_i ^{t} (g) = f_i ^{t} (g) \cdot g $ ($i=1,2$). Taking $t$ to be infinitesimal, we have the variations
\begin{align*}
\delta_i  g = \dfrac{\diff}{\diff t} \, F_i ^t (g) \big \vert _{t=0} =  \dfrac{\diff}{\diff t} \, f_i ^t (g) \big \vert _{t=0} \cdot g = \phi_i(g) \cdot g \, .
\end{align*}
The concatenation of two such variations is given by
\begin{align*}
\delta_1 \left( \delta_2 g \right) 
= \dfrac{\diff}{\diff t_2} \left( \phi_1 \left( f_2 ^{t_2}(g) \cdot g \right) \cdot f_2 ^{t_2}(g) \cdot g \right)_{t_2=0} 
= \phi_1(g) \cdot \phi_2(g) \cdot g + \left(\delta_2 \, \phi_1 \right) \cdot g \, , 
\end{align*}
and hence we have the commutator
\begin{align}
\left[ \delta_1 \, , \, \delta_2 \right] g = \left( \left[ \phi_1  ,  \phi_2 \right] + \delta_2 \phi_1 - \delta_1 \phi_2 \right) g \, .
\label{eqn:comm_generic}
\end{align}
The variations 
$\delta_{\epsilon, u} g = \chi_u ^{-1} \epsilon \chi_u g = \eta _{\epsilon, u} g $  
and $\master_u g =  \chi_u ^{-1} \dot{\chi}_u g  $ 
are constructed from the solution $\chi_u$ to the auxiliary linear problem
\begin{align}
\diff \chi_u &= \chi_u \, \ell_u \, , &
\chi_u ( z_0 , \bar{z}_0 ) &= \unit \, .  
\label{eqn:auxiliary-problem}
\end{align}
The key step in deriving their commutation relations is to compute the variation of $\chi_u$, which is achieved by solving the varied auxiliary linear problem \eqref{eqn:auxiliary-problem}:
\begin{align}
\chi_u ^{-1} \left( \diff \left( \delta  \chi_u \right) - \left( \delta  \chi_u \right) \ell_u \right) &= \delta   \ell_u \, , &
\delta  \chi_u (z_0, \bar{z}_0) &= 0 \, . 
\label{eqn:varied-auxiliary-problem}
\end{align}
A detailed derivation of the variations of $\chi_u$ can be found in appendix \ref{app:Commutators}. Here, we state the result for the variation $\delta_{\epsilon, u}$:  
\begin{align}
\delta_{\epsilon, u_1} \, \chi_{u_2} = \frac{u_2}{u_1-u_2} \left( \chi_{u_2} \eta _{\epsilon, u_1}  - \epsilon \chi_{u_2} \right) + \frac{u_1 u_2}{1+ u_1 u_2} \left( \chi_{u_2} \tilde{\eta} _{\epsilon, u_1}  - \epsilon ^\prime \chi_{u_2} \right) \, ,
\end{align}
where we defined  
\begin{align*}
\tilde{\eta} _{\epsilon, u} &= g \, \Omega \left( g^{-1} \eta _{\epsilon, u} g \right) g^{-1} \, , &
\epsilon ^\prime &= g_0 \, \Omega \left( g_0 ^{-1} \epsilon g_0 \right) g_0 ^{-1} \, , &
g_0 &= g ( z_0 , \bar{z}_0 ) \, .
\end{align*}
Once the solutions to the varied auxiliary linear problem \eqref{eqn:varied-auxiliary-problem} are found, the commutation relations can be computed from equation \eqref{eqn:comm_generic}. The first term in \eqref{eqn:comm_generic} is cancelled by the other terms and the commutator can be expressed in terms of the known variations up to gauge transformations. As an example, consider the variation
\begin{align}
\tilde{\eta} _{\epsilon, u} \, g = g \, \Omega \left( g^{-1} \eta_{\epsilon, u} g \right) 
= - \eta_{\epsilon, u} \, g  + g \, 2 P_{\alg{h}} \left( g^{-1} \eta_{\epsilon, u} g \right) 
= - \delta_{\epsilon, u} g + \delta_h g \, .
\label{eqn:gaugetrans_comm}
\end{align}
Leaving out gauge transformations such as $\delta_h$ above, we have the commutation relations
\begin{align}
\Big[ \delta_{\epsilon_1, u_1}  , \delta_{\epsilon_2, u_2}  \Big] 
&= \frac{1}{u_1-u_2} \left( u_1 \, \delta_{\left[\epsilon_1 , \epsilon_2 \right],u_1} 
- u_2 \, \delta_{\left[\epsilon_1 , \epsilon_2 \right], u_2} \right) 
+ \frac{u_1 u_2}{1 + u_1 u_2}  \left( \delta_{[\epsilon_2 , \epsilon_1 ^\prime ], u_2} - \delta_{[\epsilon_1 , \epsilon_2 ^\prime ], u_1} \right) 
\, , 
\label{eqn:comm1} \\
\Big[ \master _{u_1} , \delta_{\epsilon, u_2}  \Big]  
&=  \frac{u_2 \big( \delta_{\epsilon, u_2}  - \delta_{\epsilon, u_1}  \big)}{(u_1 - u_2 )^2} 
- \frac{u_2 \big( \delta_{\epsilon, u_2}  + \delta_{\epsilon^\prime, u_1} \big) }{(1+ u_1 u_2 )^2} 
+ \frac{u_2 (1+ u_2 ^2 ) \, \partial_{u_2} \delta_{\epsilon, u_2} }{( u_1 - u_2 ) ( 1 + u_1 u_2)  }  
\, , 
\label{eqn:comm2}
\end{align}
as well as the relation
\begin{align}
\Big[ \master _{u_1} , \master _{u_2}  \Big] 
&= 
\sum \limits _{i=1} ^2 \frac{(1 + u_i^2) \left( u_i \, \partial_{u_i} + 1 \right) \master _{u_i}}{(u_1 - u_2) (1 + u_1 u_2)} 
+ \left( \frac{2}{(u_1 - u_2)^2} - \frac{2}{(1+ u_1 u_2)^2}  \right)  \left(u_2 \, \master _{u_2} - u_1 \, \master _{u_1} \right) 
\, .
\label{eqn:comm3}
\end{align}
The underlying algebra takes a more intuitive form after performing an expansion around $u=0$. We define the coefficients $\delta_{\epsilon}^{(n)}$ and $\master ^{(n)}$ by
\begin{align}
\delta_{\epsilon, u} &= \sum \limits _{n=0} ^\infty u^n \, \delta_{\epsilon}^{(n)} \, , 
& 
\master_{u} &= \sum \limits _{n=0} ^\infty u^n \, \master ^{(n)} \,,
\end{align}
and we set $\delta_{\epsilon}^{(n)} = 0 = \master ^{(n)}$ for  $n < 0 \, $.
The expansion of all commutators can be performed by making repeated use of the identity 
\begin{align*}
u_1 ^{n+1} - u_2 ^{n+1} = \left( u_1 - u_2 \right) \sum \limits _{k=0} ^n u_1 ^{n-k} \, u_2 ^k \, . \qquad 
\end{align*}
Expanding the commutation relations of the Yangian-type symmetries leads to the relations
\begin{align}
\left[ \delta_{\epsilon_1}^{(n)} , \delta_{\epsilon_2}^{(m)} \right] &= 
\begin{cases}
\delta_{[\epsilon_1,\epsilon_2]}^{(m+n)} & \text{if} \quad n=0 \vee m=0, \\[3mm]
\delta_{[\epsilon_1,\epsilon_2]}^{(m+n)} + \left(-1 \right) ^{n}  \, \delta_{[\epsilon_2 , \epsilon_1 ^\prime ]} ^{(m-n)} -  \left(-1 \right) ^{m}  \, \delta_{[\epsilon_1 , \epsilon_2 ^\prime ]} ^{(n-m)} 
& \text{if} \quad n,m \neq 0 \, .
\end{cases} 
\label{eqn:comm1_expanded} 
\end{align}
The first term represents the commutation relations of a loop algebra, which is the symmetry algebra of principal chiral models. The additional terms can be simplified if we fix the condition $g_0 = \unit$, such that $\epsilon^\prime = P_\alg{h} \epsilon - P_\alg{m} \epsilon$. Discriminating the cases $\epsilon_i \in \alg{h}$ and $\epsilon_i \in \alg{m}$ one then reaches the commutation relations \eqref{eqn:comm1_expanded} without primes and with varying signs in front of the two additional terms.
 Based on these relations, Schwarz denotes the symmetry of symmetric space models by $\hat{\grp{G}}_\grp{H}$ \cite{Schwarz:1995td}. 

For the commutator of the master and Yangian variations, we find
\begin{align}
\left[ \master ^{(n)} , \delta_{\epsilon}^{(m)} \right] &= -m \, \delta_{\epsilon} ^{(m+n+1)} + (-1)^m m \, \delta_{\epsilon^\prime} ^{(n+1-m)} - (-1)^n m \, \delta_{\epsilon} ^{(m-n-1)} 
\, . 
\label{eqn:comm2_expanded} 
\end{align}
We observe that all of the higher master variations commute with the generators $\delta_{\epsilon}$ of the G-symmetry. 

The commutator of two generic master variations can be shown to take the form
\begin{align}
\left[ \master ^{(n)} , \master ^{(m)} \right] &= 
(n-m) \, \master ^{(n+m+1)} + (-1)^m (n+m+2) \, \master ^{(n-m-1)} 
- (-1)^n (n+m+2) \,  \master ^{(m-n-1)} 
\, .
\label{eqn:comm3_expanded} 
\end{align}
Note that the knowledge of the first two generators $\master$ and $\master ^{(1)}$ is sufficient to construct all of the higher master symmetry generators as it is the case for the Yangian variations $\delta_\epsilon$ and $\delta_\epsilon^{(1)}$ as well.   
Let us also emphasize that Schwarz gives an interpretation of the commutators \eqref{eqn:comm3_expanded}. In fact, he argues that the generators $\master^{(n)}$ can be related to half of a Virasoro algebra by considering suitable linear combinations.%
\footnote{In the present case, the algebra is obtained from the linear combinations 
$K_n = i^{(n+1)} \left( L_{n+1} -  L_{-n-1} \right)$ of the Virasoro generators $L_{n}$.  
Note that Schwarz discusses the relation for the case of the principal chiral model, where 
different linear combinations reproduce the respective algebra.}. He thus refers to this symmetry as ``Virasoro-like''; for more details on this point see \cite{Schwarz:1995td}

%%%%%%%%%%%
\subsection{Poisson Algebra of Yangian and Master Charges}
\label{sec:Poissbrack}

In this section, we discuss the Poisson algebra of the conserved charges induced by the nonlocal symmetries of symmetric space models. We find that up to ambiguous boundary terms, which commonly appear in the study of such Poisson algebras \cite{Luscher:1977rq}, the Poisson algebra of the charges $\chargeY^{(n)}$ is given by the classical analogue of a Yangian algebra. Similar results hold for principal chiral and Gross--Neveu models \cite{MacKay:1992he}. Since the form of the current algebra for symmetric space models is close to the one of principal chiral models, the analysis is simplified and many of the results of \cite{MacKay:1992he} can be transferred. As the charges associated to the master symmetry turn out to be compositions of the Yangian charges, their Poisson algebra is inherited from the latter.
For convenience, the calculations within this subsection are performed in components instead of differential forms. Also, in order to be compatible with \cite{MacKay:1992he}, we switch to a field theory point of view in this subsection. That is, we view the symmetric space model to describe the internal degrees of freedom of a two-dimensional field theory rather than the target space of a string theory. In particular, the conserved charges are integrated over an infinite line rather than a closed cycle. In order to emphasize this point, we also switch the notation from $(\tau,\sigma)$ to $(t,x)$ within this subsection.

The Poisson algebra of the Noether currents for symmetric space models was derived in \cite{Forger:1991cm}. For the components $j_0^{a}, j_1 ^{a}$ entering as 
\begin{align*}
j (t,x) = j_0 ^{a} (t , x ) \, T_a \, \diff t  +  j_1 ^{a} (t , x ) \, T_a \, \diff x ,
\end{align*}
we have the Poisson-brackets
\begin{align}
\left \lbrace j_0 ^{a} (t , x)\, , \, 
	j_0 ^{b} (t , y) \right \rbrace &= 
	- f ^{ab} {} _c \, j_0 ^c (t , x) \, \delta (x - y) \, , \\
\left \lbrace j_0 ^{a} (t , x) \, , \,  
	j_1 ^{b} (t , y) \right \rbrace &= 
	- f ^{ab} {} _c \, j_1 ^c (t , x)  \, \delta (x - y)
	+ k^{a b} (t , y) \, \partial_x \delta (x - y) \, , \\
\left \lbrace j_1 ^{a} (t , x)\, , \,  
	j_1 ^{b} (t , y) \right \rbrace &= 0 \, .
\end{align}
Here $k^{ab}$ is given by
\begin{align}
k ^{ab} = \tr \big( T^a k \big( T^b \big) \big) = \tr \big( P_\alg{m} \big( g^{-1} T^a g \big) 
P_\alg{m} \big( g^{-1} T^b g \big) \big) \, ,
\end{align}
with $k: \alg{g} \to \alg{g}$ being a map from $\alg{g}$ to itself defined as
\begin{align}
k &= \mathrm{Ad}(g) \circ P_\alg{m} \circ \mathrm{Ad}(g)^{-1} \, , &
k (X) &= g P_\alg{m} \left( g^{-1} X g \right) g^{-1} \, . 
\end{align}
The symbol $k$ is related to the Noether current $j$ by the relations
\begin{align}
\mathrm{ad}(j_\mu) &= k \circ \mathrm{ad}(j_\mu) + \mathrm{ad}(j_\mu) \circ k \, , &
j_{\mu} ^c \, f_c {} ^{a b} &=  j_{\mu} ^c \left(  f_c {} ^{ad} k_d {} ^b -  f_c {} ^{bd} k_d {} ^a \right)
\, , 
\label{kidentity}
\end{align}
which follow from a short calculation:
\begin{align*}
\left( k \circ \mathrm{ad}(j_\mu) + \mathrm{ad}(j_\mu) \circ k \right) (X) &= g P_\alg{m} \left( g^{-1} \left[ j_\mu , X \right] g \right) g^{-1} + \left[ j_\mu , g P_\alg{m} \left( g^{-1} X g \right) g^{-1} \right] \\
&= g P_\alg{m} \left( \left[ -2 a_\mu , g^{-1} X g \right]  \right) g^{-1} + g \left[ -2 a_\mu , P_\alg{m} \left( g^{-1} X g \right) \right] g^{-1} \\
&= g \left( \left[ -2 a_\mu , P_\alg{h} \left( g^{-1} X g \right) + P_\alg{m} \left( g^{-1} X g \right) \right] \right) g^{-1}
= \left[ j_\mu , X \right] \, .
\end{align*}
The Poisson algebra closes if one includes $k^{ab}(t,x)$. The additional Poisson brackets take the form
\begin{align}
\left \lbrace j_0 ^{a} (t,x) \, , \, k^{b c} (t,y) \right \rbrace &= -  \left( f^{ab} {} _d \, k^{d c} (t,x) + f^{ac} {} _d \, k^{d b} (t,x) \right) \delta ( x - y ) \, , \\
\left \lbrace j_1 ^{a} (t,x) \, , \, k^{b c} (t,y) \right \rbrace &= 0 \, , \\
\left \lbrace k^{ab} (t,x) \, , \, k^{c d} (t,y) \right \rbrace &= 0 \, .
\end{align}

\paragraph{Yangian charges.}
Given a consistent set of Poisson brackets for the Noether currents, we turn to the calculation of the Poisson brackets of the conserved charges $\chargeY^{(n)}$, for which we note the explicit expressions
\begin{align}\label{eq:Yangcharg}
\chargeY^{(0)\, a} &= \int \limits _{-\infty} ^\infty \diff x j_0 ^a (t, x) \, , & 
\chargeY^{(1)\, a} &= f^a {}_{bc} \int \limits _{-\infty} ^\infty \diff x_1 \diff x_2 \,  \theta(x_2 - x_1) j_0 ^b (x_1) j_0 ^c (x_2)
+ 2 \int \limits _{-\infty} ^\infty  \diff x j_1 ^a (x) \, .
\end{align}
We now demonstrate that the Poisson algebra of these charges represents the classical counterpart of a Yangian algebra, i.e.\ that the charges satisfy the relations
\begin{align}\label{eq:Yangcomms}
\left \lbrace \chargeY^{(0)\, a} \, , \, \chargeY^{(0)\, b} \right \rbrace &= \, f ^{ab} {} _c \, \chargeY^{(0)\, c} \, , & 
\left \lbrace \chargeY^{(0)\, a} \, , \, \chargeY^{(1)\, b} \right \rbrace &= \, f ^{ab} {} _c \, \chargeY^{(1)\, c} \, ,
\end{align}
as well as the classical counterpart of the Serre relations:
\begin{align}
f_d {} ^{[ab} \left \lbrace  \chargeY ^{(1) c]} \, , \, \chargeY ^{(1) d}\right \rbrace = \half
f ^a {} _{ip} f^b {} _{jq} f^c {}_{kr} f^{ijk} \left( \chargeY ^{(0) p} \, \chargeY ^{(0) q} \, \chargeY ^{(0) r} \right)  .
\label{Serre}
\end{align}
For the Gross--Neveu and the principal chiral model, the Serre relations for the conserved charges were shown by MacKay in \cite{MacKay:1992he}. In that case, the Poisson brackets of the respective Noether currents are similar to those given above. In the case of a principal chiral model on $\grp{G}$ we have (for the Noether current associated to left-multiplication) 
\begin{align}
\left \lbrace j_0 ^{a} (t,x) \, , \,  j_0 ^{b} (t,y) \right \rbrace_{\mathrm{PCM}} 
	&= -  f ^{ab} {} _c \, j_0 ^c (t,x) \, \delta (x-y) \, , \nn \\
\left \lbrace j_0 ^{a} (t,x) \, , \,  j_1 ^{b} (t,y) \right \rbrace_{\mathrm{PCM}} 
	&= - f ^{ab} {} _c \, j_1 ^c (t,x) \, \delta (x-y) 
	+ G^{a b} \, \partial_x \delta(x-y) \, , \\
\left \lbrace j_1 ^{a} (t,x) \, , \,  j_1 ^{b} (t,y) \right \rbrace_{\mathrm{PCM}} &= 0 \, . \nn
\end{align}
Here, $G^{ab} = \tr \left( T^a T^b \right)$ denotes the metric on the group G. In terms of the respective Noether currents, the charges $\chargeY^{(0)}$ and $\chargeY^{(1)}$ take exactly the same form \eqref{eq:Yangcharg} in principal chiral and symmetric space models. We can thus make use of the calculation of \cite{MacKay:1992he} and only need to evaluate the difference in the Poisson brackets by employing the following notation: 
\begin{align}
\left \lbrace j_0 ^{a} (t,x) \, , \,  j_0 ^{b} (t,y) \right \rbrace_{\mathrm{SSM-PCM}} &= 0 \, , \nn \\
\label{brack:ssm-pcm}
\left \lbrace j_0 ^{a} (t,x) \, , \,  j_1 ^{b} (t,y) \right \rbrace_{\mathrm{SSM-PCM}} &=  \left( k^{ab} (t,y) - G^{a b} \right) \, \partial_x \delta(x-y) \, , \\
\left \lbrace j_1 ^{a} (t,x) \, , \,  j_1 ^{b} (t,y) \right \rbrace_{\mathrm{SSM-PCM}} &= 0 \, . \nn
\end{align}
We begin by studying the Poisson bracket between a level-0 and a level-1 charge. The calculation of this bracket gives rise to boundary terms, which depend on the precise way in which the upper and lower integration boundaries are taken to infinity in \eqref{eq:Yangcharg}, cf.\ \cite{Luscher:1977rq}. They arise for both the principal chiral and symmetric space model from integrating out the $\partial_x \delta(x-y)$ contributions. The problem stems from the fact that both $G^{ab}$ and $k^{ab}$ are not suitable test functions as they do not vanish when $x$ approaches infinity. 

In order to keep the discussion general for the moment, we consider the charges $\chargeY^{(0)}$ and $\chargeY^{(1)}$ with the following boundaries:
\begin{align}
\chargeY^{(0)\, a} &= 
\int  \limits _{-L_1} ^{L_2}  \diff x j_0 ^a (x) 
\\
\chargeY^{(1)\, a} &= f^a {}_{bc} 
\int \limits _{- L_3} ^{L_4} \diff x_1 \diff x_2 \,  \theta(x_2 - x_1) j_0 ^b (x_1) j_0 ^c (x_2)
+ 2 
\int  \limits _{-L_5} ^{L_6}  \diff x j_1 ^a (x) \, .
\end{align}
Since the Poisson bracket of two 0-components of the Noether current $j$ is the same as in the case of the principal chiral model (see \eqref{brack:ssm-pcm}), the Lie algebra commutator for the level-0 charges trivially follows. For the Poisson bracket of a level-0 with a level-1 charge we find 
\begin{align}
\left \lbrace \chargeY^{(0)\, a} \, , \, \chargeY^{(1)\, b} \right \rbrace_{\mathrm{SSM-PCM}} &= 
\int \limits _{-L_1} ^{L_2}  \diff x 
\int  \limits _{-L_5} ^{L_6}  \diff y 
\left( k^{ab} (y) - G^{a b} \right) \, \partial_x \delta(x-y) \nn \\
&= \int  \limits _{-L_5} ^{L_6}  \diff y 
\left( k^{ab} (y) - G^{a b} \right) \left( \delta(L_2-y) - \delta(-L_1 -y) \right) \nn \\
&= \left( k^{ab} (L_2) - G^{a b} \right) \theta (L_6 - L_2 ) - 
\left( k^{ab} (-L_1) - G^{a b} \right) \theta (L_5 - L_1 )
\end{align}
The result shows the difference between the boundary terms for the principal chiral and symmetric space model. In the case of the principal chiral model, it has been noted \cite{MacKay:1992he} that the ambiguous boundary terms disappear if one sets the upper and lower boundaries equal, $L_1 = L_2$ and $L_5 = L_6$. This prescription is not sufficient in the case of the symmetric space model, since $k^{ab} (L_1)$ generically differs from $k^{ab} (-L_1)$. For the boundary terms to disappear --- and the Yangian algebra to be satisfied --- we have to require $L_1 > L_5$ and $L_2 > L_6$. 

In order to study the Serre relations \eqref{Serre} next, let us now turn to the Poisson bracket of two level-1 charges $\chargeY^{(1)\, a}$. Making use of equation \eqref{brack:ssm-pcm},
 we find
\begin{align}
\left \lbrace \chargeY^{(1)\, a} , \chargeY^{(1)\, b} \right \rbrace _{\mathrm{SSM-PCM}} 
& = \int \limits _{-L_3} ^{L_4} \diff x_1 \diff x_2 \, \theta(x_2-x_1) 
\int \limits _{-L_5} ^{L_6} \diff x_3 \bigg[
f^{a} {}_{cd} \left \lbrace j_0 ^c (x_1) j_0 ^d (x_2) 
, j_1 ^b (x_3) \right \rbrace   _{\mathrm{SSM-PCM}} \nonumber\\
& \hspace*{30mm}
+ f^{b} {}_{cd} \left \lbrace j_1 ^b (x_3) ,
 j_0 ^c (x_1) j_0 ^d (x_2)  \right \rbrace   _{\mathrm{SSM-PCM}}
\bigg] \nonumber\\
& 
= \int \limits _{-L_3} ^{L_4} \diff x_1 \diff x_2 \, \epsilon(x_2-x_1) 
\int \limits _{-L_5} ^{L_6} \diff x_3 \bigg[
j_0 ^c (x_1) \bigg( 
f^a {}_{cd}  \left( k^{db} (x_3) - G^{db} \right) \nonumber\\ 
& \hspace*{30mm}
-f^b {}_{cd}  \left( k^{da} (x_3) - G^{da} \right) \bigg) 
\partial_{x_2} \, \delta(x_2 - x_3 )  \bigg] \, ,
\end{align}
where we defined $\epsilon(x_2-x_1) = \theta(x_2-x_1)  - \theta(x_1-x_2)$ in the last line. Integrating by parts gives the boundary term
\begin{align}
&B^{ab} = \Big( f^a {}_{cd}  \left( k^{db} (L_4) \theta(L_6-L_4)    + k^{db} (-L_3) \theta(L_5-L_3) 
 - G^{db} \left( \theta(L_6-L_4) + \theta(L_5-L_3)  \right) \right) \nonumber\\
&- f^b {}_{cd}  \left( k^{da} (L_4) \theta(L_6-L_4)  +  k^{da} (-L_3) \theta(L_5-L_3) 
- G^{da} \left( \theta(L_6-L_4) + \theta(L_5-L_3)  \right) \right) \Big) \chargeY^{(0)\, c} \, .
\end{align}
Again, the result shows the difference between the boundary terms arising for the principal chiral and the symmetric space model. In the case of a principal chiral model, the boundary terms are not relevant for the Serre relations due to the Jacobi identity $f_b {} ^{[cd}  f^{a]b} {}_e =0$. The situation is different for a symmetric space model, where the above result shows that generically we have
\begin{equation}
f_b {} ^{[cd}  B^{a]b} \neq 0 \, ,
\end{equation} 
such that the Serre relations \eqref{Serre} are violated by the boundary terms. We must hence require $L_4 > L_6$ and $L_5 > L_3$ in order for the Yangian algebra to hold true. In this case, the boundary terms are absent and we only have the bulk term, which takes the following form after taking $L_i$ to infinity in the appropriate order: 
\begin{align}
\left \lbrace \chargeY^{(1)\, a} , \chargeY^{(1)\, b} \right \rbrace _{\mathrm{SSM-PCM}} 
&= -2 \int \diff x \, j_0 ^c (x) \left( 
2 f_c {} ^{ba} + f_c {} ^{bd} k_d {} ^a (x) -  f_c {} ^{ad} k_d {} ^b (x) \right) 
= 2 f^{ab} {} _c \, \chargeY ^{(0)\, c}  .
\end{align}
Here, we have used the relation \eqref{kidentity} to eliminate the terms involving $k$. Hence, the difference to the result for the principal chiral model is given by a term which drops out of the Serre relations. We can thus conclude that the Serre relations for symmetric space models are satisfied if a particular ordering prescription is chosen for taking the boundaries of the integration domains to infinity. Hence, the Yangian relations for a symmetric space model can be understood to fix a limit-ambiguity in the definition of the charges. The situation is thus slightly different from the principal chiral model, where the order of $L_4,L_6$ and $L_3,L_5$ is not relevant to establish the Serre relation.%
\footnote{The limit ambiguity is fixed for the principal chiral model as well, if one considers the Yangian relations over the symmetry algebra $\alg{g} \oplus \alg{g}$ rather than $\alg{g}$.}

%%%

\paragraph{Master charges.}

The conserved charges associated with the master symmetry are compositions of the Yangian charges, see e.g.\ \tabref{tab:towers}. Hence, the respective algebra relations are inherited from the Yangian algebra. For instance, the level-0 Yangian charge $\chargeY$ and the level-1 master charge $\chargeV ^{(1)} = \tr \big( \chargeY \, \chargeY ^{(1)} \big)$ given in \eqref{eq:lev1master}, commute due to \eqref{eq:Yangcomms}:
\begin{equation}
\{\chargeY_a,\chargeV^{(1)}\}=
\chargeY^b\{\chargeY_a,\chargeY^{(1)}_b\}
+
\{\chargeY_a,\chargeY^b\}\chargeY^{(1)}_b=0.
\end{equation}

%%%%%%%%%%%%%%%%%%%%%%%%%%%%%%%%%%%%%%%%%%%%%%%%%%%%%%%%%%%%%%%%%%%%%%%%%%%
\section{Master Symmetry of Holographic Wilson Loops}
\label{sec:WilsonLoops}

In this section we apply the master symmetry of symmetric space models to Maldacena--Wilson loops in $\mathcal{N}=4$ supersymmetric Yang--Mills theory. The latter were defined in \eqref{eq:MaldaWL}. The connection between the Maldacena--Wilson loop and the symmetries of symmetric space models is established via the AdS/CFT-correspondence \cite{Maldacena:1997re}. It predicts \cite{Maldacena:1998im,Rey:1998ik} that the expectation value of the Maldacena--Wilson loop at strong coupling is described by the area of a minimal surface in $\mathrm{AdS}_5$, which ends on the loop $\gamma$ on the conformal boundary of $\mathrm{AdS}_5$: 
\begin{align}
\left \langle W(\gamma) \right \rangle \overset{\lambda \gg 1}{=} \exp { \left[-\ft{\sqrt{\lambda}}{2 \pi}  A_{\mathrm{ren}}(\gamma) \right] } \, .
\label{eqn:WLstrong}
\end{align}
Here, the area functional is given by a string action and since $\mathrm{AdS}_5$ is a symmetric space, the symmetries we have discussed so far can be applied. In Poincar{\'e} coordinates, $\diff s^2 = y^{-2} \left( \diff X^\mu \diff X_\mu + \diff y ^2 \right)$, the minimal surface is described by the boundary conditions
\begin{align}
y(\tau = 0 , \sigma) &= 0 \, , & X^\mu (\tau = 0 , \sigma ) &= x^\mu (\sigma) \, , \label{eqn:boundary_conditions}
\end{align}
where $x^\mu (\sigma)$ is some parametrization of the boundary curve $\gamma$. The area of the minimal surface for these boundary conditions is divergent due to the divergence of the metric at $y=0$ and the quantity appearing in equation \eqref{eqn:WLstrong} is obtained by introducing a cut-off at $y= \eps$ and subtracting the divergence, 
\begin{align}
A_{\mathrm{ren}}(\gamma) := \lim \limits _{\eps \to 0} \left \lbrace A_{\mathrm{min}}(\gamma) \big \vert _{y \geq \eps} - \sfrac{L(\gamma)}{\eps} \right \rbrace \, . \label{eqn:Aren}
\end{align}
Here, $L(\gamma)$ denotes the length of the boundary curve $\gamma$. 

The master symmetry transformation deforms minimal surfaces ending on some contour $\gamma$ into minimal surfaces ending on a different contour $\gamma_u$. We will see below that also the calculation of the transformed boundary contour $\gamma_u$ requires knowing the minimal surface solution for the original boundary contour. This makes explicit reconstructions difficult. For the simple cases of a circle and a straight line with a cusp, one finds that the master symmetry transformation merely corresponds to a reparametrisation. Less symmetric configurations were studied based on an approach involving Riemann theta functions \cite{Kruczenski:2013bsa,Kruczenski:2014bla,Huang:2016atz} as well as in a wavy-line expansion around known solutions \cite{Dekel:2015bla}. For these configurations, the transformation of the boundary curve is nontrivial. 

In this paper, we compute the variation of a general boundary curve. The computation of the variation of the boundary coordinates of a general minimal surface shows that the transformed minimal surface still ends on the conformal boundary of $\AdS$ and we prove that the master symmetry transformation is also a symmetry of the renormalized area $A_{\mathrm{ren}}(\gamma)$. 

\subsection{Symmetry and Renormalization}

\begin{figure}
\centering
\includegraphics[width=120mm]{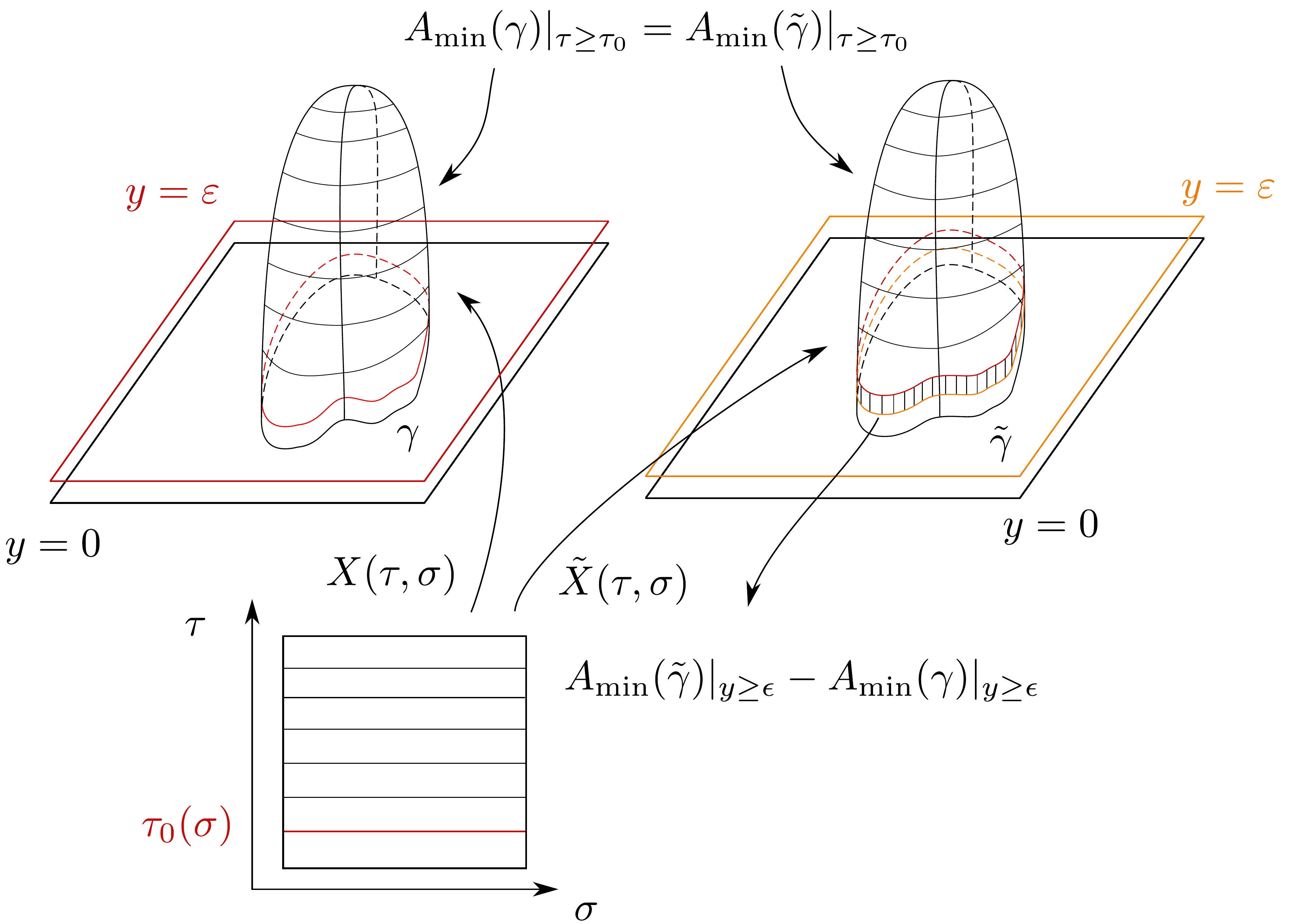}
\caption{Comparison of the minimal surfaces associated to the boundary curves $\gamma$ and $\tilde{\gamma}$.}
\label{fig:renormalization}
\end{figure}

In fact, we show that any symmetry of the area functional and the equations of motion is also a symmetry of the renormalized minimal area\footnote{We thank Harald Dorn for sharing his notes on large conformal transformations of the renormalized area with us, see also \cite{Dorn:2015bfa}.} $A_{\mathrm{ren}}(\gamma)$. Here, we rely on the expansion of the minimal surface close to the conformal boundary, which can be obtained from the equations of motion as\cite{Polyakov:2000ti}
\begin{align}
X^\mu \left( \tau , \sigma \right) &= x^\mu (\sigma)
+ \frac{\tau^2}{2} \, \dot{x}^2(\sigma) \, \partial_\sigma \left( \frac{\dot{x}^\mu(\sigma)}{\dot{x}^2(\sigma)} \right) 
- \frac{\tau^3}{3} \, \dot{x}^2(\sigma) \, \frac{\delta A_{\mathrm{ren}}(\gamma)}{\delta x_\mu (\sigma)} 
+ \mathcal{O}\left(\tau^4 \right) \label{eqn:expansion_boundary} \\
y \left( \tau , \sigma \right) &= \tau \, \lvert \dot{x}(\sigma) \rvert 
+ \mathcal{O}\left(\tau^3 \right) \, .
\label{eqn:expand_bound}
\end{align}
The minimal surface is described by a Polyakov action, 
\begin{align}
A_\text{surf} = \frac{1}{2} \int \diff\sigma \, \diff\tau \, \frac{\partial_i X^\mu \partial_i X^\mu +\partial_i y \partial_i y}{y^2} \, , 
\end{align} 
and $( \tau , \sigma )$ are conformal coordinates on the worldsheet. Since we are considering a symmetry transformation, the transformed surface $\lbrace \tilde{X}^\mu (\tau , \sigma ) , \tilde{y}(\tau , \sigma ) \rbrace$ ending on the transformed boundary contour $\tilde \gamma$ is also a solution of the equations of motion in conformal gauge and thus we have
\begin{align*}
\tilde{X}^\mu ( \tau , \sigma) &= \tilde{x}^\mu (\sigma ) + \mathcal{O}(\tau^2) \, , &
 \tilde{y} ( \tau , \sigma) &= \tau \lvert \dot{\tilde{x}}(\sigma) \rvert + \mathcal{O}(\tau^3) \, .
\end{align*}
In parameter space, the regularisation prescription $y \geq \eps$ translates to the condition $\tau \geq \tau_0 (\sigma)$, which is defined by (see \figref{fig:renormalization})
\begin{align*}
y \left( \tau_0 (\sigma) , \sigma \right) = \eps\quad \Rightarrow \quad \tau_0 (\sigma) = \frac{\eps}{\lvert \dot{x}(\sigma ) \rvert  } + \mathcal{O}(\eps^3) \, .
\end{align*}
Likewise, $\tilde{\tau_0} (\sigma)$ is defined by the requirement $ \tilde{y} \left( \tilde{\tau}_0 (\sigma) , \sigma \right) = \eps$. Using again that we are discussing a symmetry of the area functional, we have
\begin{align*}
A_{\mathrm{min}}(\gamma) \big \vert _{\tau \geq \tau_0 (\sigma)} = A_{\mathrm{min}}(\tilde{\gamma}) \big \vert _{\tau \geq \tau_0 (\sigma)} \, .
\end{align*}
Correspondingly, the difference between the minimal areas for the original curve $\gamma$ and the transformed curve $\tilde{\gamma}$ is given by
\begin{align}
A_{\mathrm{min}}(\gamma) \big \vert _{y \geq \eps} - A_{\mathrm{min}}(\tilde{\gamma}) \big \vert _{\tilde{y} \geq \eps} 
&=  \frac{1}{2} \int \limits _{0} ^{2 \pi} \diff \sigma \int \limits _{\tau_0 (\sigma)} ^{\tilde{\tau}_0 (\sigma)} \diff \tau \frac{\partial_i \tilde{X}^\mu \partial_i \tilde{X}^\mu +\partial_i \tilde{y} \partial_i \tilde{y}}{\tilde{y}^2} 
= \int \limits _{0} ^{2 \pi} \diff \sigma \int \limits _{\tau_0 (\sigma)} ^{\tilde{\tau}_0 (\sigma)} \diff \tau \left( \frac{1}{\tau^2} + \mathcal{O} ({\tau^0}) \right) \nn \\
&=   \int \limits _{0} ^{2 \pi} \diff \sigma \, \frac{\lvert \dot{x}(\sigma ) \rvert }{\eps} -   \int \limits _{0} ^{2 \pi} \diff \sigma \, \frac{ \lvert \dot{\tilde{x}}(\sigma ) \rvert}{\eps} + \mathcal{O}(\eps) = \frac{L(\gamma)}{\eps} - \frac{L(\tilde{\gamma})}{\eps} + \mathcal{O}(\eps) \, .
\end{align}
Using the definition \eqref{eqn:Aren}, this shows that the renormalized area $A_{\mathrm{ren}}(\gamma)$ is invariant under the map $\gamma \to \tilde{\gamma}$ induced by any symmetry of the model.

\subsection{Coset Description of Euclidean \texorpdfstring{$\mathrm{AdS}_5$}{AdS5}}
\label{sec:cosetAdS5}

In \secref{sec:numerics} we will study a numerical method for the computation of minimal surfaces. For this method it is advantageous to consider $\mathrm{AdS}$ with Euclidean rather than Lorentzian signature, such that those surfaces are true minima as opposed to mere saddle points of the area functional.
In order to transfer the formalism developed for general symmetric space models to the case of minimal surfaces in $\mathrm{EAdS}_5$, we describe it as the coset space
\begin{align}
 \frac{\grp{SO}(1,5)}{\grp{SO}(5)} \simeq \mathrm{EAdS}_5 \, .
\end{align}
We work with the fundamental representation of $\grp{SO}(1,5)$, see appendix \ref{app:SO15} for our conventions. The Lie algebra $\alg{so}(1,5)$ is isomorphic to the conformal algebra in a four-dimensional Euclidean space and we choose the corresponding basis $\lbrace P_\mu , M_{\mu \nu} , D , K_\mu \rbrace$. Here, the indices $\mu$ and $\nu$ run from 1 to 4. We split $\alg{so}(1,5)$ into a gauge and a coset space component,  
\begin{align}
\alg{h} &= \mathrm{span} \left \lbrace M_{\mu \nu} , P_\mu - K_\mu \right \rbrace \simeq \alg{so}(1,4) \, , &
\alg{m} &= \mathrm{span} \left \lbrace P_\mu + K_\mu , D \right \rbrace \, .
\end{align}
The coset representatives are given by
\begin{align}
g(X,y) = e^{X \cdot P} \, y^D    \quad \Rightarrow \quad U = g^{-1} \, \diff g = \frac{\diff X^\mu}{y} \, P_\mu + \frac{\diff y}{y} \, D \, , \label{eqn:UXy}
\end{align}
and we note the projections
\begin{align}
A &=  \frac{\diff X^\mu}{2y} \left( P_\mu - K_\mu \right) \, ,  &
a &= \frac{\diff X^\mu}{2y} \left( P_\mu + K_\mu \right)+ \frac{\diff y}{y} \, D \, .
\end{align}
The metric of the coset space is obtained from the group metric introduced in appendix ~\ref{app:SO15} and the Maurer--Cartan current $U$ as
\begin{align}
\half \tr \left( a \otimes a \right) = \frac{\diff y \otimes \diff y  + \delta^{\mu \nu}  \diff X_\mu \otimes \diff X_\nu }{y^2} \, .
\end{align}
This shows that the parametrization $g(X,y)$ provides Poincar{\'{e}} coordinates for Euclidean $\mathrm{AdS}_5$. Correspondingly, we may describe the area functional in these coordinates by the sigma-model action
\begin{align}
A_\text{surf} = \frac{1}{4} \int \tr \left( a \wedge \ast a \right) \, .
\end{align}
We can hence carry over the symmetries discussed so far. We are mainly interested in the variation of the boundary curve $x^\mu (\sigma )$. In order to obtain it, we translate the formal variation of the coset elements to a variation of the coordinates and take the boundary limit.

Let us begin by studying the $\grp{G}$-symmetry of the model, which amounts to the $\grp{SO}(1,5)$-isometries of $\mathrm{EAdS}_5$ in our case. The transformations are described by left-multiplication of the coset representatives by a constant $L \in \grp{G}$:
\begin{align}
g \left( X_\mu , y   \right) \mapsto L \cdot g \left( X_\mu , y  \right) = g \left( X ^\prime _\mu , y  ^\prime  \right) \cdot R \, .
\end{align}
Here, we need to allow for a general gauge transformation $R \in \grp{H}$. For an infinitesimal transformation, we replace $L$ by the generators $T_a$ of $\grp{G}$, which implies the variation
\begin{align}
\partial_\mu g \left( X_\mu , y  \right) \, \delta _a X^\mu + \partial_y g \left( X_\mu , y  \right) \, \delta _a y = T_a \cdot g \left(X_\mu , y\right) - g \left( X_\mu , y\right) \cdot h_a \, ,
\end{align}
where $h_a \in \alg{h}$ is a general element of the gauge Lie algebra. A more convenient expression in order to read off the variations $\delta_a X^M$ is given by the components of the Maurer--Cartan form:
\begin{align}
U_\mu \delta _a X^\mu + U_y \delta_a y = g \left( X_\mu , y \right)^{-1} \cdot T_a \cdot g \left( X_\mu , y \right) - h_a \, .
\end{align}
Inserting equation \eqref{eqn:UXy}, we thus have
\begin{align}
\delta_a X^\mu \, P_\mu + \delta_a y \, D = y \left( g \left( X_\mu , y \right)^{-1} \cdot T_a \cdot g \left( X_\mu , y \right) - h_a \right) \, .
\end{align}
As the right hand side of this equation only contains the generators $P_\mu$ and $D$, the gauge transformation $h_a$ can be determined from the terms proportional to $K_\mu$ and $M_{\mu \nu}$ in $g^{-1} T_a g$, which gives
\begin{align*}
h_a = \frac{1}{4} \tr \left( g^{-1} T_a \, g \, P^\mu \right) \left(K_\mu - P_\mu \right) 
- \frac{1}{4} \tr \left( g^{-1} T_a \, g \, M^{\mu \nu} \right) M_{\mu \nu} \, .
\end{align*}
Using the expressions \eqref{eqn:Metric} for the group metric we then have
\begin{align}
\delta_a X^\mu &= \frac{y}{4} \left( \tr \left(  g^{-1} T_a g \,  K^\mu \right)  
+  \tr \left( g^{-1} T_a g \, P^\mu \right) \right) \, , &
\delta_a y &= \frac{y}{2} \tr \left( g^{-1} T_a g \, D \right) \, .
\label{eqn:varcoord}
\end{align}
The variation of the boundary coordinates $x^\mu$ is obtained by taking the limit $y \to 0$. In order to take this limit, we employ the definition of $g(X,y)$ as given in \eqref{eqn:UXy} and compute the conjugation of any generator with $y^D$. This can be done by noting that our choice of basis is such that the commutation with $D$ is diagonal,
\begin{align}
\left[ D , T_a \right] = \Delta \left( T_a \right) \, T_a \, ,
\end{align}
which implies that
\begin{align}
y^D \, T_a \, y^{-D} = y^{\Delta \left( T_a \right)} \, T_a \, .
\end{align}
Using cyclicity of the trace in \eqref{eqn:varcoord} as well as $\Delta(K^\mu) =-1$, $\Delta(P^\mu) =1$ and $\Delta(D) =0$, we thus have $\delta _a y \overset{y \to 0}{\longrightarrow} 0$, which ensures that $\mathrm{AdS}$-isometries map the conformal boundary to itself and the variation of the boundary curve is given by
\begin{align}
\delta_a x^\mu = \frac{1}{4} \tr \left(  e^{-x \cdot P} T_a \, e^{x \cdot P}   K^\mu \right) =:  
\frac{1}{2} \, \xi ^\mu _a (x) \, , \qquad
\end{align}
where $\xi ^\mu _a(x)$ form a basis of conformal Killing vectors of four-dimensional Euclidean space.
%%%%
\subsection{Master and Yangian Variation of the Boundary Curve}

In this subsection we consider the variation of the boundary curve under the bilocal symmetries considered above, i.e.\ for the level-0 master and the level-1 Yangian symmetry.

\paragraph{Level-0 master symmetry.}
The discussion of the symmetry variations of the coordinates under $\AdS$-isometries can be generalized to arbitrary variations $\delta g = \eta g$ by replacing $T_a$ by $\eta$ in the variations \eqref{eqn:varcoord}, 
\begin{align}
\delta X^\mu &= \frac{y}{4} \left( \tr \left(  g^{-1} \eta g \,  K^\mu \right)  
+   \tr \left( g^{-1} \eta g \, P^\mu \right) \right) \, , &
\delta y &= \frac{y}{2} \tr \left( g^{-1} \eta g \, D \right) \, .
\label{eqn:varcoordgeneral}
\end{align}
Making use of these relations, we now turn to the master variation, which is given by
\begin{align}
\delC g &= \chi ^{(0)} \cdot g \, , & \chi ^{(0)} &= \int \ast j \, .
\end{align}
As we are aiming for the variation of the boundary curve, we only need to compute $ \chi ^{(0)} $ in the vicinity of the boundary. More precisely, we calculate $ \chi ^{(0)} $ at $y = \eps$ in an expansion in $\eps$. The expansion \eqref{eqn:expansion_boundary} of the minimal surface solution close to the conformal boundary is sufficient to do so. The $\tau$-expansion of the Noether current $j= -2 g a g^{-1}$ is of the form
\begin{align}
j_\sigma &= \frac{1}{\tau^2} \, j_{\sigma \, (-2)}
 + \mathcal{O}(\tau^0) \, , &
j_\tau &= \frac{1}{\tau} \, \partial_\sigma j_{\sigma \, (-2)}
 + j_{\tau \, (0)} 
 + \mathcal{O}(\tau) \, ,
\end{align}
where the lower index in brackets denotes the order of the $\tau$-expansion.
The above coefficients are given by
\begin{align}
j_{\sigma \, (-2)} &= - 2 \, \frac{\dot{x}_\mu}{\dot{x}^2} \, \hat{\xi}^\mu (x) 
\, , &
j_{\tau \, (0)} &= 2 \, \frac{\delta A_{\mathrm{ren}}(\gamma)}{\delta x^\mu}  \, \hat{\xi}^\mu (x) 
\, ,
\end{align}
where $\hat{\xi}^\mu(x) $ takes the form
\begin{align}
\hat{\xi}^\mu(x)  = \xi ^\mu _a (x) \, T^a = \frac{1}{2} e^{x \cdot P} K^\mu e^{-x \cdot P} \, .
\end{align}
We now calculate $ \chi ^{(0)} $ for $y= \eps$ which corresponds to the point $(\tau_0 (\sigma), \sigma )$ in parameter space. The definition of  $ \chi ^{(0)} $ requires to choose some starting point on the worldsheet, which we take to be $( \tau = c , \sigma = 0 )$. Since $ \chi ^{(0)} $ is path-independent, we may use any path connecting the points $( c , 0 )$ and $(\tau_0 (\sigma), \sigma )$. We choose the composed path $\gamma = \gamma _1 \circ \gamma_2 $, where $\gamma_1$ connects $(c,0)$ to $(\tau_0 (\sigma), 0 )$ along $\sigma = 0$ and $\gamma_2$ connects $(\tau_0 (\sigma), 0 )$  to $(\tau_0 (\sigma), \sigma )$ along the $\sigma$-direction. We find
\begin{align*}
\int _{\gamma_1} \ast j &=
\int \limits _c ^{\tau_0 (\sigma)} \diff \tau \, j_\sigma ( \tau , 0 ) 
= - \frac{1}{\tau_0 (\sigma)} j_{\sigma \, (-2)} (0)  + \zeta 
+ \mathcal{O}(\eps) \, , \\
\int _{\gamma_2} \ast j 
&= \int \limits _0 ^\sigma \diff \sigma ^\prime 
\left( - j_\tau (\tau_0 (\sigma) , \sigma ^\prime)  \right) 
= \frac{1}{\tau_0 (\sigma)} \left( j_{\sigma \, (-2)} (0) - j_{\sigma \, (-2)} (\sigma) \right) 
- \int \limits _0 ^\sigma \diff \sigma ^\prime j_{\tau \, (0)} (\sigma  ^\prime) \, .
\end{align*}
Here, $\zeta$ is some $\sigma$-independent element of $\alg{g}$ and we may neglect the conformal transformation which it parametrizes. Combining the two results we find
\begin{align}
\chi ^{(0)} (\tau_0 (\sigma) , \sigma ) =  \int _{\gamma} \ast j &= \frac{2}{\eps} \frac{\dot{x}(\sigma)_\mu}{\lvert \dot{x}(\sigma) \rvert} \, \hat{\xi}^\mu (x(\sigma)) + \zeta - 2
\int \limits _0 ^\sigma \diff \sigma ^\prime \, \frac{\delta A_{\mathrm{ren}}(\gamma)}{\delta x^\mu(\sigma ^\prime) } \,  \hat{\xi}^\mu (x(\sigma ^\prime)) + \mathcal{O}(\eps) \, .
\label{eqn:chi_expl}
\end{align}
We can then determine the variation of the coordinates. Let us first convince ourselves that the master symmetry does not move the boundary into the bulk, i.e.\ we have $\master y \to 0$ as $y$ approaches $0$. Making use of equation \eqref{eqn:varcoordgeneral} with $\eta=\chi^{(0)}$, we find
\begin{align}
\master y &= \frac{y}{2} \tr \left( g^{-1} \chi^{(0)} g D \right) = 
\frac{\dot{x}(\sigma)_\mu}{ \lvert \dot{x}(\sigma) \rvert} 
\tr \left( y^{-D} e^{-X \cdot P} \hat{\xi}^\mu e^{X \cdot P} y^D D \right) 
+ \mathcal{O}(y) \nn \\
&=  \frac{\dot{x}(\sigma)_\mu}{\lvert \dot{x}(\sigma) \rvert} 
\tr \left( K^\mu D \right) + \mathcal{O}(y) = \mathcal{O}(y) \, ,
\end{align}
which shows that the master symmetry maps the conformal boundary to itself. For the variation of the $X$-coordinates we find
\begin{align}
\delC X^\mu &= \frac{1}{4} \tr \left( e^{-X \cdot P } \chi^{(0)} e^{X \cdot P } K^\mu \right) 
+ \mathcal{O}(y) = \frac{1}{2} \tr \left( \chi^{(0)} \hat{\xi}^\mu(x) \right) + \mathcal{O}(y)  \nn \\
&= \frac{1}{4y} \,  \frac{\dot{x}(\sigma)_\nu}{\lvert \dot{x}(\sigma) \rvert} \, \tr \left( K^\nu K^\mu \right) + \frac{1}{2} \,  \delta _{\zeta} X^\mu 
-  \, G^{ab}  \int \limits _0 ^\sigma \diff \sigma ^\prime \, \frac{\delta A_{\mathrm{ren}}(\gamma)}{\delta x^\nu(\sigma ^\prime) } \,  \xi^\nu _a (x(\sigma ^\prime)) \, \xi^\mu _b( x (\sigma) )
+ \mathcal{O}(y) \, .
\end{align}
As the first term vanishes due to $\tr \left( K^\nu K^\mu \right)=0$, we can safely take the limit $y \to 0$. We neglect the conformal variation parametrized by $\zeta$ as it is independent of the point along the contour and depends on our choice of a starting point on the minimal surface. Then we obtain
\begin{align}
\master x^\mu (\sigma) = -   G^{ab}  \int \limits _0 ^\sigma \diff \sigma ^\prime \, \frac{\delta A_{\mathrm{ren}}(\gamma)}{\delta x^\nu(\sigma ^\prime) } \,  \xi^\nu _a (x(\sigma ^\prime)) \, \xi^\mu _b (x(\sigma)) \, .
\label{eqn:deltaCx}
\end{align}
The result involves the third-order term of the expansion of $X^\mu (\tau , \sigma)$, cf.\ \eqref{eqn:expansion_boundary}. While this coefficient can be identified with a functional derivative of the minimal area (as it has been above), it cannot be determined from expanding the equations of motion around the boundary, i.e.\ from treating the boundary value problem of determining the minimal surface as an initial value problem in the coordinate $\tau$. The result thus indicates that it is indeed necessary to compute the minimal surface solution in order to determine the master transformation of the boundary curve as it was done in \cite{Kruczenski:2013bsa,Kruczenski:2014bla,Cooke:2014uga,Huang:2016atz,Dekel:2015bla}.

Let us note that given the form which the master variation \eqref{eqn:deltaCx} takes for the boundary curve, one can show that it is a symmetry without referring to the formalism introduced before. We have
\begin{align}
\master A_{\mathrm{ren}}(\gamma) &= \int \limits _0 ^{2 \pi} \diff \sigma 
	\frac{\delta A_{\mathrm{ren}}(\gamma)}{\delta x^\mu(\sigma) } \, \master x^\mu (\sigma)
= -  G^{ab} \int \limits _0 ^{2 \pi} \diff \sigma \int \limits _0 ^{\sigma} \diff \sigma^\prime \,
	\frac{\delta A_{\mathrm{ren}}(\gamma)}{\delta x^\mu(\sigma) } \xi^\mu _b (x(\sigma))
	\frac{\delta A_{\mathrm{ren}}(\gamma)}{\delta x^\nu(\sigma^\prime) } \xi^\nu _a (x(\sigma^\prime)) 
\nn \\
&= -  \frac{1}{2} G^{ab} \, \delta_b \left( A_{\mathrm{ren}}(\gamma) \right)
	\delta_a \left( A_{\mathrm{ren}}(\gamma) \right)
= 0 \, .
\label{deltaAsymm}
\end{align}
It is intriguing that the invariance of the minimal area under the master symmetry, which can be employed to construct all nonlocal conserved charges and their associated symmetry transformations, does not require integrability. This corresponds to the finding that the conserved charge associated with the master symmetry itself is the Casimir of the G-symmetry charges. We should emphasize, however, that the integrability constraints are implemented in the Ward identities following from the higher-level master symmetries, which are in turn generated by the level-0 master variation $\master\equiv\master^{(0)}$.

\paragraph{Level-1 Yangian symmetry.}

The procedure described above for the master symmetry variation can in principle be carried out for any of the variations described in \secref{sec:nonlocalsymm}, although it would require to extend the expansion \eqref{eqn:expand_bound} in order to consider the higher-level variations. Here, we study the level-1 Yangian-type variation $\delta_\epsilon ^{(1)}$. 

Noting that the master variation is given by $\master g = G^{bc} \chi^{(0)}_b \, T_c g$, whereas a level-1 Yangian variation is given by
\begin{align}
\delta _a ^{(1)} g = f_a {} ^{bc} \chi^{(0)}_b \, T_c g \, ,
\end{align}
suggests that the variation of the boundary curve can be obtained from \eqref{eqn:deltaCx} by replacing $G^{bc}$ by  
$f_a {} ^{bc}$ to obtain 
\begin{align}
\delta _a ^{(1)} x^\mu (\sigma) = - f_a {} ^{bc}   \int \limits _0 ^\sigma \diff \sigma ^\prime \, \frac{\delta A_{\mathrm{ren}}(\gamma)}{\delta x^\nu(\sigma ^\prime) } \,  \xi^\nu _b (x(\sigma ^\prime)) \, \xi^\mu _c (x(\sigma)) \, .
\label{eqn:deltaax}
\end{align}
However, we still need to discuss the divergent terms contained in the expression given for $\chi^{(0)}$ in equation \eqref{eqn:chi_expl}, which did not contribute to the master variation $\master$. For the variation of the original boundary curve, we find the additional term
\begin{align}
\frac{\dot{x}_\nu}{\lvert \dot{x} \rvert \eps} 
\tr \left( \left[ T_a \, , \, \hat{\xi}^\nu (x) \right] \hat{\xi}^\mu (x) \right) 
= \frac{\dot{x}_\nu}{\lvert \dot{x} \rvert \eps} 
\tr \left( \left[\hat{\xi}^\nu (x) \, , \, \hat{\xi}^\mu (x) \right] T_a \right) = 0 \, ,
\end{align}
which shows that our expectation \eqref{eqn:deltaax} is indeed correct. For the variation of $y$, however, we have
\begin{align}
\delta_a ^{(1)} y &= \frac{\dot{x}_\mu}{\lvert \dot{x} \rvert} 
\tr \left( e^{-x \cdot P} \left[ T_a \, , \, \hat{\xi}^\mu (x) \right] e^{x \cdot P} D \right)
= \frac{\dot{x}_\mu}{\lvert \dot{x} \rvert} 
\tr \left( \left[ e^{-x \cdot P}  \hat{\xi}^\mu (x)  e^{x \cdot P} \, , \, D  \right]  e^{-x \cdot P}  T_a  e^{x \cdot P} \right) \nn \\
&= \frac{\dot{x}_\mu}{\lvert \dot{x} \rvert} \, \xi ^\mu _a (x) \, .
\end{align}
We thus see that the level-1 Yangian variation $\delta_\epsilon ^{(1)}$ generically shifts the boundary curve into the bulk. This behaviour is accompanied by a divergent boundary term arising from the application of $\delta_\epsilon ^{(1)}$ to the minimal area $A_{\mathrm{ren}}(\gamma)$. Let us also note that for certain choices of the boundary curve $\gamma$ and $\epsilon$, the boundary curve is not shifted into the bulk. The simplest example corresponds to $\epsilon = D$ and the boundary curve being a circle. In this case we find 
\begin{align*}
\delta_D ^{(1)} y = \frac{\dot{x}_\mu x^\mu }{\lvert \dot{x} \rvert} + \mathcal{O}(y)
= \mathcal{O}(y) \, .
\end{align*}
The level-1 Yangian symmetry was studied in a different approach in \cite{Muller:2013rta,Munkler:2015gja}, where the explicit evaluation of the level-1 Yangian charge $\chargeY^{(1)}$ lead to the constraint
\begin{align*}
f_a {} ^{cb} \int \limits _0 ^{L} \diff \sigma_1 \, \diff \sigma_2 \theta(\sigma_1 - \sigma_2)
	\frac{\delta A_{\mathrm{ren}}(\gamma)}{\delta x^\mu_1} 
	\frac{\delta A_{\mathrm{ren}}(\gamma)}{\delta x^\nu_2 } 
	\,  \xi^\mu _b (x_1) \, \xi^\mu _c (x_2)
+ \int \limits _0 ^{L} \diff \sigma 
	\left(\dddot{x}^\mu + \dot{x}^\mu \ddot{x}^2 \right) \xi^\mu _a (x) 
= 0 \, .
\label{eq:Yangwil}
\end{align*}
Here, the last integral is written in a parametrization where $\lvert \dot{x} \rvert \equiv 1$ for simplicity. The constraint above can be interpreted as the leading order term in the application of the level-1 Yangian generator 
\begin{align}
J_a ^{(1)} = f_a {} ^{cb} \int \limits _0 ^{L} \diff \sigma_1 \, \diff \sigma_2 \theta(\sigma_1 - \sigma_2)
	\,  \xi^\mu _b (x_1) \, \xi^\mu _c (x_2) \frac{\delta^2}{\delta x_1 ^\mu \delta x_2 ^\nu} 
+ \frac{\lambda}{2 \pi^2}	\int \limits _0 ^{L} \diff \sigma \left(\dddot{x}^\mu + \dot{x}^\mu \ddot{x}^2 \right) \xi^\mu _a (x) 
\end{align}
to the expectation value \eqref{eqn:WLstrong} of the Maldacena--Wilson loop at strong coupling. In this approach, 
we consider a specific bilocal variation of the boundary curve, which induces a variation of all points on the minimal surface given that we keep the boundary curve fixed at $y=0$. Note that this is different from considering the action of the variation $\delta_a ^{(1)}$ on the minimal surface, where we found that the boundary curve is generically shifted into the bulk. The local term appearing in the level-1 Yangian generator stems from the boundary term arising from the variation $\delta_a ^{(1)}$.

\subsection{Continuation to Arbitrary Coupling}

There are several possibilities to try to transfer the information obtained about the symmetries of the Maldacena--Wilson loop at strong coupling to arbitrary or weak coupling. The most direct approach is to compute a variation or large transformation and to consider the same transformation for the Maldacena--Wilson loop. This approach was discussed by Dekel in \cite{Dekel:2015bla}, who considered the spectral-parameter transformation introduced in \cite{Ishizeki:2011bf} by employing an expansion in the 'waviness' of the contour. In this approach Dekel found that the symmetry observed at strong coupling seems to be broken at weak coupling beyond a certain order in the waviness. 

A different approach is to employ the generator $J_a ^{(1)}$ given in \eqref{eq:Yangwil} at any value of the coupling constant $\lambda$. In our context, this approach suffers from the base-point dependence of the \mbox{level-1} Yangian generator: If we choose a different starting point $x(\Delta)$ instead of $x(0)$, we obtain a different level-1 generator $\tilde{J}_a^{(1)}$. The difference to the original generator $J_a ^{(1)}$ contains the term  
\begin{align*}
f_a {} ^{cb}  \, f_{bc}{}^d  \, \int  _0 ^\Delta  \diff \sigma \,
 	\xi^\mu _d (x) \, \dfrac{\delta}{\delta x^\mu} \, , 
\end{align*}
which cannot be a symmetry for arbitrary $\Delta$. The problem does not appear in the strong-coupling discussion since the term resulting from the above part of the generator is subleading in $\lambda$. At weak coupling however, the combination of periodic boundary conditions with the underlying algebra prohibits the generator $J_a ^{(1)}$ from becoming a symmetry. The situation changes, if the underlying symmetry algebra is $\alg{psu}(2,2 \vert 4)$, for which the contraction $f_a {} ^{cb}  \, f_{bc}{}^d $ vanishes, such that the above obstruction for the presence of Yangian symmetry is absent. In this case, Yangian symmetry has indeed been observed for tree-level scattering amplitudes \cite{Drummond:2009fd} as well as for Wilson loops in superspace at weak and strong coupling \cite{Beisert:2015jxa,Beisert:2015uda,Munkler:2015gja}. It is conceivable that the problem of base-point dependence does not occur for generators associated to the tower of master symmetries, such that they might be carried over to weak coupling in the above way. 

Here we suggest a new approach to understand the above symmetries at arbitrary coupling. We allow the variations derived in the last subsection to depend on the coupling constant $\lambda$. In fact, the master symmetry variation may serve as inspiration. The analysis of \cite{Dekel:2015bla} shows that generically we have 
\begin{align*}
\master \left \langle   W(\gamma) \right \rangle = 
-  G^{ab} \int \limits _0 ^{2 \pi} \diff \sigma \int \limits _0 ^{\sigma} \diff \sigma^\prime \,
	\frac{\delta \left \langle   W(\gamma) \right \rangle}{\delta x^\mu(\sigma) } \xi^\mu _b (x(\sigma))
	\frac{\delta A_{\mathrm{ren}}(\gamma)}{\delta x^\nu(\sigma^\prime) } \xi^\nu _a (x(\sigma^\prime)) 
\neq 0 \ .
\end{align*}
However, it is clear that by the same reasoning as in \eqref{deltaAsymm} the variation defined as
\begin{align}
\master _{(\lambda)} x^\mu (\sigma) = -   G^{ab}  \int \limits _0 ^\sigma \diff \sigma ^\prime \, 
\frac{\delta \log \left \langle   W(\gamma) \right \rangle}{\delta x^\nu(\sigma ^\prime) } \,  \xi^\nu _a (x(\sigma ^\prime)) \, \xi^\mu _b (x(\sigma)) 
\end{align}
constitutes a symmetry of the Maldacena--Wilson loop,
\begin{align}
\master _{(\lambda)} \left \langle   W(\gamma) \right \rangle = 
-  \frac{1}{2 \left \langle   W(\gamma) \right \rangle } G^{ab} 
	\, \delta_b  \left \langle   W(\gamma) \right \rangle 
	\delta_a \left \langle   W(\gamma) \right \rangle 
= 0 \, .
\end{align}
The Yangian and master variations could be adapted in the same way, but for those transformations it is not clear whether they become symmetries. It is interesting to note, however, that the problem of the dependence on the starting point is not present for the variations, if they are carried over in this way.

%%%%%%%%%%%%%%%%%%%%%%%%%%%%%%%%%%%%%%%%%%%%%%%%%%%%%%%%%%%%%%%%%%%%%%%%%%%
\section{Relation to Previous Literature}
\label{sec:prevlit}

In this section we present more details on the relation of the above master symmetry to some closely related previous papers.

%%%%%%%%%%%%%

\paragraph{The dual symmetry of Eichenherr and Forger.}

In 1979, Eichenherr and Forger constructed an infinite set of conserved charges for symmetric space models. Their construction is based on the \emph{dual symmetry} $g \mapsto g_u = \chi_u g$, which has been called \emph{master symmetry} within the present paper.
Since the transformation $g \mapsto g_u$ preserves the equations of motion, they find that the transformed current
\begin{align}
j_u = - 2 g_u \, a_u \, g_u ^{-1} = \frac{1-u^2}{1+u^2} \, \chi_u \, j \, \chi_u ^{-1} - \frac{2u}{1+u^2} \, \chi_u \, \ast j \, \chi_u ^{-1} \, 
\end{align}
is again conserved if the equations of motion are satisfied. Consequently, one has a one-parameter family of conserved charges $\int \ast j_u$, and the expansion around $u=0$ gives the tower of conserved charges,  
\begin{align}
\chargeY &= \int \ast j \, , &
\chargeY^{(1)} & = 2 \int j + \int \limits_{\sigma_1 < \sigma_2} [\ast j_1, \ast j_2 ] \, , 
& \ldots
\end{align}
These charges are equivalent to those derived from the BIZZ recursion as we show in \secref{sec:nonlocalsymm}. They go on to show that a principal chiral model can be rewritten as a symmetric space model on $\grp{G} \times \grp{G} / \Delta(\grp{G})$, where $\Delta(\grp{G})$ denotes the diagonal subgroup of $\grp{G} \times \grp{G}$. This equivalence allows them to transfer the dual symmetry to principal chiral models and to construct the corresponding conserved charges there as well. For completeness, we discuss the master symmetry for principal chiral models in an analogous fashion in \appref{app:PCM}.

%%%%%%%%%%%%%
\paragraph{Schwarz' nonlocal Virasoro symmetries.}

In his paper \cite{Schwarz:1995td} of 1995 Schwarz gave an extensive discussion of the nonlocal symmetries of symmetric space and principal chiral models. In particular, he described two types of nonlocal symmetries. The first is the one referred to as Yangian symmetries in the present paper. The second is named ``Virasoro-like'' by Schwarz, since the algebraic properties of its generators resemble those of linear combinations of Virasoro generators.

In order to put Schwarz' Virasoro symmetry into the context of our master symmetry, we note that he uses the following variation for the Virasoro-like symmetries (translated to our conventions)
\begin{align}\label{eq:Schwcompare}
\delta_{\mathrm{V}, u} \, g &= \left( (1+u^2) \chi_u ^{-1} \dot{\chi}_u - \chi^{(0)} \right) g  
= \chi_u ^{-1} \Big(\, \master \chi_u \Big) g 
=: \eta  _\mathrm{V, u} \, g \, .
\end{align}
The above symmetry action is the natural symmetric space generalization of the nonlocal symmetries of the principal chiral model, which Schwarz discussed in the same paper. Let us compare this to our variation \eqref{eq:mastervars}, which is given by
\begin{equation}\label{eq:wecompare}
\master_u g =  \chi_u ^{-1} \dot{\chi}_u \, g =: \eta _u \, g\,.
\end{equation}
We see that the two variations are related by
\begin{align}
\delta_{\mathrm{V}, u} \, g = \left( \left( 1 + u^2 \right) \master_u - \master \right) g \, .
\end{align}
The master symmetry $\delC$ is thus absent in the discussion of Schwarz since the variation $\delta_{\mathrm{V}, u} $ becomes trivial in the limit $u \to 0$. Note also that the variation $\master$ cannot be extracted for $u \in \mathbb{C}$ at $u^2 = -1$ since $\chi_u$ has poles at these points. In fact, there is a simple argument which shows that $\master$ is not contained in the family of variations given by \eqref{eq:Schwcompare}. Since the variation $\delta_{\mathrm{V}, u} $  is of the form $\chi_u ^{-1} \delta_0 \chi_u$, the proof given in section \ref{sec:nonlocalsymm} shows that
\begin{align}
\diff \ast \left( \diff \eta  _\mathrm{V,u} + \left[ j , \eta  _\mathrm{V,u} \right] \right) = 0 \, ,
\end{align}
which can also be seen from the proof given in \cite{Schwarz:1995td}. In contrast, the variation $\master$ only satisfies the necessary condition \eqref{eqn:criterion}, since
\begin{align}
g^{-1} \diff \ast \lrbrk{ \diff \chi^{(0)} + \left[ j , \chi^{(0)}  \right]  } g = - 8 \, a \wedge a \in \alg{h} \, .
\end{align}

%%%%%%%%%%%%%
\paragraph{Beisert--L\"ucker construction of Lax connections.}

The 2012 paper \cite{Beisert:2012ue} by Beisert and L\"ucker discusses a method to construct flat Lax connections in integrable coset sigma models. The method is based on an operator $\Sigma$, whose action is defined on the Maurer--Cartan form $U$ and its dual $\ast U$. Written in our conventions,%
\footnote{The main difference to their conventions is that we use a Euclidean worldsheet metric here, which implies that $\ast^2=-\unit$ on one-forms. In the Minkowski case one has $\ast^2=\unit$.}
the spectral-parameter dependence of the flat Lax connection is generated via the relation
%In their framework, the spectral parameter dependence of the flat Lax connection is generated via the relation (here $\lambda(u)$ denotes the spectral parameter employed in \cite{Beisert:2012ue})
\begin{align}
L_u &= e^{\theta(u) \Sigma}\, U \, ,
&
e^{i \theta(u)} = \frac{1-iu}{1+iu} \, .
\end{align}
In the present paper we lift this auxiliary operator $\Sigma$ to a nonlocal symmetry $\delC$ acting on the fields $g(\tau,\sigma)$. On the Maurer--Cartan form $U$, the auxiliary operator acts as 
%the two operations coincide up to a conventional factor of $2$:
\begin{align}
\Sigma(U)=\half \delC U&= -\ast a\,,
&
\Sigma(\ast U)=\half \delC \ast U&= a\, .
\end{align}
We note that Beisert and L\"ucker discuss a multitude of theories for which their construction is applicable. These include supersymmetric coset models with $\mathbb{Z}_4$ grading as well as $\mathcal{N}=16$ supergravity in two dimensions. However, they also note that all models to which they successfully applied their construction are integrable models of rational type. The application to trigonometric generalizations did not succeed.

%%%%%%%%%%%%%%%%%%%%%%%%%%%%%%%%%%%%%%%%%%%%%%%%%%%%%%%%%%%%%%%%%%%%%%%%%%%
\section{Discrete Geometry and Numerical Surfaces}
\label{sec:numerics}

\begin{figure}
\centering
\includegraphics[width=70mm]{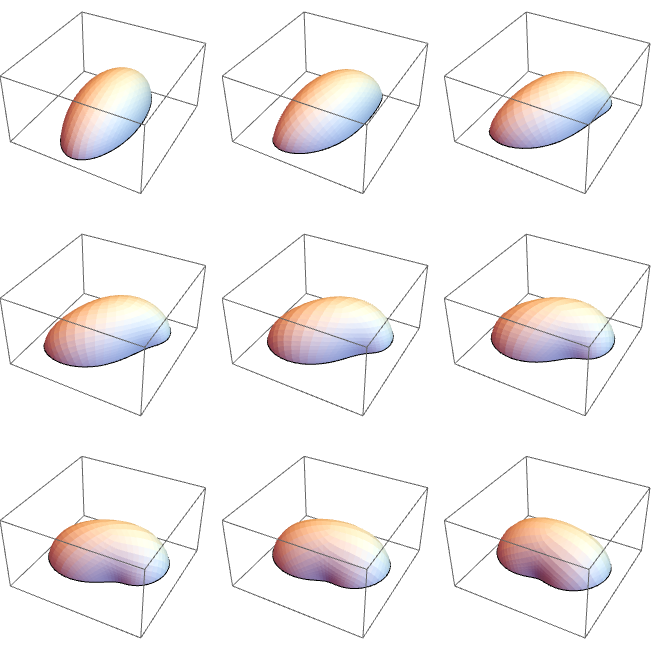}
\hspace{15mm}
\includegraphics[width=70mm]{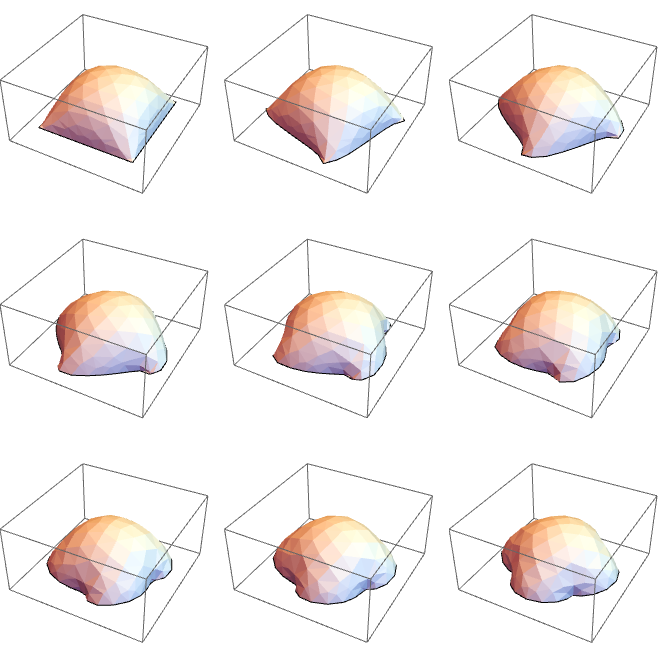}
\caption{\textbf{Ellipse and square.} Some representatives of the family of equiareal minimal surfaces associated to the elliptical and rectangular Wilson loops, respectively, for values of the spectral parameter from $\theta = 0$ to $\pi$ in uniform steps. For values from $\pi$ to $2\pi$, the figure keeps ``turning'' until it reaches at the original shape again.}
\label{fig:ellipse-family}
\end{figure}

In this section, we will develop a numerical approach to calculating the deformation of minimal surfaces due to the master symmetry. Being a nonlocal symmetry, it is clear that the deformation at some location of the surface depends in general on the shape of the \emph{entire} surface and not just on the shape near that point. This fact makes it quite nontrivial to predict the deformation. However, one might gain some intuition from the explicit examples presented here. For instance, \figref{fig:ellipse-family} shows the deformation of the minimal surface with elliptical boundary and the flipbook figures at the bottom of each page show the deformation of a triangular boundary.

Since not many minimal surfaces in hyperbolic space are known analytically, we will follow a numerical approach and work with discretized (approximate) minimal surfaces. While there is a dedicated branch of mathematics dealing with ``discrete geometry,'' which even comprises the study of integrability in a discrete setting, we will neither invoke any nontrivial theorems nor do we intend to generalize the master symmetry to discrete surfaces.

In order to find the minimal surfaces by the method of steepest descent, we have to work in Euclidean signature, as this method does not allow us to detect saddle points. 
 The dimensionality of the target space is in principle arbitrary, but for visualization purposes, we consider three dimensions, i.e.\ the target space is given by
\begin{align}
  \mathrm{EAdS}_3 \simeq \frac{\grSL(2,\Complex)}{\grSU(2)} \, .
\end{align}
Using the generators $P_{1} = \frac{1}{2} (\sigma^{1} + i \sigma^{2})$, $P_{2} = \frac{1}{2} (\sigma^{2} - i \sigma^{1})$, and $D = \frac{1}{2} \sigma^{3}$, the coset representative has the form
\be
  g = e^{X^1 P_1 + X^2 P_2} y^D = \matr{cc}{\sqrt{y} & \frac{X^1 - i X^2}{\sqrt{y}} \\[2mm] 0 & \frac{1}{\sqrt{y}} } \; ,
\ee
where $\vec{X} = (X^1,X^2)$ and $y$ are ordinary Poincar\'e coordinates. The Maurer--Cartan form reads 
\be
  U = g^{-1}\diff g = \matr{cc}{ \frac{\diff y}{2y} & \frac{\diff X^1 - i\diff X^2}{y} \\[2mm] 0 & -\frac{\diff y}{2y} } \; .
\ee

\begin{figure}
%\begin{wrapfigure}{r0}{85mm}
\centering
\includegraphics[width=85mm]{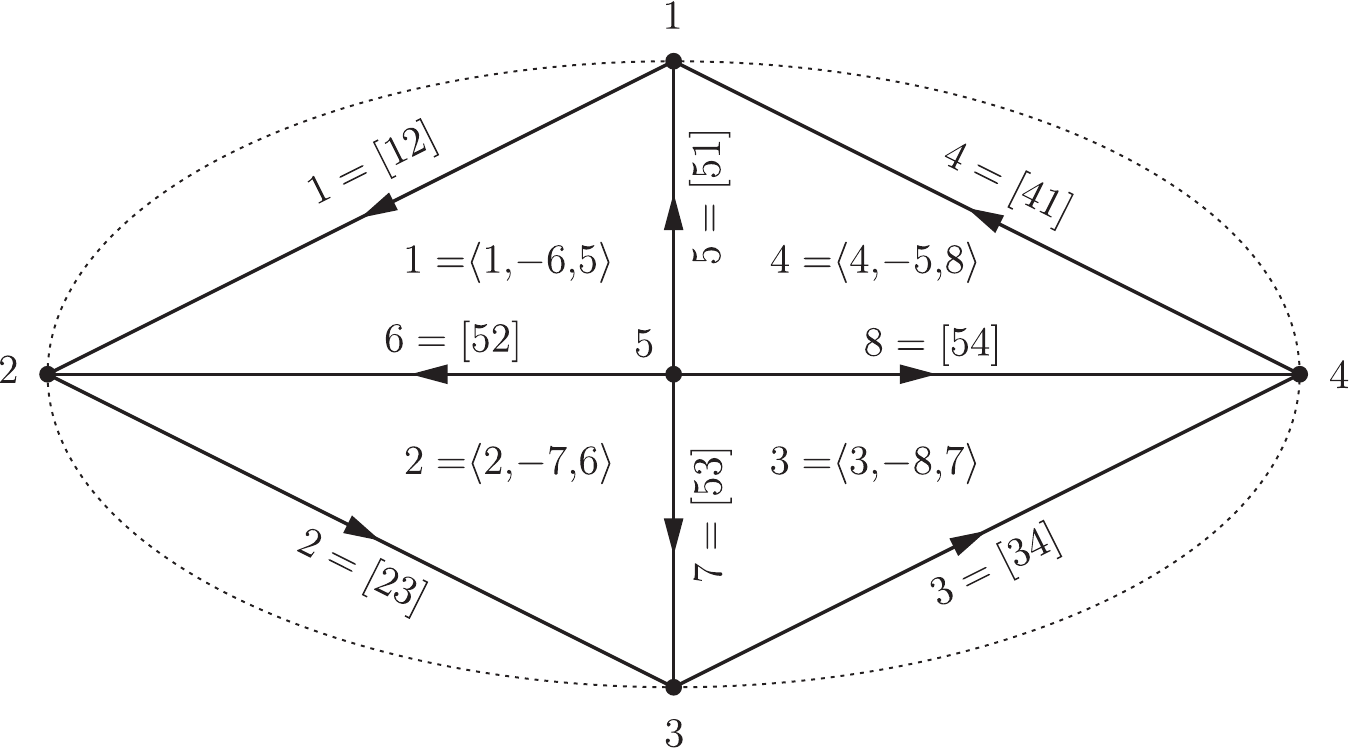}
\caption{Mesh of triangulated surface with $N_{\mathrm{V}} = 5$ vertices, $N_{\mathrm{E}} = 8$ edges, and $N_{\mathrm{F}} = 4$ faces.}
\label{fig:ellipse-ini}
%\end{wrapfigure}
\end{figure}
A discrete surface is conveniently represented by a triangular mesh which is specified by 
\begin{itemize}
\item a set of $N_{\mathrm{V}}$ vertices and their locations $(\vec{X}_k,y_k)$ in Poincare coordinates for $k=1,\ldots,N_{\mathrm{V}}$,
\item a set of $N_{\mathrm{E}}$ oriented edges $[kl] = -[lk]$ for every pair of vertices that is connected by an edge from vertex $k$ to vertex $l$,
\item a set of $N_{\mathrm{F}}$ oriented faces $\langle efg\rangle$ for each triplet of edges $e=[kl]$, $f=[lm]$, $g=[mk]$ that form a triangular piece of the surface. (The orientation of the edges defines the surface normal according to the right-hand-rule.)
\end{itemize}

An utterly crude approximation to a surface with elliptical boundary is shown in \figref{fig:ellipse-ini}. This mesh is, in fact, our input for the \textit{Surface Evolver} --- a conveniently versatile, open-source software that allows one to interactively determine surfaces by minimizing user-defined functionals under user-defined constraints \cite{brakke1992}. The actual input file defining this surface is reproduced in \figref{fig:Evolver-Input}. By providing the vertices of the boundary in a parametric form, we ensure that new vertices, which arise from refining the mesh during the surface evolution, lie on a proper ellipse. As the target space metric diverges at the boundary, we need to introduce the cutoff \texttt{epsilon}. Initially, we start with a large cutoff and then reduce it during the evolution to values of around $0.1$.

The output of \textit{Surface Evolver} is a list of vertices, a list of edges and a list of faces --- much like the input was. The number of vertices, edges, and faces may have changed due to splitting large triangles or weeding out small ones during runtime. The positions of the vertices of the (aproximate) minimal surface are given by their Poincar\'e coordinates. From these coordinates, we calculate the values $\underline{g}_k$ as estimates for the smooth field $g(z)$ at each vertex $k$:
\be \label{eqn:discrete-g}
  \underline{g}_k = e^{X_k^1 P_1 + X_k^2 P_2} y_k^D = \matr{cc}{\sqrt{y_k} & \frac{X_k^1 - i X_k^2}{\sqrt{y_k}} \\[2mm] 0 & \frac{1}{\sqrt{y_k}} } \; .
\ee

\begin{wrapfigure}{r0}{65mm}
\raggedleft
\begin{minipage}{0.38\textwidth}
\scriptsize 
\begin{verbatim}
metric
1/z^2  0      0
0      1/z^2  0
0      0      1/z^2
\end{verbatim}
\begin{verbatim}
PARAMETER xmax = 1     // small semi axis
PARAMETER ymax = 2     // large semi axis
PARAMETER epsilon = 1  // cutoff
\end{verbatim}
\begin{verbatim}
boundary 1 parameters 1  // ellipse
x: xmax * cos(p1)
y: ymax * sin(p1)
z: epsilon
\end{verbatim}
\begin{verbatim}
vertices
1    0       boundary 1    fixed
2    1*pi/2  boundary 1    fixed
3    2*pi/2  boundary 1    fixed
4    3*pi/2  boundary 1    fixed
5    0    0    epsilon+1
\end{verbatim}
\begin{verbatim}
edges
1    1 2     boundary 1    fixed
2    2 3     boundary 1    fixed
3    3 4     boundary 1    fixed
4    4 1     boundary 1    fixed
5    5 1
6    5 2
7    5 3
8    5 4
\end{verbatim}
\begin{verbatim}
faces
1    1 -6 5
2    2 -7 6
3    3 -8 7
4    4 -5 8
\end{verbatim}
\begin{verbatim}
read
\end{verbatim}
\end{minipage}
\caption{Input file for \textit{Surface Evolver} for a surface with elliptical boundary. The resulting minimal surface is shown in the upper left inset of \protect\figref{fig:ellipse-family}.}
\label{fig:Evolver-Input}
\end{wrapfigure}
Here and below, we denote the discretized counterpart of a continuous quantity by underlining it. The Maurer--Cartan form $U = g^{-1}\diff g$ contains a derivative, hence its discretization should be some kind of difference of the values of $g$ at adjacent vertices. The Maurer--Cartan form is therefore more naturally associated to edges, rather than to individual vertices. In fact, in discrete geometry, a discrete one-form $\underline{U}$ is obtained from a continuous one-form $U$ by calculating the integrated projection
\be
  \underline{U}_{[kl]} := \int_{[kl]} U \; .
\ee
In our case, we do not know the continuous $U$ as we are only able to calculate the discretized minimal surface. Still, we can get some ``inspiration'' by looking at the continuous case, where we have $g(z) = g(z_0) \mathcal{P} \exp \int_{z_0}^z U$. Writing this equation for the edge $[kl]$ and assuming that $U$ was constant along this edge, we obtain $\underline{g}_l = \underline{g}_k \exp \int_{[kl]} U = \underline{g}_k \exp \underline{U}_{[kl]}$, or 
\be
  \underline{U}_{[kl]} \stackrel{!}{=} \ln \bigbrk{ \underline{g}_k^{-1} \underline{g}_l} \, ,
\ee
which we will take as the definition of the discrete Maurer--Cartan form. Using the explicit form of $\underline{g}$ in \eqref{eqn:discrete-g}, we obtain
\be \label{eqn:discrete-U}
  \underline{U}_{[kl]} = \matr{cc}{ \frac{1}{2} & \frac{(X^1_l - X^1_k) - i ( X^2_l - X^2_k)}{y_l - y_k} \\[2mm] 0 & -\frac{1}{2} } \ln\frac{y_l}{y_k} \; .
\ee
Interestingly, the same formula is obtained when we assume a linear interpolation of the Poincar\'{e} coordinates along the edges. To see this, set $\vec{X}(\gamma) = \vec{X}_k + \bigbrk{ \vec{X}_l - \vec{X}_k } \gamma$ and $y(\gamma) = y_k + \bigbrk{ y_l - y_k } \gamma$ where $\gamma$ is a parameter that runs from 0 to 1. When these formulas are inserted into $g^{-1}\diff g$ and the result is integrated over $\gamma$, we obtain
\be
  \underline{U}_{[kl]} = \int \matr{cc}{ \frac{\diff y}{2y} & \frac{\diff X^1 - i\diff X^2}{y} \\[2mm] 0 & -\frac{\diff y}{2y} }
           = \int_{\gamma = 0}^1 \matr{cc}{ \frac{y_l - y_k}{2} & (X_l^1 - X_k^1) - i (X_l^2 - X_k^2) \\[2mm] 0 & -\frac{y_l - y_k }{2} } \frac{\diff\gamma}{y(\gamma)} \; ,
\ee
which yields exactly \eqref{eqn:discrete-U}. Now, calculating the even and odd parts, \eqref{eqn:Ucomp}, is done trivially according to the formulas
\be
  \underline{A}_{[kl]} = \frac{1}{2} \bigbrk{ \underline{U}_{[kl]} - \underline{U}^\dagger_{[kl]} }
  \comma
  \underline{a}_{[kl]} = \frac{1}{2} \bigbrk{ \underline{U}_{[kl]} + \underline{U}^\dagger_{[kl]} } \; .
\ee

The calculation of the discretization $\underline{b}_{[kl]}$ of the Hodge dual $b := \ast a$ from the discrete $\underline{a}_{[kl]}$ is the main challenge of the algorithm. By definition, we have $\ast \diff z = - i \diff z$, which implies that $b_z = -ia_z$ and $b_{\bar z} = ia_{\bar z}$. So, instead of the components of $a$ along the edges of the mesh ($a_{[kl]}$), we rather need its decomposition with respect to a coordinate system that conformally parametrizes the surface ($a_z$, $a_{\bar{z}}$). This is where the conformal gauge choice made in writing the master symmetry transformation enters the calculation in the discrete setting. So, let us turn to the calculation of this conformal parametrization next.

We want to construct a map $z$ from the surface to the complex plane which is conformal, i.e.\ which preserves angles (and since the surface is curved, this map will not preserve distances). Conformality is most conveniently expressed through the Cauchy--Riemann equations. Using the local surface coordinates $(\xi,\eta)$ induced by the Poincar\'{e} coordinates of the surface embedded in $\mathrm{EAdS}_3$, the condition of conformality is written as
\be \label{eqn:Cauchy-Riemann-condition}
  \frac{\partial z}{\partial \xi} + i \frac{\partial z}{\partial \eta} = 0 \; ,
\ee
where $z(\xi,\eta) \in \Complex$ is the searched-for conformal parametrization. In the discretized setting, this map is commonly approximated by a piecewise linear function which is specified by a set of complex numbers $\{z_k\}$ giving the values of the conformal parametrization at the vertices $k=1,\ldots,N_{\mathrm{V}}$. Within each triangle, the map thus has the form
\be
  z = A\,\xi + B\,\eta + C \; ,
\ee
where $A$, $B$, $C$ are constants specific to each triangle. Denoting the local coordinates for each vertex of a given triangle by $(\xi_k,\eta_k)$ such that $z(\xi_k,\eta_k) = z_k$, we can solve for the constants $A$, $B$, and $C$ and obtain
\begin{align}
  A & = \frac{\partial z}{\partial \xi}  = \frac{ (\eta_2-\eta_3)z_1 + (\eta_3-\eta_1)z_2 + (\eta_1-\eta_2)z_3 }{ (\xi_2 \eta_3 - \eta_2 \xi_3) + (\xi_3 \eta_1 - \eta_3 \xi_1) + (\xi_1 \eta_2 - \eta_1 \xi_2) } \; , \\[1mm]
  B & = \frac{\partial z}{\partial \eta} = \frac{ (\xi_3 -\xi_2 )z_1 + (\xi_1 -\xi_3 )z_2 + (\xi_2 -\xi_1 )z_3 }{ (\xi_2 \eta_3 - \eta_2 \xi_3) + (\xi_3 \eta_1 - \eta_3 \xi_1) + (\xi_1 \eta_2 - \eta_1 \xi_2) } \; .
\end{align}
The constant $C$ is also fixed 
but in light of \eqref{eqn:Cauchy-Riemann-condition}, we only care about the partial derivatives. In principle, we should fix the numbers $z_k$ such that \eqref{eqn:Cauchy-Riemann-condition} is satisfied everywhere on the surface. However, for the discretized map, this condition is in general too strong and cannot be satisfied by any choice of $z_k$'s. The appropriate weaker condition, resulting in what is referred to as quasi-conformal map, is to find $z_k$'s that minimize the so-called conformal energy
\be \label{eqn:conformal-energy}
  E_{\mathrm{C}} = \int \lrabs{ \frac{\partial z}{\partial \xi} + i \frac{\partial z}{\partial \eta} }^2 \,\diff\mathcal{A} = \sum_{t} \lrabs{ \frac{\partial z}{\partial \xi} + i \frac{\partial z}{\partial \eta} }^2 \mathcal{A}_t \, .
\ee
The integrand in \eqref{eqn:conformal-energy} is piecewise constants such that the integral over the surface simplifies to a sum over all triangles weighted by the areas $\mathcal{A}_t$ of each triangle $t$. The conformal energy is manifestly non-negative, and it vanishes if and only if the map is exactly conformal. Furthermore, it is a quadratic form of the numbers $\{z_k\}$ and can be written in the form
\be
  E_{\mathrm{C}} = Z^\dagger M^\dagger M Z = \abs{\abs{MZ}}^2 \; ,  
\ee
where $Z = (z_1\; z_2 \; \cdots \; z_{N_{\mathrm{V}}})^\trans$ and $M$ is a constant rectangular matrix of dimension $N_{\mathrm{F}} \times N_{\mathrm{V}}$ whose entries are calculated from the local surface coordinates $(\xi_k,\eta_k)$. Now, our task is to find a nontrivial vector $Z$ for which $E_C$ is minimal. The solution $Z=0$, which clearly minimizes $E_C$, should not allowed as it corresponds to a trivial map. In order to properly remove this option, one imposes some appropriate constraint on $Z$, which we take to be $\abs{\abs{Z}} = 1$. Now, the constrained minimization problem is solved by the eigenvector of $K =  M^\dagger M$ with the smallest eigenvalue. This can be argued as follows. Say $Z_i$ with $i=1,\ldots,N_{\mathrm{V}}$ is the ordered list of orthonormalized eigenvectors of $K$ with eigenvalues $\lambda_1 \le \lambda_2 \le \ldots \le \lambda_{N_{\mathrm{V}}}$. As the basis of eigenvectors is complete, we can expand any vector as $Z = \sum_i \alpha_i Z_i$, such that
\be
  E_{\mathrm{C}} = \sum_i \lambda_i \abs{\alpha_i}^2 \; .
\ee
Note that the constraint on Z implies $\sum_i \abs{\alpha_i}^2 = 1$. This shows that the minimum is attained when only $\alpha_1$ is turned on, i.e.\ when $Z = Z_1$. Hence, the calculation of the discrete conformal parametrization has boiled down to the calculation of an eigenvector of a large, but rather sparse matrix, which is something that e.g.\ \textit{Mathematica} can do easily.

Equipped with the conformal parametrization, we can now resume the calculation of the Hodge dual $b = *a$. Specifically, our goal is to derive a formula for the projections $\underline{b}_{[kl]}$ for all edges $[kl]$ from the projections $\underline{a}_{[mn]}$. To this end, we will be thinking of the algebra valued one-forms $a$ and $b$ as arrows $\vec{a}$ and $\vec{b}$, i.e.\ we understand them literally as co-\emph{vectors}. The real and imaginary parts, $\diff \sigma$ and $\diff \tau$, of the complex one-form $\diff z$ represent the Cartesian basis vectors. The Hodge star operation acts as $*\diff \sigma = \diff \tau$ and $*\diff \tau = -\diff \sigma$ and is therefore nothing but a counter-clockwise $90^\circ$ rotation. Thus, in this language, $\vec{b}$ is obtained from $\vec{a}$ by such a rotation.

\begin{wrapfigure}{r0}{65mm}
\centering
\includegraphics[width=65mm]{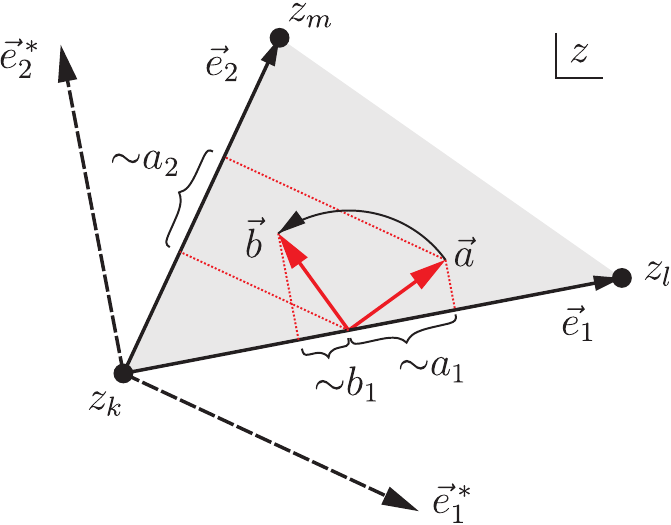}
\caption{\textbf{Discrete Hodge Dual.} The edges of the mesh in the conformal parametrization define the basis vectors $\vec{e}_1 \hat{=} z_l - z_k$ and $\vec{e}_2 \hat{=} z_m - z_k$. The projection $b_1$ of the Hodge dual $\vec{b} = *\vec{a}$ is determined by the projections $a_1$ and $a_2$ onto two adjacent edges.}
\label{fig:Hodge-Dual}
\end{wrapfigure}
Since the projection of $\vec{a}$ onto a single edge of the mesh does not allow us to reconstruct uniquely the vector $\vec{a}$, a single projection will not suffice to determine the rotated vector $\vec{b}$ either. Thus, we also need to include information about $\vec{a}$ from the projection onto another, adjacent edge. To be specific, let us say we wish to calculate $b_1$ for edge $1 = [kl]$ and we use this edge and the edge $2=[km]$ to reconstruct $\vec{a}$ from the respective components
\be
  a_1 = \vec{a}\cdot\vec{e}_1  \comma   a_2 = \vec{a}\cdot\vec{e}_2 \; ,
\ee
where the basis vectors $\vec{e}_1$ and $\vec{e}_2$ point from vertex $k$ to $l$ and $m$, respectively, see figure \figref{fig:Hodge-Dual}. This figure shows one face of the mesh pulled back to the complex plane where each vertex is drawn according to its value in the conformal parametrization. As $\{\vec{e}_i\}$ is not an orthonormal basis, we need to introduce the dual basis $\{\vec{e}_i^{\:\star}\}$ according to
\be \label{eqn:dual-basis}
  \vec{e}_i^{\:\star} \cdot \vec{e}_j = \delta_{ij}
\ee
in order to write $\vec{a} = a_1 \vec{e}_1^{\:\star} + a_2 \vec{e}_2^{\:\star}$. Under a $90^\circ$ counter-clockwise rotation, $\vec{e}_1^{\:\star}$ ends up parallel to $\vec{e}_2$, and $\vec{e}_2^{\:\star}$ ends up anti-parallel to $\vec{e}_1$. Taking the different lengths into account, we obtain the formulas
\be
  *\vec{e}_1^{\:\star} =  \frac{\abs{\vec{e}_1^{\:\star}}}{\abs{\vec{e}_2}} \vec{e}_2
  \comma
  *\vec{e}_2^{\:\star} = -\frac{\abs{\vec{e}_2^{\:\star}}}{\abs{\vec{e}_1}} \vec{e}_1
  \; ,
\ee
which allow us to write down the Hodge dual explicitly as
\be
  \vec{b} = *\vec{a} = a_1 \frac{\abs{\vec{e}_1^{\:\star}}}{\abs{\vec{e}_2}} \vec{e}_2 - a_2 \frac{\abs{\vec{e}_2^{\:\star}}}{\abs{\vec{e}_1}} \vec{e}_1 \; .
\ee
The projection onto edge $1$ is then given by taking the scalar product 
\be
  b_1 = \vec{b}\cdot \vec{e}_1 = a_1 \abs{\vec{e}_1^{\:\star}}\abs{\vec{e}_1} \cos\sphericalangle(1,2) - a_2 \abs{\vec{e}_2^{\:\star}}\abs{\vec{e}_1} \; ,
\ee
where $\sphericalangle(1,2)$ denotes the angle between the vectors $\vec{e}_1$ and $\vec{e}_2$. Using \eqref{eqn:dual-basis}, we can replace the lengths of the dual vectors and obtain
\be
  b_1 = a_1 \frac{\cos\sphericalangle(1,2)}{\cos\sphericalangle(1,1^\star)} - a_2 \frac{\abs{\vec{e}_1}}{\abs{\vec{e}_2}} \frac{1}{\cos\sphericalangle(2,2^\star)} \; .
\ee
Finally, we note that $\sphericalangle(1,1^\star) = \sphericalangle(2,2^\star) = \frac{\pi}{2} - \sphericalangle(1,2^\star)$, such we can get rid of the dual edges altogether
\be
  b_1 = a_1 \cot\sphericalangle(1,2) - a_2 \frac{\abs{\vec{e}_1}}{\abs{\vec{e}_2}} \csc\sphericalangle(1,2) \; .
\ee

The angle $\sphericalangle(1,2)$ can be calculated directly from the complex coordinates $\Delta z_1 = z_l - z_k$ and $\Delta z_2 = z_m - z_k$
\be
  \beta_{12} := \cos\sphericalangle(1,2) = \frac{1}{2} \frac{\Delta z_1 \overline{\Delta z_2} + \Delta z_2 \overline{\Delta z_1}}{\abs{\Delta z_1}\abs{\Delta z_2}} \: .
\ee
\be \label{eqn:discrete-hodge-formula}
  b_1 = \frac{1}{\sqrt{1-\beta_{12}^2}} \biggsbrk{ \beta_{12} \, a_1 - \frac{\abs{\Delta z_1}}{\abs{\Delta z_2}} a_2 } \; .
\ee
\mbox{}\begin{wrapfigure}{r0}{50mm}
\centering
\includegraphics[width=50mm]{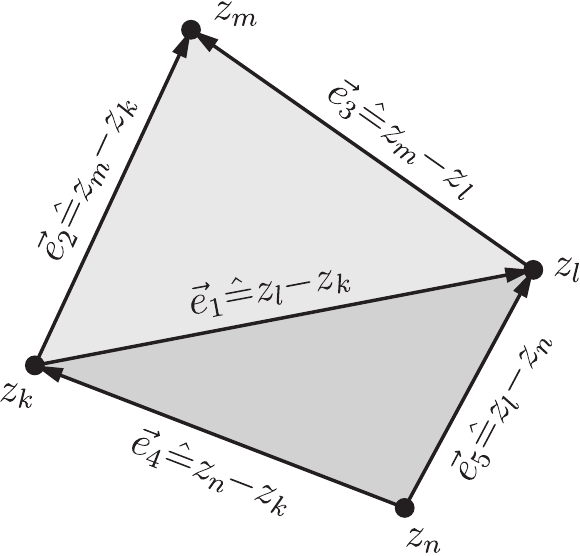}
\caption{\textbf{\mathversion{bold}Neighbors of edge $[kl]$.} The calculation of the discrete Hodge dual $(*a)_{[kl]}$ requires the knowledge of $a_{[kl]}$ and one of $a_{[km]}$, $a_{[lm]}$, $a_{[nk]}$, and $a_{[nl]}$.}
\label{fig:Edges}
\end{wrapfigure} %
Similar formulas for $b_1$ can be obtained if one uses any of the three other edges that are adjacent to $1=[kl]$. In fact, the formula is literally the same, including all signs, as in \eqref{eqn:discrete-hodge-formula}, if the orientation of the edges is chosen as in \figref{fig:Edges}. This means, we get an equivalently good estimate for $b_1$, if we replace edge $2=[km]$ in \eqref{eqn:discrete-hodge-formula} by any of the edges $3=[lm]$ or  $4=[nk]$ or  $5=[nl]$. We call them ``estimates'' because the discretization assumes that the (co)vector $\vec{a}$ is constant within the faces. Thus, we should get the best estimate if we average over all four ways\footnote{If edge $1=[kl]$ is a boundary edge, then there are only two ways to calculate the Hodge dual.} in which $b_1$ can be calculated. Although we employed a vector notation in the derivation, formula \eqref{eqn:discrete-hodge-formula} also applies to matrix-valued one-forms.

Eventually, we are in the position to define the deformed discrete Maurer--Cartan form as 
\be
  \underline{L_u}_{[kl]} \eq A_{[kl]} + \cos\theta \, a_{[kl]} + \sin{\theta} \, b_{[kl]} \; ,
\ee
which is nothing but the definition \eqref{eqn:definition-Lax} with \eqref{eq:theta} written for each edge $[kl]$.

This discrete one-form needs to be ``integrated'' along the edges of the mesh to find the deformed surface $\underline{g_u}_k$. To this end, we pick an arbitrary vertex $k_0$ to set $\underline{g_u}_{k_0} = \underline{g}_{k_0}$ as initial condition for the integration. Because of the divergence in the metric near the boundary of $\mathrm{EAdS}$, it has proved to be advantages for greater numerical accuracy to impose the initial condition at a vertex that is far from the boundary. Then, an unknown $\underline{g_u}_l$ at some vertex $l$ can be calculated from a known $\underline{g_u}_k$ at some adjacent vertex $k$ by
\be
  \underline{g_u}_l = \underline{g_u}_k \exp\bigbrk{\underline{L_u}_{[kl]}} \; .
\ee
If $\underline{g_u}$ is known at more than one vertex adjacent to vertex $l$, then the average of all results obtained form calculating $\underline{g_u}_l$ in all possible ways should be used for best numerical precision.

The last step is to extract the Poincar\'e coordinates of the vertices of the deformed surface from the matrices $\underline{g_u}_k$. To remove the gauge freedom in the representation, we first calculate at each vertex the matrix 
\be
  \underline{\mathbb{G}}_k = \underline{g_u}_k \underline{g_u^\dagger}_k
  = \matr{cc}{\frac{(X_{u,k}^1)^2 + (X_{u,k}^2)^2 + y_{u,k}^2}{y_{u,k}} & \frac{X_{u,k}^1 - i X_{u,k}^2}{y_{u,k}} \\[2mm] \frac{X_{u,k}^1 + i X_{u,k}^2}{y_{u,k}} & \frac{1}{y_{u,k}} }
\ee
and then read off the Poincar\'e coordinates from the components as follows:
\be
  X_{u,k}^1 = \frac{1}{2} \frac{\underline{\mathbb{G}}_{k,12} + \mathbb{G}_{k,21}}{\mathbb{G}_{k,22}}
  \comma
  X_{u,k}^2 = \frac{i}{2} \frac{\mathbb{G}_{k,12} - \mathbb{G}_{k,21}}{\mathbb{G}_{k,22}}
  \comma
  y_{u,k} = \frac{1}{\mathbb{G}_{k,22}} \; .
\ee

\begin{figure}
\centering
\includegraphics[width=0.9\textwidth]{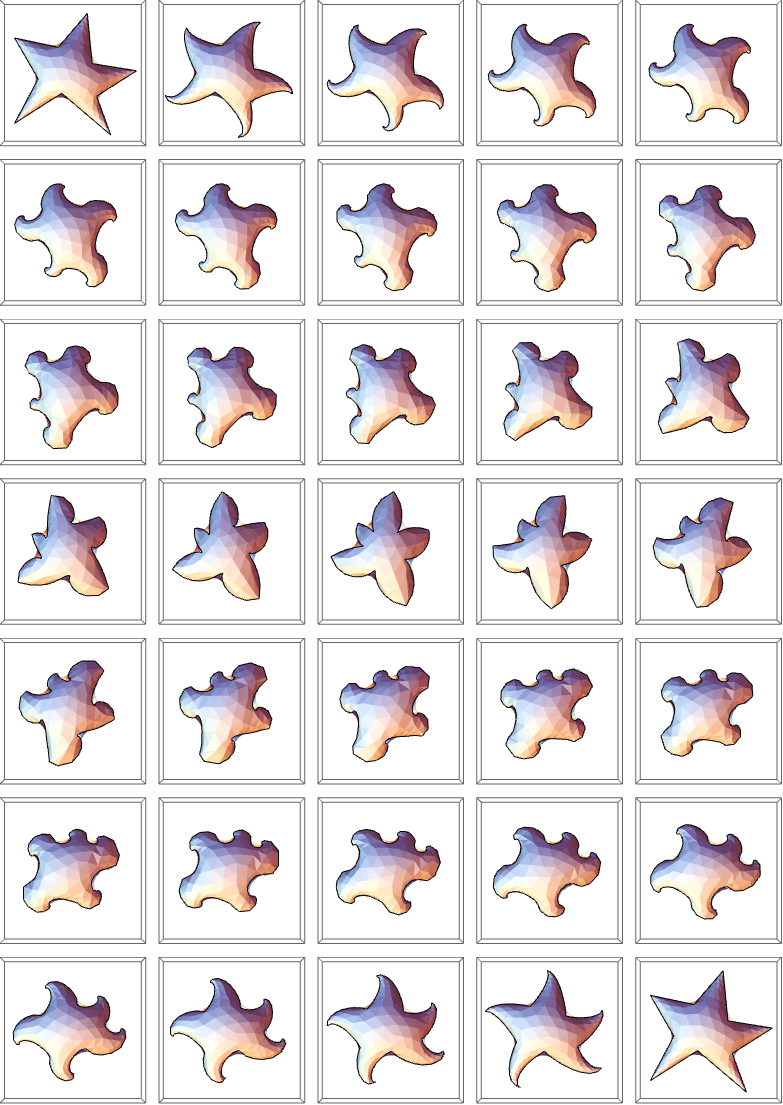}
\caption{\textbf{Irregular shaped star.} Family of associated surfaces plotted for spectral parameter values $\theta = 0, \frac{\pi}{17}, \ldots, 2\pi$ as viewed from below the boundary of $\mathrm{EAdS}$.}
\label{fig:Star}
\end{figure}

\begin{figure}
\centering
\includegraphics[width=0.9\textwidth]{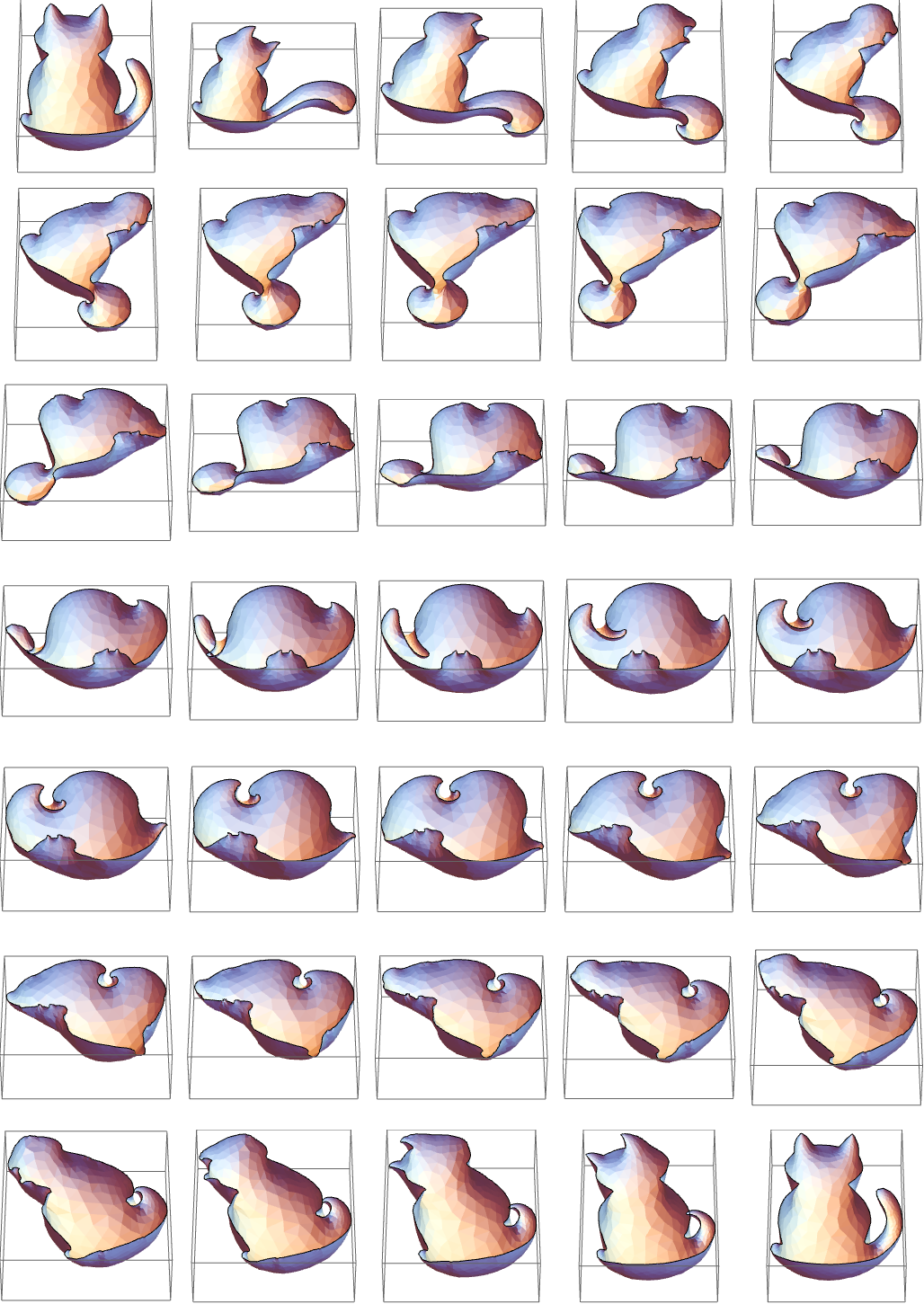}
\caption{\textbf{Silhouette of a cat.} As seen in this example, there is no guarantee that the deformed surface is not self-intersecting.}
\label{fig:Kitty}
\end{figure}

We conclude this section with two nontrivial examples given in \figref{fig:Star} and \figref{fig:Kitty}.

%%%%%%%%%%%%%%%%%%%%%%%%%%%%%%%%%%%%%%%%%%%%%%%%%%%%%%%%%%%%%%%%%%%%%%%%%%%
\clearpage
\section{Conclusions and Outlook}
\label{sec:conclusions}

In this paper we have given a generic discussion of nonlocal symmetries of symmetric space models. In particular, we laid the focus on the nonlocal master symmetry which generates the spectral parameter and thereby implies the model's integrability, manifested in the Yangian algebra. We should emphasize that the Yangian is the expected symmetry structure underlying an integrable model of rational type. On the other hand, a master symmetry that generates the spectral parameter appears to feature only in some integrable systems. 
\medskip

An intriguing question is whether this master symmetry is merely a property of the integrable sigma model describing the AdS/CFT correspondence at \emph{strong coupling}, or whether it is a feature of the planar AdS/CFT duality \emph{in general}. In particular, it will be very interesting to clarify whether a similar symmetry can be formulated for the weakly coupled $\superN=4$ SYM theory. If possible, this should pave the way to establish the role of the spectral parameter in the context of this four-dimensional gauge theory. Notably, traces of a spectral parameter at weak coupling have been identified in the context of tree-level scattering amplitudes \cite{Ferro:2012xw}, and it would be very enlightening to establish a connection to the present work.
\medskip

As demonstrated explicitly in \secref{sec:WilsonLoops} and \secref{sec:numerics}, the master symmetry generates a one-parameter family of Maldacena--Wilson loops with constant expectation value.
A natural starting point for extending these considerations would be to apply this symmetry to polygonal Wilson loops with lightlike edges. It is well-established by now that this class of Wilson loops is dual to scattering amplitudes in $\superN=4$ SYM theory \cite{Alday:2007hr,Drummond:2008aq}, which underlines the importance of understanding their behavior. Another question in this direction is the relation between the master symmetry and dual conformal symmetry. The latter furnishes an AdS/CFT-specific alternative to understand the nonlocal Yangian symmetry \cite{Drummond:2010km}. At strong coupling, dual conformal symmetry was understood to originate from the self-T-duality of the $\mathrm{AdS}_5\times \mathrm{S}^5$ (super)string model \cite{Berkovits:2008ic,Beisert:2008iq}. What is the meaning of the above master symmetry in the context of this self-duality?
\medskip

In order to approach the above points, one might first want to study the considered master symmetry in the context of the full $\mathrm{AdS}_5\times \mathrm{S}^5$ sigma model with $\mathbb{Z}_4$ grading. While a $\mathbb{Z}_2$ grading is sufficient to establish the symmetry, a generic study should be based on the understanding of all algebraic constituents. For recent progress in this direction see \cite{Chandia:2016ueo}, which employs the pure spinor formalism to study the master symmetry for the $\mathrm{AdS}_5\times \mathrm{S}^5$ superstring model; see also \cite{Beisert:2012ue} for a generic construction for models with $\mathbb{Z}_4$ grading, which starts from the Maurer--Cartan form.
\medskip

The original motivation to study the nonlocal symmetries in this paper was the spectral-parameter deformation of Wilson loops studied in \cite{Ishizeki:2011bf}. In that context, the so-called Schwarzian derivative has proven to be an extremely useful mathematical tool. The Schwarzian derivative has a plethora of applications in mathematics and features for instance in the solution theory of differential equations. In particular, it was shown that the Schwarzian derivative furnishes an instrument  to efficiently compute the area of minimal surfaces in $\mathrm{AdS}_3$; see e.g.\ the recent works \cite{Irrgang:2015txa,Huang:2016atz}. We believe that understanding the connection between this concept and the master symmetry should result in a deeper understanding of the mathematics underlying holographic Wilson loops.
\medskip

From an algebraic point of view, the considered master symmetry furnishes a raising operator for the Yangian levels. A respective lowering operator, an automorphism of the Yangian algebra, was actually introduced with Drinfel'd's original definition of the Yangian and its relation to the quantum Yang--Baxter equation. In 1$+$1-dimensional models, the latter is often realized through the Lorentz boost or its discrete analogue, the spin chain boost operator, see e.g. \cite{Loebbert:2016cdm}. It seems worthwhile to better understand the relation between this boost automorphism and the above master symmetry and to clarify the algebraic role of the latter for the Yangian. Another curious question is whether the Casimirs of the underlying Lie algebra, which were identified above as the conserved charges associated with the master symmetry, play a natural role here.
\medskip

For the $\mathrm{AdS}_5/\mathrm{CFT}_4$ duality there is in fact another distinguished generator which raises the level of the Yangian. The so-called bonus or secret symmetry represents a bilocal operator (i.e.\ an operator of the same type as the Yangian level-1 generators) that is a particular feature of the Yangian over the superconformal algebra $\alg{psu}(2,2|4)$ or the algebra $\alg{su}(2|2)$, respectively, and has been identified as a symmetry of  the two-dimensional AdS/CFT worldsheet S-matrix \cite{Matsumoto:2007rh}, scattering amplitudes \cite{Beisert:2011pn} and Wilson loops \cite{Munkler:2015xqa}. While the above master symmetry appears to be independent of the choice of the underlying Lie algebra, the bonus symmetry is not. In the context of dual conformal symmetry, the bonus generator is special since it is bilocal in both the ordinary and the dual coordinates, while this does not apply to the rest of the Yangian generators. In particular, it should be worked out how the master variation acts on the bonus charge.
\medskip

Another open question is whether it is useful to investigate the master symmetry on observables at strong coupling that are different from Wilson loops. For instance one may think of including this symmetry into the integrability formalism for three-point correlation functions or of utilizing them in the bootstrap program for form factors at strong coupling.
\medskip 

A very interesting but more technical question concerns the Poisson algebra of the Yangian generators. As demonstrated in \secref{sec:Poissbrack} one can show that the respective generators indeed satisfy the classical analogue of the Yangian algebra on the infinite line. However, to arrive at the correct relations, one has to tune the order of limits of the boundaries of integration, such that unwanted contributions from boundary terms are avoided. On the one hand, one may argue that the Yangian commutation relations thereby fix the limit ambiguity in the definition of the generators \cite{MacKay:1992he}. On the other hand this means that, to arrive at the correct Yangian relations, one has to make a particular choice. It would be desirable to bypass this choice and to find a reasoning that yields the classical Yangian as the unique structure following from the above definition of the generators. We note that a similar ambiguity exists in the case of the principal chiral model \cite{MacKay:1992he}, which can be expressed in the language of symmetric space models, cf.\ \secref{app:PCM}.
\medskip

Obviously, it would be very interesting to see whether a similar master symmetry can be defined in the context of other integrable AdS/CFT dualities. In the case of the $\mathrm{AdS}_4/\mathrm{CFT}_3$ duality (ABJM theory), this might reveal surprises since its sigma model formulation appears to have some significant qualitative differences to that of $\mathrm{AdS}_5/\mathrm{CFT}_4$. It is still an open problem, for instance, whether the observed traces of a dual conformal (alias Yangian) symmetry can be explained via an analogue of the fermionic self-T-duality \cite{Adam:2010hh,Colgain:2016gdj}. Another interesting connection point could be the q-deformed boost operator that was recently studied in the context of the $\mathrm{AdS}_3/\mathrm{CFT}_2$ duality \cite{Fontanella:2016opq}. Considering the analogy to the boost automorphism of the Yangian algebra, this observation suggests to look for a q-deformed counterpart of the master symmetry.
\medskip 

Finally, we are tempted to speculate whether the considered master symmetry can be defined in other integrable field theories in two dimensions. Attractive cases to analyze are the (chiral) Gross--Neveu or the Landau--Lifschitz model for instance. An extension to the latter would be interesting in the context of the AdS/CFT correspondence, since the Landau--Lifschitz model represents the thermodynamic limit of the weakly coupled and the ultrarelativistic limit of the strongly coupled AdS/CFT model in the $\alg{su}(2)$ sector \cite{Kruczenski:2003gt}.

%%%%%%%%%%%%%%%%%%%%%%%%%
\subsection*{Acknowledgements}
We owe thanks to Till Bargheer, Amit Dekel, Harald Dorn, Ben Hoare, Martin Kruczenski, Tristan McLoughlin, Jan Plefka, Arkady Tseytlin and Gang Yang for useful discussions.
This research is supported in part by the SFB 647 \textit{``Space-Time-Matter. Analytic and Geometric Structures''} and the Research Training Group GK 1504
\textit{``Mass, Spectrum, Symmetry''}.

%%%%%%%%%%%%%%%%%%%%%%%%%
%%%%%%%%%%%%%%%%%%%%%%%%%
\appendix

\section{Fundamental Representation of SO(1,5)}
\label{app:SO15}
In this appendix, we provide our conventions for the fundamental representation of $\grp{SO}(1,5)$. We take the metric $\eta_{IJ} = \mathrm{diag}(1, \ldots , 1 , -1 )$ with indices $I,J$ running from 1 to 6. The basis elements $M_{IJ}$ are given by
\begin{align}
\left( M_{I J} \right) ^a {} _b := \eta_{J b} \, \delta _I ^a - \eta_{I b} \, \delta _J ^a  \, .
\end{align}
They satisfy the commutation relations
\begin{align}
\left[M_{IJ} , M_{KL} \right] = \eta_{I L} M_{J K} - \eta_{J L} M_{I K} + \eta_{J K} M_{I L} - \eta_{I K} M_{J L} \, . 
\label{comm:MIJ}
\end{align}
We define a metric on the Lie algebra using the trace in the fundamental representation. This metric is related to the Killing metric by a simple rescaling. For the basis elements $M_{IJ}$, the metric is given by
\begin{align}
\tr \left( M_{I J} M_{K L} \right) = 2\, \eta_{I L} \, \eta_{J K} - 2\, \eta_{I K} \, \eta_{J L} \, .
\label{metric:MIJ}
\end{align}
It is well-known that the conformal algebra associated to four-dimensional Euclidean space is isomorphic to $\mathfrak{so}(1,5)$. The basis elements introduced above may be related to the usual basis elements of the conformal algebra by 
\begin{align}
P_\mu &= M_{\mu 6} - M_{\mu 5} \, , &
K_\mu &= M_{\mu 6} + M_{\mu 5} \, , &
D &= M_{5 6} \, .
\end{align}
Here, the index $\mu$ extends from 1 to 4. Apart from the commutation relations \eqref{comm:MIJ}, the non-vanishing commutators are given by
\begin{alignat}{3}
\left[ D, P_{\mu} \right] &=  P_{\mu} \, ,& \qquad  
\left[ M_{\mu \nu}, P_{\lambda} \right] &= \delta_{\nu \lambda} P_{\mu}    -  \delta_{\mu \lambda} 	 	P_{\nu} \, ,& \qquad 
\left[ P_{\mu}, K_{\nu} \right] &=  2 M_{\mu \nu} +  2 \delta_{\mu \nu} \,  D \, , \nn \\
\left[ D, K_{\mu} \right] &= - K_{\mu}\, ,& \qquad  
\left[ M_{\mu \nu}, K_{\lambda} \right] &= \delta_{\nu \lambda} K_{\mu}    -  \delta_{\mu \lambda} 	 	K_{\nu} \, .   \label{conf_algebra}
\end{alignat}
In addition to \eqref{metric:MIJ}, we note the remaining non-vanishing elements of the metric 
\begin{align}
\tr \left(P_\mu \, K_\nu \right) &= 4 \, \delta_{\mu \nu} \, , & 
\tr \left( D \, D \right) &= 2 \, .
\label{eqn:Metric}
\end{align}
In order to describe the decomposition of $\alg{so}(1,5)$ into a gauge and a coset part, we introduce the automorphism ${\Omega: \alg{so}(1,5) \to \alg{so}(1,5)}$,
\begin{align}
\Omega(X) = M X M^{-1} \, , \quad \text{where} \quad M = \mathrm{diag}(1,1,1,1,1,-1) \, .
\end{align}
We then have
\begin{align}
\alg{h} &= \mathrm{span} \left \lbrace M_{\mu \nu} , P_\mu - K_\mu \right \rbrace \simeq \alg{so}(1,4) \, , &
\alg{m} &= \mathrm{span} \left \lbrace P_\mu + K_\mu , D \right \rbrace \, .
\end{align}

\section{Calculation of Commutation Relations}
\label{app:Commutators}
In this appendix, we derive the commutation relations stated in \secref{sec:Algebra}. For two generic variations $\delta_i g = \phi_i g$, we found the commutation relation
\begin{align}
\left[ \delta_1 \, , \, \delta_2 \right] g = \left( \left[ \phi_1  ,  \phi_2 \right] + \delta_2 \phi_1 - \delta_1 \phi_2 \right) g \, .
\end{align}
All arising terms of the form $\delta_i \phi_j$ can be inferred from the variations $\delta_{\epsilon_1, u_1} \chi_{u_2}$ and $\master _{u_1} \chi_{u_2}$. To reduce the amount of writing, we introduce the abbreviations
\begin{align*}
 \delta_{\epsilon_i, u_i} g &= \delta _i \, g = \eta _i \, g =  \chi _i  ^{-1} \epsilon_i \, \chi _i \, g \, , &  \master_{u_i} g &= \psi_i g = \chi_i^{-1} \dot{\chi}_i g \, .
\end{align*}
For both $ \delta_{\epsilon, u}$ and $\master_u$ the variation of $\chi_2$ can be obtained by varying the auxiliary linear problem \eqref{eqn:auxiliary-problem}, which leads to the relations \eqref{eqn:varied-auxiliary-problem}, 
\begin{align}
\chi_2 ^{-1} \left( \diff \left( \delta  \chi_2 \right) - \left( \delta  \chi_2 \right) \ell_2 \right) &= \delta \ell_2 \, , & \delta \chi_2 (z_0, \bar{z}_0) &= 0 \, . 
\label{eqn:varied-auxiliary-problem2}
\end{align}
We begin by calculating $\delta_1 \ell_2 $. We recall that $\ell_i = \left(1 + u_i^2 \right) ^{-1} \left( u_i^2 \, j +  u_i \ast j \right)$ and use equation \eqref{eqn:delta-j} to find
\begin{align}
\delta_1 j & = - \diff \eta_1 - \left[ j , \eta_1 \right] + g \Omega \left( g^{-1} \diff \eta_1 g \right) g^{-1} 
= \left[ \ell_1 - j , \eta _1 \right] + \left[ \ell_1 , \tilde{\eta} _1 \right] \nn \\
&=  \frac{1}{1+ u_1^2} \left(  \left[- j +  u_1 \ast j , \eta_1 \right] + \left[u_1^2 \, j + u_1 \ast j , \tilde{\eta}_1 \right] \right) \, .
\end{align}
Here, we introduced the abbreviation $\tilde{\eta} = g \, \Omega \left( g^{-1} \eta g \right) g^{-1} $. The above relation implies that the variation $\delta_1 \ell_2 $ may be written as
\begin{align}
\delta_1 \ell_2 = \frac{1}{1+u_2^2} \left( u_2 ^2 \, \delta_1 j + u_2 \ast \delta_1 j \right) 
= \frac{u_2}{u_1-u_2} \left[ \ell_2 - \ell_1 , \eta_1 \right] + \frac{u_1 u_2}{1 + u_1 u_2} \left[ \ell_2 + \ell_1 - j , \tilde{\eta}_1 \right] \, .
\end{align}
We now construct the solution to equation \eqref{eqn:varied-auxiliary-problem2} from the ansatz
\begin{align}
\delta_1 \chi_2 &= \gamma _1 \left(\delta_1 \chi_2 \right)_1 + \gamma _2 \left(\delta_1 \chi_2 \right)_2 \, , & \left(\delta_1 \chi_2 \right)_1 &= \chi_2 \eta_1 - \epsilon_1 \chi_2  \, , & \left(\delta_1 \chi_2 \right)_2 &= \chi_2 \tilde{\eta}_1 - \epsilon_1 ^\prime \chi_2 \, .
\end{align}
Here, $\epsilon^\prime$ is defined by
\begin{align*}
\epsilon ^\prime &= \tilde{\eta}( z_0 , \bar{z}_0 )  = g_0 \, \Omega \left( g_0 ^{-1} \epsilon g_0 \right) g_0 ^{-1} \, , &
g_0 &= g ( z_0 , \bar{z}_0 ) \, ,
\end{align*}
such that both $\left(\delta_1 \chi_2 \right)_1$ and $\left(\delta_1 \chi_2 \right)_2$ satisfy the boundary condition $\delta_1 \chi_2 (z_0, \bar{z}_0) = 0$. Making use of the relation
\begin{align}
\diff \tilde{\eta} = g \, \Omega \left(g^{-1}\left( \diff \eta + \left[j, \eta \right] \right) g \right) g^{-1} = \left[\ell - j , \tilde{\eta} \right] \, , 
\end{align}
we then obtain 
\begin{align*}
\chi_2 ^{-1} \left( \diff \left( \delta_1  \chi_2 \right) - \left( \delta_1  \chi_2 \right) \ell_2 \right) = \gamma _1 \left[ \ell_2 - \ell_1 , \eta_1 \right] + \gamma_2 \left[ \ell_2 + \ell_1 - j , \tilde{\eta}_1 \right] \, , 
\end{align*}
from which we can read of the solution to equation \eqref{eqn:varied-auxiliary-problem2} as 
\begin{align}
\delta_1 \chi_2 = \frac{u_2}{u_1-u_2} \left( \chi_2 \eta_1 - \epsilon_1 \chi_2 \right) + \frac{u_1 u_2}{1+ u_1 u_2} \left( \chi_2 \tilde{\eta}_1 - \epsilon_1 ^\prime \chi_2 \right) \, .
\label{eqn:delta1chi2}
\end{align}
We proceed similarly for the calculation of $\master_{u_1} \chi_2$ and begin by calculating $\master_{u_1} \ell_2$. Using equation \eqref{eqn:delta-j} as well as the relation 
$\diff \psi = [ \psi , \ell ] + \dot{\ell} $, we have
\begin{align}
\master_{u_1} \,  j = \big[\ell_1 -j , \psi_1 \big] +  \big[\ell_1 , \tilde{\psi}_1 \big] - 2 \, \dot{\ell}_1 \, , 
\end{align} 
which implies that
\begin{align}
\master_{u_1} \, \ell_2 = \frac{u_2}{u_1 - u_2} \, \big[\ell_2 - \ell_1 , \psi_1 \big] + \frac{u_1 u_2}{1 + u_1 u_2} \, \big[\ell_2 + \ell_1 - j , \tilde{\psi}_1 \big] - \frac{2}{1+ u_2^2} \left( u_2 ^2 \, \dot{\ell}_1 + u_2 \ast \dot{\ell}_1 \right) .
\end{align}
Making use of the relation 
\begin{align}
\diff \tilde{\psi} = g \Omega \left( g^{-1} \left( \diff \psi + \left[ j , \psi \right] \right) g \right) g^{-1} = \big[ \ell - j , \tilde{\psi} \big] - \dot{\ell} \, ,
\end{align}
one may then show that the defining relation \eqref{eqn:varied-auxiliary-problem2} for $ \master_{u_1} \chi_2$ is solved by
\begin{align}
\master_{u_1} \, \chi_2 = \frac{u_2}{u_1 - u_2} \, \chi_2 \psi_1 + \frac{u_1 u_2}{1 + u_1 u_2} \, \chi_2 \tilde{\psi}_1 - \frac{u_2(1+u_2^2)}{(u_1 - u_2)(1+ u_1 u_2) } \, \chi_2 \psi_2 \, . 
\label{eqn:deltaVchi} 
\end{align}
Note that the boundary condition $ \master_{u_1}  \chi_2 ( z_0 ) = 0$ is automatically satisfied since we have $\psi_i  ( z_0 ) = 0$. These results allow us to compute the commutators \eqref{eqn:comm1}-\eqref{eqn:comm3}. In order to compute the commutator $\left[\delta_1 , \delta_2 \right]$ we note that 
\begin{align*}
\delta_1 \eta_2 = \left[ \eta_2 , \chi_2 ^{-1} \delta_1 \chi_2 \right] = \frac{u_2}{u_1-u_2}  \left( \left[\eta_2 , \eta_1 \right] - \chi_2 ^{-1} \left[\epsilon_2 , \epsilon_1 \right] \chi_2 \right) + \frac{u_1 u_2}{1 + u_1 u_2} \left( \left[\eta_2 , \tilde{\eta}_1 \right] - \chi_2 ^{-1} \left[\epsilon_2 , \epsilon_1 ^\prime \right] \chi_2 \right) \, ,
\end{align*}
such that 
\begin{align}
\left[\delta_1 , \delta_2 \right] g &= \left( \left[\eta_1 , \eta_2 \right] + \delta_2 \eta_1 - \delta_1 \eta_2 \right) g \nn \\
&=  \frac{u_1 \, \delta_{\left[\epsilon_1 , \epsilon_2 \right], u_1}  - u_2 \, \delta_{\left[\epsilon_1 , \epsilon_2 \right], u_2} }{u_1-u_2} \, g + \frac{u_1 u_2 \left( \left[\eta_1 , \tilde{\eta}_2 \right] - \left[\eta_2 , \tilde{\eta}_1 \right] - \delta_{[\epsilon_1 , \epsilon_2 ^\prime ], u_1}  + \delta_{[\epsilon_2 , \epsilon_1 ^\prime ], u_2}\right)}{1+u_1 u_2} \, g 
\end{align}
Noting that
\begin{align}
\left( \left[\eta_1 , \tilde{\eta}_2 \right] - \left[\eta_2 , \tilde{\eta}_1 \right] \right) g = g P_{\alg{h}} \left[ g^{-1} \eta_1 g , \Omega \left(g^{-1} \eta_2 g \right) \right] 
\label{eqn:gaugetrans}
\end{align}
we thus have
\begin{align}
\left[ \delta_{\epsilon_1, u_1} , \delta_{\epsilon_2, u_2} \right] = \frac{1}{u_1-u_2} \left( u_1 \, \delta_{\left[\epsilon_1 , \epsilon_2 \right], u_1} - u_2 \, \delta_{\left[\epsilon_1 , \epsilon_2 \right], u_2} \right) + \frac{u_1 u_2}{1 + u_1 u_2}  \left( \delta_{[\epsilon_2 , \epsilon_1 ^\prime ], u_2} - \delta_{[\epsilon_1 , \epsilon_2 ^\prime ], u_1} \right) \, , 
\end{align}
up to gauge transformations. In order to compute the commutator $\left[ \master_{u_1},  \master_{u_2} \right]$, we employ equation \eqref{eqn:deltaVchi} to find
\begin{align}
\master_{u_1} \psi_2 &= - \chi_2 ^{-1} \big( \master_{u_1} \chi_2 \big) \psi_2 + \chi_2 ^{-1} \frac{\partial}{\partial u_2} \big( \master_{u_1} \chi_2 \big)   \nn \\
&= \frac{u_2}{u_2 - u_1} \big[ \psi_1 , \psi_2 \big] 
- \frac{u_1 u_2}{1 + u_1 u_2} \big[ \tilde{\psi}_1 , \psi_2 \big] 
- \frac{u_2 (1 + u_2 ^2 ) }{(u_1 - u_2 ) (1+ u_1 u_2)} \, \dot{\psi}_2 \nn \\
&  - \left( \frac{\partial}{\partial u_2} \, \frac{u_2}{u_2 - u_1} \right) \psi_1 
+ \left( \frac{\partial}{\partial u_2} \, \frac{u_1 u_2}{1 + u_1 u_2} \right) \tilde{\psi}_1 
- \left( \frac{\partial}{\partial u_2} \, \frac{u_2 (1 + u_2 ^2 ) }{(u_1 - u_2 ) (1+ u_1 u_2)} \right) \psi_2 \, .
\end{align}
Using equations similar to \eqref{eqn:gaugetrans} as well as \eqref{eqn:gaugetrans_comm} and again leaving out gauge transformations we find the commutator
\begin{align}
\Big[ \master _{u_1} , \master _{u_2}  \Big] 
&= 
\sum \limits _{i=1} ^2 \frac{(1 + u_i^2) \left( u_i \, \partial_{u_i} + 1 \right) \master _{u_i}}{(u_1 - u_2) (1 + u_1 u_2)} 
+ \left( \frac{2}{(u_1 - u_2)^2} - \frac{2}{(1+ u_1 u_2)^2}  \right)  \left(u_2 \, \master _{u_2} - u_1 \, \master _{u_1} \right) 
\, .
\end{align}
In order to compute the commutator $\left[ \master_{u_1}, \delta_{\epsilon_2, u_2} \right]$, we use equations \eqref{eqn:deltaVchi} and \eqref{eqn:delta1chi2}, to find the variations
\begin{align}
\master_{u_1} \eta_2 &= 
\frac{u_2}{u_1 - u_2} \, \big[ \eta_2, \psi_1 \big] 
+ \frac{u_1 u_2}{1 + u_1 u_2} \, \big[ \eta_2 , \tilde{\psi}_1 \big] 
- \frac{u_2 (1 + u_2^2)}{(u_1 - u_2) (1 + u_1 u_2)} \, \partial_{u_2} \eta_2 \, , \\
\delta_{\epsilon_2, u_2} \psi_1 &=
\frac{u_1}{u_2 - u_1} \, \big[ \psi_1 , \eta_{\epsilon_2, u_2} \big]
+ \frac{u_1 u_2}{1 + u_1 u_2} \big[ \psi_1 , \tilde{\eta}_{\epsilon_2, u_2}, \big] 
+ \left( \frac{\partial}{\partial u_1}  \, \frac{u_1}{u_2 - u_1} \right) \left( \eta_{\epsilon_2, u_2} - \eta_{\epsilon_2, u_1} \right) \nn \\
& \; + \left( \frac{\partial}{\partial u_1}  \, \frac{u_1 u_2}{1 + u_1 u_2} \right) \left( \tilde{\eta}_{\epsilon_2, u_2} - \eta_{\epsilon ^\prime _2, u_1} \right) \, .
\end{align} 
By making use of identities similar to \eqref{eqn:gaugetrans} and \eqref{eqn:gaugetrans_comm} we then obtain the commutator
\begin{align}
\left[ \master_{u_1} , \delta_{\epsilon, u_2} \right] 
&=  \frac{u_2 \big( \delta_{\epsilon, u_2} - \delta_{\epsilon, u_1} \big)}{(u_1 - u_2 )^2} 
- \frac{u_2 \big( \delta_{\epsilon, u_2} + \delta_{\epsilon^\prime, u_1} \big) }{(1+ u_1 u_2 )^2} 
+ \frac{u_2 (1+ u_2 ^2 ) \, \partial_{u_2} \delta_{\epsilon, u_2}}{( u_1 - u_2 ) ( 1 + u_1 u_2)  }  
\, ,
\end{align}
where we have again not spelled out the gauge transformation.

\section{Master Transformation for Principal Chiral Models}
\label{app:PCM}

In this appendix we describe the master symmetry transformation for principal chiral models. The basic variables of a principal chiral model are matrices $g_p(\tau , \sigma )$ taking values in some representation of a Lie Group $\grp{G}$. We define the left and right currents
\begin{align}
U_p ^r &= g_p ^{-1} \diff g_p \, , & U_p ^l &= - \diff g_p g_p ^{-1} \, .    
\end{align}  
Both currents are flat by construction, $\diff U_p ^{r/l} + U_p ^{r/l} \wedge U_p ^{r/l} = 0$. The classical principal chiral model is defined by the action
\begin{align}
S = \int \tr \big( U_p ^r \wedge \ast U_p ^r \big) = \int \tr \big( U_p ^l \wedge \ast U_p ^l \big) \, .   \label{eqn:PCM_action}
\end{align}
The action is invariant under left- and right-multiplication of $g$ by elements of $\grp{G}$ and the currents $U_p ^r$ and $U_p ^l$ can be identified as the respective Noether currents. The equations of motion for the action \eqref{eqn:PCM_action} are given by
\begin{align}
\diff \ast U_p  ^r = 0 \quad \Leftrightarrow \quad \diff \ast U_p  ^l = 0 \, .
\end{align}
Both currents can be deformed to obtain a Lax connection, which is flat if the equations of motion are satisfied,
\begin{align}
L_{p\, u} ^{r/l} = \frac{u^2}{1+u^2} \,  U_p ^{r/l} + \frac{u}{1+u^2} \, \ast U_p ^{r/l} \, . \label{eqn:Lax_PCM}
\end{align}
For the symmetric space model, the master symmetry transformation is obtained from deforming $g$ to $g_u$ in such a way that the Maurer--Cartan current associated to $g_u$ is the Lax connection. If one defines the master transformation for the principal chiral model in the same way, one does not obtain a symmetry of the action since
\begin{align*}
\tr \big( L_{p\, u} ^r \wedge \ast L_{p\, u} ^r \big) \neq \tr \big( U_p ^r \wedge \ast U_p ^r \big) \, .
\end{align*}
The appropriate definition for the master symmetry transformation for a principal chiral model can be obtained by recasting it as a symmetric space model \cite{Eichenherr:1979ci}. Consider a symmetric space model with Lie group $\grp{G}_s = \grp{G} \times \grp{G}$. The basic variables are matrices $g_s$, which we represent as
\begin{align*}
g_s = \begin{pmatrix}
g_1 & 0 \\ 0 & g_2
\end{pmatrix} \, ,
\end{align*} 
where $g_1$ and $g_2$ take values in G. We introduce an automorphism $\sigma: \grp{G}_s \to \grp{G}_s$ given by
\begin{align}
\sigma \left( g_s \right) &= M g_s M^{-1} = \begin{pmatrix}
g_2 & 0 \\ 0 & g_1 
\end{pmatrix} \, , & 
M &= \begin{pmatrix}
0 & \unit \\ \unit & 0 
\end{pmatrix} \, .
\end{align}
The set of fixed points of $\sigma$ is the diagonal subgroup 
\begin{align*}
\grp{H} = \left \lbrace \begin{pmatrix}
g & 0 \\ 0 & g 
\end{pmatrix} \, : g \in \grp{G} \right \rbrace \, ,
\end{align*}
which appears as the gauge group in this context. We have
\begin{align*}
a_s = \frac{1}{2} \begin{pmatrix}
U_1 - U_2 & 0 \\ 0 & U_2 - U_1 
\end{pmatrix} \, , 
\end{align*}
and correspondingly the Lagrangian of the symmetric space model is given by
\begin{align*}
\mathcal{L}_s = \tr \big( a_s \wedge \ast a_s \big) = \frac{1}{2} \tr \big( \left(U_1 - U_2 \right) \wedge \ast \left(U_1 - U_2 \right) \big) \, .
\end{align*}
We can hence identify the symmetric space model with a principal chiral model by setting
\begin{align}
g_p = g_2 g_1 ^{-1} \, .
\end{align}
This leads to $U_p ^r = g_1 \left( U_2 - U_1 \right) g_1 ^{-1}$ and thus we have
\begin{align}
\mathcal{L}_p = \tr \big( U_p^r \wedge \ast U_p ^r \big) = 2 \, \mathcal{L}_s \, .
\end{align}
For symmetric space models, large master symmetry transformations can be formulated as
\begin{align*}
g_{s , u} = \chi_u \cdot g_s \, , \quad \text{with} \quad  \diff \chi_u  = \chi_u  \ell_u \, .
\end{align*}
Here, $\ell_u$ is the Lax connection in the moving frame,
\begin{align}
\ell_u &= \left( \frac{u^2}{1+u^2} + \frac{u}{1+u^2} \ast \right) \begin{pmatrix}
g_1 \left(U_2 - U_1 \right) g_1 ^{-1} & 0 \\
0 & g_2 \left(U_1 - U_2 \right) g_2 ^{-1} 
\end{pmatrix} 
= \begin{pmatrix}
L^r _p & 0 \\
0 & L^l _p
\end{pmatrix} \, .
\end{align}
Correspondingly we have
\begin{align*}
g_{1 \, u} &= \chi ^r _u \cdot g_1 \, , & g_{2 \, u} &= \chi ^l _u \cdot g_2 \, , 
\end{align*}
where $\chi ^{r/l}_u $ are defined by the auxiliary linear problems
\begin{align*}
\diff \chi ^r_u  &= \chi _u ^r \cdot L^r _{p \, u} \, , & \diff \chi _u ^l &= \chi_u ^l \cdot L^l _{p \, u} \, .
\end{align*}
The master symmetry transformation for the principal chiral model is thus given by
\begin{align}
g_{p \, u} = \chi _u ^l \cdot g_p \cdot \chi ^r_u  {} ^{-1} \, ,
\end{align}
and for the associated variation we have
\begin{align}
\delC \, g_p = \chi ^{l , (0)} \cdot g_p - g_p \cdot \chi ^{r , (0)} \, ,
\end{align}
where $\chi ^{r/l , (0)}$ are the potentials for the left and right Noether currents, 
\begin{align}
\chi ^{r/l , (0)} = \int \ast U_p ^{r/l} \, .  
\end{align} 

%%%%%%%%%%%%%%%%%%%%%%%%%%%%%%%%%%%%%%%%%%%%%%%%%%%%%%%%%%%%%%%%%%%%%%%%%%%
%%%%%%%%%%%%%%%%%%%%%%%%%%%%%%%%%%%%%%%%%%%%%%%%%%%%%%%%%%%%%%%%%%%%%%%%%%%
\bibliographystyle{nb}
\bibliography{Boost}

%bibliography generated by nb.bst v1.06 (C) 2003-2011 Niklas Beisert
\begin{thebibliography}{10}
\providecommand{\href}[2]{#2}
\providecommand{\arxivref}[2]{\href{http://arxiv.org/abs/#1}{#2}}
\providecommand{\doiref}[2]{\href{http://dx.doi.org/#1}{#2}}
\providecommand{\nbbstauthor}[1]{#1}
\providecommand{\nbbstjournal}[1]{\textsf{#1}}
\providecommand{\nbbsttitle}[1]{\textit{#1}}
\providecommand{\nbbsturl}[1]{\texttt{#1}}
\providecommand{\nbbsteprint}[1]{\texttt{#1}}
\providecommand{\nbbststyle}{\raggedright\small\parskip0pt}
\nbbststyle

\bibitem{Maldacena:1997re}
\nbbstauthor{J.~M.~Maldacena},
\nbbsttitle{``{The Large N limit of superconformal field theories and
  supergravity}''},
\nbbstjournal{\doiref{10.1023/A:1026654312961}{Int.~J.~Theor.~Phys.~38,~1113~(1999)}},
\nbbsteprint{\arxivref{hep-th/9711200}{hep-th/9711200}},
[Adv. Theor. Math. Phys.2,231(1998)].
%\%CITATION = HEP-TH/9711200;\%\%

\bibitem{Beisert:2010kp}
\nbbstauthor{N.~Beisert},
\nbbsttitle{``{Review of AdS/CFT Integrability, Chapter VI.1: Superconformal
  Symmetry}''},
\nbbstjournal{\doiref{10.1007/s11005-011-0479-8}{Lett.~Math.~Phys.~99,~529~(2012)}},
\nbbsteprint{\arxivref{1012.4004}{arxiv:1012.4004}}.
%\%CITATION = ARXIV:1012.4004;\%\%

\bibitem{Beisert:2010jq}
\nbbstauthor{N.~Beisert},
\nbbsttitle{``{On Yangian Symmetry in Planar N=4 SYM}''},
\nbbsteprint{\arxivref{1004.5423}{arxiv:1004.5423}},
in: \nbbsttitle{``{Quantum chromodynamics and beyond: Gribov-80 memorial
  volume. Proceedings, Memorial Workshop devoted to the 80th birthday of V.N.
  Gribov, Trieste, Italy, May 26-28, 2010}''},
pp.~175--203,
\href{https://inspirehep.net/record/853644/files/arXiv:1004.5423.pdf}{\nbbsturl{https://inspirehep.net/record/853644/files/arXiv:1004.5423.pdf}}.
%\%CITATION = ARXIV:1004.5423;\%\%

\bibitem{Torrielli:2011gg}
\nbbstauthor{A.~Torrielli},
\nbbsttitle{``{Yangians, S-matrices and AdS/CFT}''},
\nbbstjournal{\doiref{10.1088/1751-8113/44/26/263001}{J.~Phys.~A44,~263001~(2011)}},
\nbbsteprint{\arxivref{1104.2474}{arxiv:1104.2474}}.
%\%CITATION = ARXIV:1104.2474;\%\%

\bibitem{Spill:2012qe}
\nbbstauthor{F.~Spill},
\nbbsttitle{``{Yangians in Integrable Field Theories, Spin Chains and
  Gauge-String Dualities}''},
\nbbstjournal{\doiref{10.1142/S0129055X12300014}{Rev.~Math.~Phys.~24,~1230001~(2012)}},
\nbbsteprint{\arxivref{1201.1884}{arxiv:1201.1884}},
[Rev. Math. Phys.24,0001(2012)].
%\%CITATION = ARXIV:1201.1884;\%\%

\bibitem{Loebbert:2016cdm}
\nbbstauthor{F.~Loebbert},
\nbbsttitle{``{Lectures on Yangian Symmetry}''},
\nbbstjournal{\doiref{10.1088/1751-8113/49/32/323002}{J.~Phys.~A49,~323002~(2016)}},
\nbbsteprint{\arxivref{1606.02947}{arxiv:1606.02947}}.
%\%CITATION = ARXIV:1606.02947;\%\%

\bibitem{Drummond:2008vq}
\nbbstauthor{J.~M.~Drummond, J.~Henn, G.~P.~Korchemsky and E.~Sokatchev},
\nbbsttitle{``{Dual superconformal symmetry of scattering amplitudes in N=4
  super-Yang-Mills theory}''},
\nbbstjournal{\doiref{10.1016/j.nuclphysb.2009.11.022}{Nucl.~Phys.~B828,~317~(2010)}},
\nbbsteprint{\arxivref{0807.1095}{arxiv:0807.1095}}.
%%CITATION = ARXIV:0807.1095;%%

\bibitem{Drummond:2010km}
\nbbstauthor{J.~M.~Drummond},
\nbbsttitle{``{Review of AdS/CFT Integrability, Chapter V.2: Dual
  Superconformal Symmetry}''},
\nbbstjournal{\doiref{10.1007/s11005-011-0519-4}{Lett.~Math.~Phys.~99,~481~(2012)}},
\nbbsteprint{\arxivref{1012.4002}{arxiv:1012.4002}}.
%\%CITATION = ARXIV:1012.4002;\%\%

\bibitem{Alday:2007hr}
\nbbstauthor{L.~F.~Alday and J.~M.~Maldacena},
\nbbsttitle{``{Gluon scattering amplitudes at strong coupling}''},
\nbbstjournal{\doiref{10.1088/1126-6708/2007/06/064}{JHEP~0706,~064~(2007)}},
\nbbsteprint{\arxivref{0705.0303}{arxiv:0705.0303}}.
%\%CITATION = ARXIV:0705.0303;\%\%

\bibitem{Brandhuber:2007yx}
\nbbstauthor{A.~Brandhuber, P.~Heslop and G.~Travaglini},
\nbbsttitle{``{MHV amplitudes in N=4 super Yang-Mills and Wilson loops}''},
\nbbstjournal{\doiref{10.1016/j.nuclphysb.2007.11.002}{Nucl.~Phys.~B794,~231~(2008)}},
\nbbsteprint{\arxivref{0707.1153}{arxiv:0707.1153}}.
%%CITATION = ARXIV:0707.1153;%%

\bibitem{Drummond:2008aq}
\nbbstauthor{J.~M.~Drummond, J.~Henn, G.~P.~Korchemsky and E.~Sokatchev},
\nbbsttitle{``{Hexagon Wilson loop = six-gluon MHV amplitude}''},
\nbbstjournal{\doiref{10.1016/j.nuclphysb.2009.02.015}{Nucl.~Phys.~B815,~142~(2009)}},
\nbbsteprint{\arxivref{0803.1466}{arxiv:0803.1466}}.
%\%CITATION = ARXIV:0803.1466;\%\%

\bibitem{Maldacena:1998im}
\nbbstauthor{J.~M.~Maldacena},
\nbbsttitle{``{Wilson loops in large N field theories}''},
\nbbstjournal{\doiref{10.1103/PhysRevLett.80.4859}{Phys.~Rev.~Lett.~80,~4859~(1998)}},
\nbbsteprint{\arxivref{hep-th/9803002}{hep-th/9803002}}.
%\%CITATION = HEP-TH/9803002;\%\%

\bibitem{Rey:1998ik}
\nbbstauthor{S.-J.~Rey and J.-T.~Yee},
\nbbsttitle{``{Macroscopic strings as heavy quarks in large N gauge theory and
  anti-de Sitter supergravity}''},
\nbbstjournal{\doiref{10.1007/s100520100799}{Eur.~Phys.~J.~C22,~379~(2001)}},
\nbbsteprint{\arxivref{hep-th/9803001}{hep-th/9803001}}.
%\%CITATION = HEP-TH/9803001;\%\%

\bibitem{Muller:2013rta}
\nbbstauthor{D.~M{\"u}ller, H.~M{\"u}nkler, J.~Plef\-ka, J.~Pollok and
  K.~Za\-rembo},
\nbbsttitle{``{Yangian Symmetry of smooth Wilson Loops in $\mathcal{N} = $ 4
  super Yang-Mills Theory}''},
\nbbstjournal{\doiref{10.1007/JHEP11(2013)081}{JHEP~1311,~081~(2013)}},
\nbbsteprint{\arxivref{1309.1676}{arxiv:1309.1676}}.
%\%CITATION = ARXIV:1309.1676;\%\%

\bibitem{Munkler:2015gja}
\nbbstauthor{H.~M{\"u}nkler and J.~Pollok},
\nbbsttitle{``{Minimal surfaces of the ${{AdS}}_{5}\times {S}^{5}$ superstring
  and the symmetries of super Wilson loops at strong coupling}''},
\nbbstjournal{\doiref{10.1088/1751-8113/48/36/365402}{J.~Phys.~A48,~365402~(2015)}},
\nbbsteprint{\arxivref{1503.07553}{arxiv:1503.07553}}.
%\%CITATION = ARXIV:1503.07553;\%\%

\bibitem{Alday:2009dv}
\nbbstauthor{L.~F.~Alday, D.~Gaiotto and J.~Maldacena},
\nbbsttitle{``{Thermodynamic Bubble Ansatz}''},
\nbbstjournal{\doiref{10.1007/JHEP09(2011)032}{JHEP~1109,~032~(2011)}},
\nbbsteprint{\arxivref{0911.4708}{arxiv:0911.4708}}.
%\%CITATION = ARXIV:0911.4708;\%\%

\bibitem{Alday:2010vh}
\nbbstauthor{L.~F.~Alday, J.~Maldacena, A.~Sever and P.~Vieira},
\nbbsttitle{``{Y-system for Scattering Amplitudes}''},
\nbbstjournal{\doiref{10.1088/1751-8113/43/48/485401}{J.~Phys.~A43,~485401~(2010)}},
\nbbsteprint{\arxivref{1002.2459}{arxiv:1002.2459}}.
%\%CITATION = ARXIV:1002.2459;\%\%

\bibitem{Alday:2010ku}
\nbbstauthor{L.~F.~Alday, D.~Gaiotto, J.~Maldacena, A.~Sever and P.~Vieira},
\nbbsttitle{``{An Operator Product Expansion for Polygonal null Wilson
  Loops}''},
\nbbstjournal{\doiref{10.1007/JHEP04(2011)088}{JHEP~1104,~088~(2011)}},
\nbbsteprint{\arxivref{1006.2788}{arxiv:1006.2788}}.
%\%CITATION = ARXIV:1006.2788;\%\%

\bibitem{Drukker:2012de}
\nbbstauthor{N.~Drukker},
\nbbsttitle{``{Integrable Wilson loops}''},
\nbbstjournal{\doiref{10.1007/JHEP10(2013)135}{JHEP~1310,~135~(2013)}},
\nbbsteprint{\arxivref{1203.1617}{arxiv:1203.1617}}.
%\%CITATION = ARXIV:1203.1617;\%\%

\bibitem{Basso:2013vsa}
\nbbstauthor{B.~Basso, A.~Sever and P.~Vieira},
\nbbsttitle{``{Spacetime and Flux Tube S-Matrices at Finite Coupling for N=4
  Supersymmetric Yang-Mills Theory}''},
\nbbstjournal{\doiref{10.1103/PhysRevLett.111.091602}{Phys.~Rev.~Lett.~111,~091602~(2013)}},
\nbbsteprint{\arxivref{1303.1396}{arxiv:1303.1396}}.
%\%CITATION = ARXIV:1303.1396;\%\%

\bibitem{Toledo:2014koa}
\nbbstauthor{J.~C.~Toledo},
\nbbsttitle{``{Smooth Wilson loops from the continuum limit of null
  polygons}''},
\nbbsteprint{\arxivref{1410.5896}{arxiv:1410.5896}}.
%%CITATION = ARXIV:1410.5896;%%

\bibitem{Klose:2016uur}
\nbbstauthor{T.~Klose, F.~Loebbert and H.~Munkler},
\nbbsttitle{``{Master Symmetry for Holographic Wilson Loops}''},
\nbbstjournal{\doiref{10.1103/PhysRevD.94.066006}{Phys.~Rev.~D94,~066006~(2016)}},
\nbbsteprint{\arxivref{1606.04104}{arxiv:1606.04104}}.
%\%CITATION = ARXIV:1606.04104;\%\%

\bibitem{Ishizeki:2011bf}
\nbbstauthor{R.~Ishizeki, M.~Kruczenski and S.~Ziama},
\nbbsttitle{``{Notes on Euclidean Wilson loops and Riemann Theta functions}''},
\nbbstjournal{\doiref{10.1103/PhysRevD.85.106004}{Phys.~Rev.~D85,~106004~(2012)}},
\nbbsteprint{\arxivref{1104.3567}{arxiv:1104.3567}}.
%\%CITATION = ARXIV:1104.3567;\%\%

\bibitem{Kruczenski:2013bsa}
\nbbstauthor{M.~Kruczenski and S.~Ziama},
\nbbsttitle{``{Wilson loops and Riemann theta functions II}''},
\nbbstjournal{\doiref{10.1007/JHEP05(2014)037}{JHEP~1405,~037~(2014)}},
\nbbsteprint{\arxivref{1311.4950}{arxiv:1311.4950}}.
%\%CITATION = ARXIV:1311.4950;\%\%

\bibitem{Kruczenski:2014bla}
\nbbstauthor{M.~Kruczenski},
\nbbsttitle{``{Wilson loops and minimal area surfaces in hyperbolic space}''},
\nbbstjournal{\doiref{10.1007/JHEP11(2014)065}{JHEP~1411,~065~(2014)}},
\nbbsteprint{\arxivref{1406.4945}{arxiv:1406.4945}}.
%\%CITATION = ARXIV:1406.4945;\%\%

\bibitem{Cooke:2014uga}
\nbbstauthor{M.~Cooke and N.~Drukker},
\nbbsttitle{``{From algebraic curve to minimal surface and back}''},
\nbbstjournal{\doiref{10.1007/JHEP02(2015)090}{JHEP~1502,~090~(2015)}},
\nbbsteprint{\arxivref{1410.5436}{arxiv:1410.5436}}.
%\%CITATION = ARXIV:1410.5436;\%\%

\bibitem{Dekel:2015bla}
\nbbstauthor{A.~Dekel},
\nbbsttitle{``{Wilson Loops and Minimal Surfaces Beyond the Wavy
  Approximation}''},
\nbbstjournal{\doiref{10.1007/JHEP03(2015)085}{JHEP~1503,~085~(2015)}},
\nbbsteprint{\arxivref{1501.04202}{arxiv:1501.04202}}.
%\%CITATION = ARXIV:1501.04202;\%\%

\bibitem{Huang:2016atz}
\nbbstauthor{C.~Huang, Y.~He and M.~Kruczenski},
\nbbsttitle{``{Minimal area surfaces dual to Wilson loops and the Mathieu
  equation}''},
\nbbsteprint{\arxivref{1604.00078}{arxiv:1604.00078}}.
%\%CITATION = ARXIV:1604.00078;\%\%

\bibitem{Brezin:1979am}
\nbbstauthor{E.~Brezin, C.~Itzykson, J.~Zinn-Justin and J.~B.~Zuber},
\nbbsttitle{``{Remarks About the Existence of Nonlocal Charges in
  Two-Dimensional Models}''},
\nbbstjournal{\doiref{10.1016/0370-2693(79)90263-6}{Phys.~Lett.~B82,~442~(1979)}}.
%\%CITATION = PHLTA,B82,442;\%\%

\bibitem{Eichenherr:1979ci}
\nbbstauthor{H.~Eichenherr and M.~Forger},
\nbbsttitle{``{On the Dual Symmetry of the Nonlinear Sigma Models}''},
\nbbstjournal{\doiref{10.1016/0550-3213(79)90276-1}{Nucl.~Phys.~B155,~381~(1979)}}.
%\%CITATION = NUPHA,B155,381;\%\%

\bibitem{Schwarz:1995td}
\nbbstauthor{J.~H.~Schwarz},
\nbbsttitle{``{Classical symmetries of some two-dimensional models}''},
\nbbstjournal{\doiref{10.1016/0550-3213(95)00276-X}{Nucl.~Phys.~B447,~137~(1995)}},
\nbbsteprint{\arxivref{hep-th/9503078}{hep-th/9503078}}.
%\%CITATION = HEP-TH/9503078;\%\%

\bibitem{Schwarz:1995af}
\nbbstauthor{J.~H.~Schwarz},
\nbbsttitle{``{Classical symmetries of some two-dimensional models coupled to
  gravity}''},
\nbbstjournal{\doiref{10.1016/0550-3213(95)00455-2}{Nucl.~Phys.~B454,~427~(1995)}},
\nbbsteprint{\arxivref{hep-th/9506076}{hep-th/9506076}}.
%\%CITATION = HEP-TH/9506076;\%\%

\bibitem{Beisert:2012ue}
\nbbstauthor{N.~Beisert and F.~Luecker},
\nbbsttitle{``{Construction of Lax Connections by Exponentiation}''},
\nbbstjournal{\doiref{10.1063/1.4769824}{J.~Math.~Phys.~53,~122304~(2012)}},
\nbbsteprint{\arxivref{1207.3325}{arxiv:1207.3325}}.
%\%CITATION = ARXIV:1207.3325;\%\%

\bibitem{Wu:1982jt}
\nbbstauthor{Y.-S.~Wu},
\nbbsttitle{``{Extension of the Hidden Symmetry Algebra in Classical Principal
  Chiral Models}''},
\nbbstjournal{\doiref{10.1016/0550-3213(83)90190-6}{Nucl.~Phys.~B211,~160~(1983)}}.
%\%CITATION = NUPHA,B211,160;\%\%

\bibitem{Dolan:1980kz}
\nbbstauthor{L.~Dolan and A.~Roos},
\nbbsttitle{``{Nonlocal Currents as Noether Currents}''},
\nbbstjournal{\doiref{10.1103/PhysRevD.22.2018}{Phys.~Rev.~D22,~2018~(1980)}}.
%\%CITATION = PHRVA,D22,2018;\%\%

\bibitem{Hou:1981hn}
\nbbstauthor{B.-y.~Hou, M.-l.~Ge and Y.-s.~Wu},
\nbbsttitle{``{Noether Analysis for the Hidden Symmetry Responsible for
  Infinite Set of Nonlocal Currents}''},
\nbbstjournal{\doiref{10.1103/PhysRevD.24.2238}{Phys.~Rev.~D24,~2238~(1981)}}.
%\%CITATION = PHRVA,D24,2238;\%\%

\bibitem{Luscher:1977rq}
\nbbstauthor{M.~Luscher and K.~Pohlmeyer},
\nbbsttitle{``{Scattering of Massless Lumps and Nonlocal Charges in the
  Two-Dimensional Classical Nonlinear Sigma Model}''},
\nbbstjournal{\doiref{10.1016/0550-3213(78)90049-4}{Nucl.~Phys.~B137,~46~(1978)}}.
%\%CITATION = NUPHA,B137,46;\%\%

\bibitem{MacKay:1992he}
\nbbstauthor{N.~J.~MacKay},
\nbbsttitle{``{On the classical origins of Yangian symmetry in integrable field
  theory}''},
\nbbstjournal{\doiref{10.1016/0370-2693(92)90280-H}{Phys.~Lett.~B281,~90~(1992)}},
[Erratum: Phys. Lett.B308,444(1993)].
%\%CITATION = PHLTA,B281,90;\%\%

\bibitem{Forger:1991cm}
\nbbstauthor{M.~Forger, J.~Laartz and U.~Schaper},
\nbbsttitle{``{Current algebra of classical nonlinear sigma models}''},
\nbbstjournal{\doiref{10.1007/BF02102634}{Commun.~Math.~Phys.~146,~397~(1992)}},
\nbbsteprint{\arxivref{hep-th/9201025}{hep-th/9201025}}.
%\%CITATION = HEP-TH/9201025;\%\%

\bibitem{Dorn:2015bfa}
\nbbstauthor{H.~Dorn},
\nbbsttitle{``{Wilson loops at strong coupling for curved contours with
  cusps}''},
\nbbstjournal{\doiref{10.1088/1751-8113/49/14/145402}{J.~Phys.~A49,~145402~(2016)}},
\nbbsteprint{\arxivref{1509.00222}{arxiv:1509.00222}}.
%%CITATION = ARXIV:1509.00222;%%

\bibitem{Polyakov:2000ti}
\nbbstauthor{A.~M.~Polyakov and V.~S.~Rychkov},
\nbbsttitle{``{Gauge field strings duality and the loop equation}''},
\nbbstjournal{\doiref{10.1016/S0550-3213(00)00183-8}{Nucl.~Phys.~B581,~116~(2000)}},
\nbbsteprint{\arxivref{hep-th/0002106}{hep-th/0002106}}.
%\%CITATION = HEP-TH/0002106;\%\%

\bibitem{Drummond:2009fd}
\nbbstauthor{J.~M.~Drummond, J.~M.~Henn and J.~Plefka},
\nbbsttitle{``{Yangian symmetry of scattering amplitudes in N=4 super
  Yang-Mills theory}''},
\nbbstjournal{\doiref{10.1088/1126-6708/2009/05/046}{JHEP~0905,~046~(2009)}},
\nbbsteprint{\arxivref{0902.2987}{arxiv:0902.2987}},
in: \nbbsttitle{``{Strangeness in quark matter. Proceedings, International
  Conference, SQM 2008, Beijing, P.R. China, October 5-10, 2008}''},
pp.~046.
%\%CITATION = ARXIV:0902.2987;\%\%

\bibitem{Beisert:2015jxa}
\nbbstauthor{N.~Beisert, D.~M{\"u}ller, J.~Plefka and C.~Vergu},
\nbbsttitle{``{Smooth Wilson loops in $ \mathcal{N}=4 $ non-chiral
  superspace}''},
\nbbstjournal{\doiref{10.1007/JHEP12(2015)140}{JHEP~1512,~140~(2015)}},
\nbbsteprint{\arxivref{1506.07047}{arxiv:1506.07047}}.
%\%CITATION = ARXIV:1506.07047;\%\%

\bibitem{Beisert:2015uda}
\nbbstauthor{N.~Beisert, D.~M{\"u}ller, J.~Plefka and C.~Vergu},
\nbbsttitle{``{Integrability of smooth Wilson loops in $ \mathcal{N}=4 $
  superspace}''},
\nbbstjournal{\doiref{10.1007/JHEP12(2015)141}{JHEP~1512,~141~(2015)}},
\nbbsteprint{\arxivref{1509.05403}{arxiv:1509.05403}}.
%\%CITATION = ARXIV:1509.05403;\%\%

\bibitem{brakke1992}
\nbbstauthor{K.~A.~Brakke},
\nbbsttitle{``{The surface evolver}''},
\nbbstjournal{Experiment.~Math.~1,~141~(1992)},
\href{http://projecteuclid.org/euclid.em/1048709050}{\nbbsturl{http://projecteuclid.org/euclid.em/1048709050}}.

\bibitem{Ferro:2012xw}
\nbbstauthor{L.~Ferro, T.~Lukowski, C.~Meneghelli, J.~Plefka and
  M.~Staudacher},
\nbbsttitle{``{Harmonic R-matrices for Scattering Amplitudes and Spectral
  Regularization}''},
\nbbstjournal{\doiref{10.1103/PhysRevLett.110.121602}{Phys.~Rev.~Lett.~110,~121602~(2013)}},
\nbbsteprint{\arxivref{1212.0850}{arxiv:1212.0850}}.
%\%CITATION = ARXIV:1212.0850;\%\%

\bibitem{Berkovits:2008ic}
\nbbstauthor{N.~Berkovits and J.~Maldacena},
\nbbsttitle{``{Fermionic T-Duality, Dual Superconformal Symmetry, and the
  Amplitude/Wilson Loop Connection}''},
\nbbstjournal{\doiref{10.1088/1126-6708/2008/09/062}{JHEP~0809,~062~(2008)}},
\nbbsteprint{\arxivref{0807.3196}{arxiv:0807.3196}}.
%\%CITATION = ARXIV:0807.3196;\%\%

\bibitem{Beisert:2008iq}
\nbbstauthor{N.~Beisert, R.~Ricci, A.~A.~Tseytlin and M.~Wolf},
\nbbsttitle{``{Dual Superconformal Symmetry from AdS(5) x S**5 Superstring
  Integrability}''},
\nbbstjournal{\doiref{10.1103/PhysRevD.78.126004}{Phys.~Rev.~D78,~126004~(2008)}},
\nbbsteprint{\arxivref{0807.3228}{arxiv:0807.3228}}.
%\%CITATION = ARXIV:0807.3228;\%\%

\bibitem{Chandia:2016ueo}
\nbbstauthor{O.~Chandia, W.~D.~Linch and B.~C.~Vallilo},
\nbbsttitle{``{Master symmetry in the $AdS_5\times S^5$ pure spinor string}''},
\nbbsteprint{\arxivref{1607.00391}{arxiv:1607.00391}}.
%\%CITATION = ARXIV:1607.00391;\%\%

\bibitem{Irrgang:2015txa}
\nbbstauthor{A.~Irrgang and M.~Kruczenski},
\nbbsttitle{``{Euclidean Wilson loops and Minimal Area Surfaces in Minkowski
  $AdS_3$}''},
\nbbsteprint{\arxivref{1507.02787}{arxiv:1507.02787}}.
%\%CITATION = ARXIV:1507.02787;\%\%

\bibitem{Matsumoto:2007rh}
\nbbstauthor{T.~Matsumoto, S.~Moriyama and A.~Torrielli},
\nbbsttitle{``{A Secret Symmetry of the AdS/CFT S-matrix}''},
\nbbstjournal{\doiref{10.1088/1126-6708/2007/09/099}{JHEP~0709,~099~(2007)}},
\nbbsteprint{\arxivref{0708.1285}{arxiv:0708.1285}}.
%%CITATION = ARXIV:0708.1285;%%

\bibitem{Beisert:2011pn}
\nbbstauthor{N.~Beisert and B.~U.~W.~Schwab},
\nbbsttitle{``{Bonus Yangian Symmetry for the Planar S-Matrix of N=4 Super
  Yang-Mills}''},
\nbbstjournal{\doiref{10.1103/PhysRevLett.106.231602}{Phys.~Rev.~Lett.~106,~231602~(2011)}},
\nbbsteprint{\arxivref{1103.0646}{arxiv:1103.0646}}.
%%CITATION = ARXIV:1103.0646;%%

\bibitem{Munkler:2015xqa}
\nbbstauthor{H.~Munkler},
\nbbsttitle{``{Bonus Symmetry for Super Wilson Loops}''},
\nbbstjournal{\doiref{10.1088/1751-8113/49/18/185401}{J.~Phys.~A49,~185401~(2016)}},
\nbbsteprint{\arxivref{1507.02474}{arxiv:1507.02474}}.
%%CITATION = ARXIV:1507.02474;%%

\bibitem{Adam:2010hh}
\nbbstauthor{I.~Adam, A.~Dekel and Y.~Oz},
\nbbsttitle{``{On the fermionic T-duality of the $AdS_4 x CP^3$
  sigma-model}''},
\nbbstjournal{\doiref{10.1007/JHEP10(2010)110}{JHEP~1010,~110~(2010)}},
\nbbsteprint{\arxivref{1008.0649}{arxiv:1008.0649}}.
%%CITATION = ARXIV:1008.0649;%%

\bibitem{Colgain:2016gdj}
\nbbstauthor{E.~Colg\'ain and A.~Pittelli},
\nbbsttitle{``{Comments on the origin of dual superconformal symmetry in ABJM
  theory}''},
\nbbsteprint{\arxivref{1609.03254}{arxiv:1609.03254}}.
%%CITATION = ARXIV:1609.03254;%%

\bibitem{Fontanella:2016opq}
\nbbstauthor{A.~Fontanella and A.~Torrielli},
\nbbsttitle{``{Massless sector of $\mathrm{AdS}_3$ superstrings: a geometric
  interpretation}''},
\nbbsteprint{\arxivref{1608.01631}{arxiv:1608.01631}}.
%\%CITATION = ARXIV:1608.01631;\%\%

\bibitem{Kruczenski:2003gt}
\nbbstauthor{M.~Kruczenski},
\nbbsttitle{``{Spin chains and string theory}''},
\nbbstjournal{\doiref{10.1103/PhysRevLett.93.161602}{Phys.~Rev.~Lett.~93,~161602~(2004)}},
\nbbsteprint{\arxivref{hep-th/0311203}{hep-th/0311203}}.
%\%CITATION = HEP-TH/0311203;\%\%

\end{thebibliography}

\end{document}